\documentclass{IEEEtran}
\IEEEoverridecommandlockouts

\usepackage{graphicx}
\usepackage{makeidx}  
\usepackage{acronym}
\usepackage{subfigure}
\usepackage{booktabs}
\usepackage{algorithm}
\usepackage{algpseudocode}
\usepackage{url}
\usepackage{amstext}
\usepackage{amsmath}

\acrodef{P2P}{Peer-to-Peer}
\acrodef{TTL}{Time-To-Live}
\acrodef{ECC}{Edge Clustering Coefficient}

\begin{document}

\title{Self-Healing Protocols for Connectivity Maintenance in Unstructured Overlays
}

\author{\IEEEauthorblockN{Stefano Ferretti}
\IEEEauthorblockA{Department of Computer Science and Engineering, University of Bologna\\
Mura Anteo Zamboni 7, Bologna, Italy\\
s.ferretti@unibo.it}
}

\maketitle

\begin{abstract}
In this paper, we discuss on the use of self-organizing protocols to improve the 
reliability of dynamic Peer-to-Peer (P2P) overlay networks. Two similar 
approaches are studied, which are based on local knowledge of the nodes' $2$nd 
neighborhood. The first scheme is a simple protocol requiring interactions among 
nodes and their direct neighbors. The second scheme adds a check on the Edge 
Clustering Coefficient (ECC), a local measure that allows determining edges 
connecting different clusters in the network. The performed simulation 
assessment evaluates these protocols over uniform networks, clustered networks 
and scale-free networks. Different failure modes are considered. Results 
demonstrate the effectiveness of the proposal.
\end{abstract}

\section{Introduction}

A significant part of the research in \ac{P2P} systems of the last years has been in the design of overlay
networks. Overlay networks operate at the application layer, on top of the traditional Internet transport protocols.
Each node in the overlay is a peer that has an unique ``id''. Messages are routed to a node based on that application level id and through the overlay links, rather than on a communication based on IP addresses.

A main outcome of these studies was the introduction of P2P structured architectures. 
In essence, these are architectural solutions where links among nodes are created based on the contents hold by nodes. Distributed Hash Tables (DHTs) are peculiar examples of these systems \cite{Basu05nodewiz,Cai03maan,Hidalgo:2011}. 

Conversely, unstructured P2P overlays represent networks where links a\-mong nodes are established arbitrarily. 
Peers locally manage their connections to build some general desired topology and links do not depend on the contents being disseminated \cite{simplex}. They are particularly simple to build and manage, with little maintenance costs, yet at the price of a non-optimal organization of the overlay. 
Unstructured overlays can be used as a building block in a variety of distributed applications, especially when the environment, where the application is run, is highly dynamic.
Examples are concerned with system monitoring \cite{gems}, failure detection \cite{vanRenesse:2009}, messaging, resource discovery \cite{Costa:2003,Ferretti-fgcs,Melliar-Smith:2012,SimontonCS06}, management of flash crowd crises over gossip-based information dissemination \cite{Baraglia:2013,simplex}.
The use of unstructured overlays enables scalable and efficient solutions that obviate the need for a structure \cite{Melliar-Smith:2012,Terpstra:2007,Wong:2008}.

Unstructured P2P systems aim at exploiting randomness to disseminate information across a
large set of nodes. A key issue is to keep the overlay connected even in the event
of major disasters, without maintaining any global information or requiring any
sort of administration. Connections between nodes in these systems are highly
dynamic.

This work focuses on a decentralized self-healing algorithm that aims providing resilience of unstructured overlay networks.
The approach exploits local knowledge that each node has about its neighborhood, i.e., nodes that are linked to it in the overlay.
In particular, each node $n$ maintains and actively manages the list of nodes directly connected to it (i.e.~its neighbors), and the neighbors of its neighbors (the so called $2$nd neighbors). 
In a network overlay the failure of a neighbor can disrupt, or at least worsen, the communication capabilities of a node with the rest of the network.
To avoid this, the node $n$ reacts to these failures by running a self-healing procedure, so as to get back those connections with $2$nd neighbors which were lost.
A contention among $n$ and its $1$st neighbors is performed to replace the lost 
connection. Thus, only one among these nodes creates such a link; this way, 
nodes share the load for the creation and management of these novel 
links \cite{simplex13}. 

Together with this basic self-healing protocol, a variation is proposed that exploits the notion of \ac{ECC} \cite{radicchi2004}.
This metrics is a local measure that identifies those edges connecting different clusters. In fact, the ECC associated to a link counts the number of triangles it belongs, with respect to the number of triangles that might potentially include it.
The lower the ECC of a link the lower the short paths connecting the two nodes that share that link (since they are in few common triangles).
Since many triangles exist within clusters, ECC is a measure of how inter-communitarian a link is.

Based on this ECC, a second version of the protocol is presented, according to which a node $n$ decides to activate the self-healing procedure with a probability which is inversely proportional to the ECC of the link lost upon a neighbor failure. In other words, the more the link was part of triangles, the lower the probability of triggering the recovery procedure. 
The recovery procedure consists in creating links with the lost $2$nd neighbors, as described above.
The idea is that in this case, a node might avoid to activate the self-healing procedure for those lost links with higher ECC values.
Not only, with the aims of preserving the network topology and of limiting the potential growth on the number of links in the network, a link removal phase is included in the protocol. Basically, it removes (with a certain probability) links with higher ECC values, associated to nodes with a degree exceeding their target degree.

A simulation assessment is presented that studies the protocols over uniform networks (where links are created by randomly choosing nodes as neighbors), clustered networks and scale-free networks. 
These different network topologies are exemplars that model different P2P 
systems. 
Uniforms networks (with links created as random graphs) resemble typical data 
sharing P2P systems, where usually peers connect to a almost static (and quite 
often pre-configured) amount of peers, to share data with. This number of 
neighbors is a trade-off to avoid, on one hand, that a low number of 
connections limits the sharing capabilities, and on the other hand, that a too 
high amount of neighbors causes an unbearable communication and computation 
overhead for a peer.
Clustered networks allow to consider those situations where there are clusters 
of nodes that share several connections while there are fewer connections among 
different clusters. This is a typical situation in social networks and the 
like.
Scale-free networks are considered the main network topology that models most 
real networks \cite{simutools,newmanHandbook}. For example, it has been recognized 
that the well-known Gnutella overlay is a scale-free network. 
Moreover, there is evidence that the overlay created in Skype has several hubs (i.e.~nodes 
with many connections much higher than the majority of other nodes), suggesting 
that this type of network is a scale-free \cite{baset}.

Different types of simulations are considered with different types of node 
removals. 
The first mode was based on a random selection of nodes that fail, in a 
situation where the amount of failed nodes is equal to the amount of joining 
nodes.
Second, a ``targeted attack'' was simulated, meaning that at each step 
the ``important'' nodes with some specific characteristics were selected to 
fail. In particular, as concerns uniform and scale-free networks, nodes with the 
higher degrees were selected to fail. Instead, in clustered networks the 
selected nodes were those with higher number of links connecting different 
clusters (the rationale was to augment the probability of disconnecting the 
clusters).
A variation of the targeted attack is considered, where removed nodes are those with the highest betweenness centrality value.
Finally, another mode was set where only failures occurred. 

Results demonstrate that the presented self-healing approaches preserve networks 
connectivity, coping with node churn and targeted attacks. 
Moreover, the use of the ECC can lower the clustering coefficient on the 
overlay (depending on its topology).

The remainder of this paper is organized as follows. Section \ref{sec:related} 
discusses on some background and related studies available in the literature. 
Section \ref{sec:prot} presents the P2P protocol. Section \ref{sec:eval} 
describes the simulation environment, while Section \ref{sec:res} discusses the obtained results. 
Finally, Section \ref{sec:conc} provides some concluding remarks.

\section{Related Work}\label{sec:related}

Several works have been presented in the literature, which focus on 
self-organization of P2P systems and their robustness to failures and node 
departures.
One of the most fascinating aspects of the presented distributed approaches is 
that peers can execute local strategies in order to maintain some global 
properties of the overall network through decentralized interactions. These 
global properties are usually referred as self-* properties (e.g., 
self-organization, self-adaptation, self-management). 
Peers might interact in order to self-organize the contents they maintain 
(e.g., \cite{forestiero,Giordanelli:2012}), or even the connections each peer 
maintains with other peers (i.e., links in the overlay).
Among all these possibilities, self-healing figures as a key characteristics to 
improve the dependability of the managed infrastructure.
Self-healing is not novel in networks. It is an interesting approach to cope with the general problem of providing network resilience \cite{doerr}.
It has been a long time since self-healing ring topologies have been introduced.
In the domain of P2P (and networks), several works concerned with this issue have been proposed \cite{chaudhry,simplex,Pournarasb}.

However, in P2P systems, certain network properties are guaranteed usually on the steady state. Thus, 
it may happen that they disappear in case of multiple node departures. For 
instance, the overlay might get partitioned upon failure of links connecting 
different clusters.  
Alternatively, some important links might be lost that were playing a main role 
to keep a low network diameter. For instance, in small worlds there are links 
among distant nodes that strongly reduce the average shortest path length.
Although the P2P network is unstructured, it has certain characteristics that 
should be maintained, at least up to a certain extent, in order to provide some 
guarantees and the ability of the network to spread contents. 
The purpose of this work is to understand if some decentralized self-healing 
algorithm can guarantee the resilience and the communication capabilities of a 
P2P system. 

In the literature, some works make a distinction between reactive 
and proactive approaches. In essence, with reactive approaches novel links are 
created only when nodes join, leave, or when a failure is detected. This 
is different from proactive approaches, where nodes try periodically to find 
new neighbors to link at \cite{qiu}.

Basically, reactive overlay recovery mechanisms may work by resorting to 
either centralized or decentralized approaches to identify novel peers. 
According to a centralized approach, a peer that ''needs neighbors`` contacts a 
set of well-known nodes that answer with a list of nodes. This 
approach is exploited in general P2P systems; for instance in BitTorrent this 
role is played by the tracker. Also Gnutella exploits this kind of strategy.
This method is adequate when the P2P overlay is unstructured or loosely 
organized; however, its weakness relies on the robustness of these well-known 
nodes. If they are reliable nodes in the network (such as public trackers in 
BitTorrent, that are in charge of this service only), the system stays up. 
Failures of such nodes may cause the whole system to partition or crash. 

In a reactive decentralized approach, a peer locally asks its neighbors to 
provide 
information on nodes it is not connected to. This method is widely used in
structured P2P systems \cite{zhuang}. These schemes require
information on how to make connections between independent components when 
an overlay partition occurs \cite{qiu}.

SCAMP is a prominent example of a reactive recovery approach \cite{Ganesh:2003}.
It is a gossip-based protocol where the neighborhood size of each node adapts 
w.r.t.~a-priori unknown size of the whole system.
Thus, each node can modify its set of neighbors when system size changes.

Similarly, Phenix is an approach that creates robust topologies with a 
low-diameter \cite{phenix}. In particular, it creates scale-free networks, which 
are well known to be tolerant to node random removals. 
The approach specifically focuses on the particular case where malicious nodes 
try to collect information on the network in order to devise targeted 
attacks. Such a scenario is avoided by hiding information to those nodes that 
are in local black lists.

As concerns proactive strategies, in the literature seminal works have been 
proposed that build a peer-sampling service. Such a service provides nodes with 
a randomly chosen set of neighbors to exchange information with.
Typically, this information exchange is realized through gossip approaches \cite{complenet}.
The set of neighbors creates a dynamic unstructured overlay.
These approaches mainly differ in the way new nodes' neighbor lists are
built, after merging and/or truncating the neighbor lists of communicating peers.

For instance, Cyclon is a popular scheme that allows to construct gossip-based unstructured P2P systems that have low diameter, low clustering, highly symmetric node degrees, and that are highly resilient
to massive node failures \cite{voulgaris.jnsm.2005}.
Is is a quite inexpensive membership management, where nodes maintain a small, 
partial view of the entire network. According to this protocol, nodes 
periodically perform a shuffling protocol which ensures that peers maintain a 
list of active neighbors.  
The difference with our scheme is that this approach builds a specific and 
robust overlay, with given topology characteristics. Instead, the aim of the 
approach described in this paper is to have a decentralized protocol that, given 
a certain unstructured P2P overlay with any possible characteristics, reacts to 
important failures to avoid further network partitioning. 

An approach that is conceptually similar to Cyclon is that proposed in \cite{stavrou}.
It uses a randomized overlay construction method to provide network robustness.

Newscast is a gossip-based protocol that builds and maintains a continuously changing random
overlay \cite{jelasity2003newscast}. The generated topology is built to ensure stability and connectivity. The idea is that each node modifies periodically its set of neighbors by randomly exchanging information with nodes it is connected with.
Thus, a continuous rewiring strategy is performed.

With respect to this reactive/proactive classification, it is worth mentioning 
that our proposed approach enables nodes to react to node disconnections, by 
creating novel links with nodes that have been proactively discovered before 
the failure. Thus, the peer discovery is proactive (and local), 
while the link creation is reactive. A similar philosophy is exploited in 
\cite{qiu}. Moreover, our proposed approach requires local information only, 
hence maintaining the amount of information to be exchanged in background quite 
limited.

Several interesting works look at ways to form ``good'' topologies. One
example is \cite{pandurangan}, which focuses on building randomized
topologies with bounds on the overlay graph diameter.
In general, the topology of the overlay has a strong influence on the performance of the information dissemination, nodes workload and on the overlay robustness.
For instance, if a scale-free network is employed, then the network has a low diameter and it is robust to random node failures.
However, a scale-free net contains a non-negligible fraction of peers which 
maintain a high number of active connections, and hence they sustain a workload 
higher than low-degree nodes.
Conversely, if a network has a more uniform degree distribution, then the workload is equally shared among all peers. However, the diameter of the network increases, and so does the number of hops needed to
cover the whole network with a broadcast~\cite{gridpeer}. 
Therefore, some approaches in the literature force the use of a 
specific topo\-logy.
The scheme presented in this work has a different goal. It copes with locally 
important failures that might partitionate the overlay, without affecting that 
much the original topology of the overlay. Thus, our scheme aims at augmenting 
network resilience and it can be coupled with other approaches that create some 
overlay with certain features. 
Indeed, in the performance assessment section, the proposed algorithm is 
evaluated over different overlay topologies. 

\section{Self-Healing Protocols}\label{sec:prot}

\subsection{System Model}

We consider \ac{P2P} systems built on top of an unstructured overlay network. 
(Note that in the following the terms ``peer'' and ``node'' are employed as synonyms.)
No assumptions are made on the topology of the overlay. In fact, it is not the 
aim of the protocol to build an overlay with specific characteristics. Rather, 
the idea is to provide a simple protocol that augments the reliability of an 
overlay, whatever its starting topology, during its evolution with nodes that 
enter and leave the overlay, dynamically. 
For simplicity, we consider networks with undirected links. Actually, this 
setting is quite common in many P2P systems, e.g., BitTorrent, Gnutella 
\cite{tarkoma}.

Each node $n$ has a certain degree, i.e.~the amount of $1$st neighbors or, in 
other words, the nodes directly connected with $n$ in the overlay. 
The list of these $n$'s $1$st neighbors is denoted with $\Pi_n$, while the 
degree of $n$ is denoted with $|\Pi_n|$. $n$ maintains also the list of its 2nd 
neigh\-bors, $\Pi^2_n$, i.e.~nodes distant $2$ hops from $n$. 
Every time the list $\Pi_n$ changes, due to some node arrival or departure, $n$ 
informs its other 1st neighbors of this update.
With $\Pi^2_{n|m} = \Pi_m - \Pi_n$, we identify the $n$'s 2nd neighbors which can be reached through $m$. Hence, $\Pi^2_n = \cup_{k \in \Pi_n} \Pi^2_{n|k}$.
The discussed protocols employ a threshold on the maximum 
node degree. 
In Section \ref{sec:sim_details} a discussion on such a threshold is reported, 
and in Section \ref{sec:eval_degree} we show a study on the impact of this 
threshold. 

As concerns failures, for the sake of a simpler discussion, we assume that only 
nodes can fail, while it cannot happen that single links are removed from the 
overlay. 
This is a common simplification made in most P2P system models. Anyway, the 
protocol can be easily upgraded (without any substantial modifications) to 
handle single link failures.
We assume that a failure detection service is employed, that informs a 
node upon a $1$st neighbor failure. This service can be implemented using some 
sort of ''keep alive`` mechanism, such as \cite{huan,zhuang}.

Nodes can join and leave the network dynamically. 
We assume that network changes (in a given neighborhood) are slower than a 
given execution of the communication protocol 
\cite{ferretti_trans.cs.2012.10-12.e2}. 
Thus, in general, upon failure of a node $f$, its neighbor $n$ is 
enabled to send messages to $\Pi^2_{n|f}$.
When not differently stated, we will consider cases when nodes arrivals and 
departures occur at the same rate.\footnote{This will be the scenarios of the so 
called ``evolution'' and ``targeted attack'' simulation modes, while in the 
``failures only'' the 
arrival rate is set to $0$, mimicking a worst churn scenario.}

\subsection{Protocol $P_{2n}$: Use of the $2$-Neighborhood}

Upon a neighbor $f \in \Pi_n$ departure, by looking at $\Pi^2_{n|f}$ each node $n$ is able to understand if some 2nd neighbor is no more reachable. 
If this is the case, this protocol ensures that $n$, or one of its neighbors, creates a link with it.
Algorithms \ref{alg:active}--\ref{alg:passive} sketch the related pseudo-code.
In particular, when a node $f$ fails, $\forall p \in \Pi_f$ there are three 
possible cases.
\begin{enumerate}
 \item $p \in \Pi_n$ : $n$ and $p$ are neighbors. In this case there is nothing to do (at $n$).
 \item $p \notin \Pi_n$, but $p \in \Pi^2_n$ since $p \in \Pi^2_{n|q}$ for some $q \in \Pi_n, q \neq f$: $p$ is still a 2nd neighbor of $n$; also in this case there is nothing to do.
 \item $p \notin \Pi_n, p \notin \Pi^2_n$ : after the failure $p$ is no more a 
$1$st or 2nd neighbor of $n$. In this case, $n$ takes part to the distributed 
procedure to create a link with $p$ (see Algorithm \ref{alg:active}).
\end{enumerate}

\begin{figure}[t]
   \centering
     \includegraphics[width=.5\linewidth]{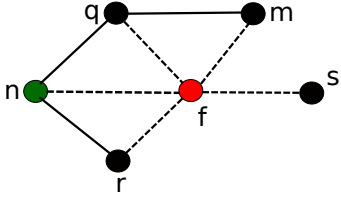}
   \caption{Example of a node failure, as managed at a node $n$ (green node). 
Upon failure of a neighbor $f$, $n$ needs to replace the lost 
$2$-hop connection with $s$. Nothing has to be accomplished at $n$ for other 
nodes, since $q$, $r$ are $n$ $1$st neighbors, while $m$ can be reached through 
$q$.}
   \label{fig:failures}
\end{figure}

In essence, links are created among nodes which were connected through $f$ only.
This list is computed by analyzing the old view $n$ had of its 
$2$nd-neigh\-bor\-hood, before removing its connection information about $f$ 
and $\Pi^2_{n|f}$ (Algorithm \ref{alg:active}, line \ref{code:P}).
Take as an example the situation reported in Figure \ref{fig:failures}. In this 
case, upon failure of $f$, all dashed links are removed. 
Focusing on node $n$, this node will need to replace its lost $2$-hop 
connection with $s$, while 
other nodes remain still $1$st neighbors (node $q, r$) or $2$nd neighbors 
(node 
$m$).

As already mentioned, each node $n$ keeps a threshold value for its degree, to 
avoid that its degree grows out of control (Algorithm \ref{alg:active}, line 
\ref{code:control}). (This threshold should not be too low, otherwise it might 
contrast the creation of additional links, and this might generate network 
partitions.)
Moreover, in order to diminish the probability that multiple nodes of the same 
cluster attempt to create a novel link with the same node $p$ at the same time, 
a classic contention-based approach is used, so that each node $n$ waits for a 
random time before transmitting messages (Algorithm \ref{alg:active}, line 
\ref{code:wait}). 
Such a random waiting time is generated within a predefined time interval, 
using a uniform distribution. This way, each node has the same 
probability of triggering the creation of a novel link. This provides load 
balancing among nodes.

Then, upon reception of a message from a node $p$ asking $n$ to become 
neighbors, $n$ accepts the request only if $p$ is not a 1st or 2nd neighbor of 
$n$ (it is possible that some of its neighbors just created a link with $p$; see 
Algorithm \ref{alg:passive}).
Then, $n$ answers this request through a direct message to $p$ (Algorithm 
\ref{alg:passive}, lines \ref{code:l_creat_no}, \ref{code:l_creat_ok}). 

Upon creation of a novel link between two nodes, these nodes inform 
all their $1$st neighbors that a novel link has been created 
(Algorithm \ref{alg:passive}, lines \ref{code:ans_multi}, 
\ref{code:l_creat_multi}).

Finally, when a node $n$ receives a message from a neighbor (say $q$) 
confirming the creation of a link between $q$ and $m$, then $n$ can remove
$m$ from the list of lost nodes in its $2$nd neighborhood, since after this 
novel connections, $n$ and $m$ are $2$ hops away (Algorithm 
\ref{alg:passive}, line \ref{code:l_creation}).

\begin{algorithm}[htbp]
\caption{$P_{2n}$: Active behavior at $n$ upon failure of $f$}
\label{alg:active}
\begin{algorithmic}[1]
\Statex \Comment{$P$ contains old $2$nd neighbors, reachable through $f$ only, 
hence no more reachable in 2 hops after $f$ failure}
\State $P \gets \{ p \in \Pi^2_{n|f} | \ p \notin \Pi_n,\ p \notin \Pi^2_{n|q}, 
q \in \Pi_p, q \neq f\}$\label{code:P}
\State update neighbor lists in view of $f$ failure
\Statex
\While {($P \neq \emptyset) \wedge (| \Pi_n | \leq \text{thresholdDegree})$}\label{code:control}
  \State wait random time \label{code:wait}
  \State $p \gets$ extract random node from $P$ \label{code:extract}
  \State send link creation request to $p$\label{code:req}
\EndWhile
\end{algorithmic}
\end{algorithm}
\begin{algorithm}
\caption{$P_{2n}$: Passive behavior at $n$}
\label{alg:passive}
\begin{algorithmic}[1]
\Require message from $p$ answering a link creation request
  \If {answer is OK}\label{code:ans_b}
    \State sendAll($\Pi_n$, ``novel link $(n, p)$'')\label{code:ans_multi}
    \State add $p$ to $\Pi_n$
  \EndIf\label{code:ans_e}
\Statex 
\Require message from $q \in \Pi_n$: novel link $(q, m), m \in P$
  \State extract $m$ from $P$\label{code:l_creation}
\Statex 
\Require message from $p$ with a link creation request
  \If{$p \in \Pi^2_n$}\label{code:l_creat_req_b}
    \State send refuse message\label{code:l_creat_no}
  \Else
    \State send accept message\label{code:l_creat_ok}
    \State sendAll($\Pi_n$, ``novel link $(n, p)$'')\label{code:l_creat_multi}
    \State add $p$ to $\Pi_n$
  \EndIf\label{code:l_creat_req_e}
\end{algorithmic}
\end{algorithm}

\subsection{Protocol $P_{ECC}$: Edge Clustering Coefficient}

This protocol is an extension of $P_{2n}$, and it is based on the idea of exploiting the importance of failed links, so as to identify those that, once failed, must be replaced with novel ones. 
In complex network theory, several centrality measures have been introduced to characterize the importance of a node or a link in a network, e.g.~betweenness centrality, or to detect different communities and identify their boundaries in the net \cite{Bader:2007,girvan,Goncalves:2012,Newman200539}.
The calculation of these metrics usually involves a full (or partially full) knowledge about the whole network. 
Conversely, the aim of this work is to preserve connectivity without such a global knowledge \cite{simplex13,networking14,massoulie,voulgaris.jnsm.2005}.

The Edge Clustering Coefficient (ECC) has been defined in analogy with the usual node clustering coefficient, but it is referred to an edge of the network \cite{radicchi2004}. It measures the number of triangles
to which a given edge belongs, divided by the number of triangles that might potentially include it, given the degrees of the adjacent nodes. More formally, given a link $(n, m)$ connecting node $n$ with node $m$, the edge clustering coefficient $ECC_{n,m}$ is
$$ECC_{n,m} = \frac{T_{n,m}}{min((|\Pi_n| - 1), (|\Pi_m| - 1))},$$
where $T_{n,m}$ is the number of triangles built on that edge $(n,m)$, and $min((|\Pi_n | -1), (|\Pi_m | -1))$ is the amount of triangles that might potentially include it. We add the constraint that this measure is $0$ when there are no possible triangles at one of the nodes, i.e.~when $min((|\Pi_n | - 1), (|\Pi_m | - 1))=0$.

The idea behind the use of this quantity is that edges connecting nodes in 
different communities are included in few or no triangles, and tend to have 
small values of $ECC_{n,m}$. On the other hand, many triangles exist within 
clusters. Hence the coefficient $ECC_{n,m}$ is a measure of how 
inter-communitarian a link is.

Thus, based on this notion of $ECC_{n,m}$, the protocol $P_{ECC}$ works as 
follows. (Algorithm \ref{alg:ecc_active} shows the pseudo-code of the active 
behavior only, since the passive behavior is equivalent to Algorithm 
\ref{alg:passive}).
Each node $n$ knows its $2$nd neighbors, i.e.~$1$st neighbors of its neighbors; 
thus, it can understand if some triangle exists that includes itself.  
Indeed, let say that three nodes $n, m, p$ create a triangle. Then, $n$ has 
$m,p$ in its neighbor list $\Pi_n$ (and the same happens for the two other 
nodes). When $n$ sends its list $\Pi_n$ to $m$ and $p$, they recognize that 
there is a common neighbor that creates a triangle.
If one of the three nodes would fail in the future, the other two nodes will 
understand automatically that the triangle no longer exists. 

\begin{algorithm}[thbp]
\caption{$P_{ECC}$: Active behavior at $n$ upon failure of $f$}
\label{alg:ecc_active}
\begin{algorithmic}[1]
\Statex \Comment{$P$ contains old $2$nd neighbors, reachable through $f$ only, 
hence no more reachable in 2 hops after $f$ failure}
\State $P \gets \{ p \in \Pi^2_{n|f} | \ p \notin \Pi_n,\ p \notin \Pi^2_{n|q}, 
q \in \Pi_p, q \neq f\}$\label{code:P}
\State update neighbor lists in view of $f$ failure
\Statex
\If{random() $> ECC_{n,f}$}\label{code:ecc}
 \While {($P \neq \emptyset) \wedge (| \Pi_n | \leq \text{thresholdDegree})$}
  \State wait random time 
  \State $p \gets$ extract random node from $P$ 
    \State send link creation request to $p$
 \EndWhile
\EndIf
\Statex 
\Require $(|\Pi_n| \gg |\Pi_n|_{target})\ \wedge \ (L_{n} \gg L_{n, target})$
\State Remove at most $r$ links with $ECC > T_{ECC}$
\end{algorithmic}
\end{algorithm}

When a node $f$ fails, each neighbor $n \in \Pi_f$ checks the value 
$ECC_{n,f}$.
Depending on this value, a reconfiguration phase may be executed. The idea is that the higher the ECC the lower the need to create novel links to keep the network connected, since that link was part of multiple triangles. This decision is taken probabilistically, i.e.~the lower $ECC_{n,f}$ the more probable that the rest of the procedure is executed (line \ref{code:ecc}, Algorithm \ref{alg:ecc_active}).
If this is the case, $n$ checks if its $2$nd neighbors ($\Pi^2_{n|f}$), reached 
formerly through $f$, still remain in its $2$nd neighborhood; otherwise it 
creates links with them, as in $P_{2n}$. 

Due to the overlay reconfiguration, it is expected that the degree of a node 
changes (suddenly, in some cases). Indeed, the goal of the self-healing 
reconfiguration scheme is that the network should evolve to react to nodes 
arrivals and departures. For instance, if a hub goes down for some reason, it is 
likely that its past neighbors will create more links in order to maintain the 
overlay connected.
Thus, it might happen that the total number of links augments, due to the parallel activity of nodes, and this can alter the network topology. In $P_{ECC}$, this is more probable when there is a low network clustering, with few triangles.  

To overcome this possible problem, a periodical check is accomplished on the growth of links at each node and its neighborhood. 
Thus, periodically each node $n$ checks its actual degree $|\Pi_n|$ and the actual number of links in its neighborhood $L_{n}$, i.e.~the sum of all different links departing from $\Pi_{n} \cup \{n\}$.
These values are compared with two values that $n$ stores, related to the target degree $|\Pi_n|_{target}$ and a target number of links in the $n$'s neighborhood $L_{n, target}$.
By monitoring the amount of links in its neighborhood, $n$ obtains an approximate understanding of how the network is evolving.
(These two values are periodically updated, based on values assumed in a window time interval.)

In case of an important increment on the amount of links in some portion of the 
network, then the nodes with the higher variations on their degrees check if 
some links (i.e.~those with higher ECC values) can be removed.
Indeed, if the difference between the target values and the actual ones surpasses a given threshold, then the node $n$ invokes a procedure that removes its $r$ links with higher ECC values (larger than a threshold value $T_{ECC}$), if there are any. (In the simulations, we consider $r=1$ since it suffices to control the rate of the periodical check to increase/decrease the number of links that can be removed.)


\section{Evaluation Assessment}\label{sec:eval}

Simulation was used to assess the performances of the proposed algorithms. 
In these simulations, we varied:
\begin{itemize}
 \item the topology of the unstructured overlays over which the approaches were executed. In particular, we employed uniform networks, clustered networks and scale-free networks. It is worth mentioning that simulations were made also on classic random graphs, but we omit results here, since they are similar to those obtained for uniform networks.
 \item the types of simulation. We simulated (i) the classic scenario where the 
network evolves with an equal amount of joining nodes and leaving nodes (i.e., 
equal join and fail rate probabilities); (ii) a case similar to the previous 
one, but nodes to be removed are those nodes that might have some important role 
in the network, i.e., we performed two types of simulations where removed nodes 
were those with highest degrees in one case, and those with highest betweenness 
in the other case; (iii) the case when only failures occur.
\end{itemize}
The considered approaches are $P_{2n}$, $P_{ECC}$ and ``none'', which represents the (typical) situation when peers do not react to node disconnections, simply assuming that other links will be created upon arrivals of novel nodes. Details of the simulation are discussed in the next subsection.

\subsection{Simulation Details}\label{sec:sim_details}

The simulator was a discrete event simulator, implemented using the GNU Octave 
language and the Octave-network-toolbox, a set of graph/networks analysis 
functions in Octave \cite{nettoolbox}.
Based on it, we assume that the communication among peers is reliable, 
with a latency that is negligible, with respect to the inter-arrival times of 
overlay related events (e.g., node arrivals and departures), times 
required by the failure detector to identify a node failure, and so on.
Hence, once a message is sent from a node to another, the communication can be 
thought as instantaneous and completely reliable.
This is a common approximation that relieves the simulation dealing with all the 
underlying communication related issues, simply focusing on the overlay 
parameters. 
Other P2P discrete event simulators offer similar abstractions, e.g. PeerSim 
\cite{peersim}, P2PSim \cite{p2psim}, PlanetSim \cite{planetsim}, LUNES 
\cite{lunes}, SimGrid \cite{simgrid}.

We present results averaged from a corpus of $20$ simulations for the same 
scenario. In each simulation, we started with an overlay network 
with a specified degree distribution and network characteristics, and let the 
simulation advance for an amount ($\sim100$) of simulation steps. 
All the configuration parameters were varied; we present here results for some particular configuration settings, since those obtained for different ones were comparable to those we will show.

Upon a node failure, all its links with other nodes are removed. Then, the node passes to an inactive state; it can be selected further on to simulate a novel node arrival.
Thus, a node arrival is realized by changing the state of a randomly selected inactive node to pass to the active state. This event triggers the creation of novel links with other randomly selected nodes. 
Different joining procedures were executed, depending on the network topology under investigation. The idea was to adopt a join mechanism that would maintain the topology unaltered.

Both protocols $P_{2n}$, $P_{ECC}$ employ a threshold on the maximum degree. 
In Section 5.6, we show the impact of varying this value; 
when not differently stated, the threshold 
was set equal to $100$.
As a matter of fact, the threshold strongly depends on the P2P system one wants 
to build, on the specific application run on top of of the overlay, and on the 
typical number of connections 
a peer maintains during its lifetime in the network. Thus, it should be tuned 
with this in view.
For instance, BitTorrent sets the maximum degree for peers equal to 80 (then, 
each peer limits the amount of connections contemporaneously active, using the 
choke algorithm) \cite{tarkoma}.
Gnutella has a degree distribution that follows a power law function; a 
snapshot made in 2000 revealed that nodes had a maximum degree equal 136, 
with a median value of 2, and an average of 5.5 \cite{tarkoma}.
In PPlive, the average node degree varies in a small range between 28 to 42 over
the course of the day, with no correlation between the variation of average 
degree and the channel size. The overlay  resembles a random graph when net 
size is small (around 500 nodes) but becomes more clustered when net size 
grows \cite{Vu:2010}. 
For this reason, a specific static value is not proposed in this work; 
however, results will show that changing the threshold on the maximum degree 
can lower significantly the amount of 1st and 2nd neighbors, without evident 
differences on the size of the main component.

\subsection{Network Topologies}

As already mentioned, we employed three different kinds of overlay topologies, varying their specific parameters.
In the following, the general characteristics of such topologies are described, together with the method employed to simulate the arrival of a novel node in the network, that is accomplished to respect the typical attachment process of that topology. 

As concerns node removals, a related subsection is reported in the following of 
this section.

\subsubsection{Uniform Networks}
Uniform networks are those where all nodes start with the same degree. 
Then, due to node failures and arrivals (and the reconfiguration imposed by the P2P protocol), the node degree might change.
We varied the initial degree of nodes. 
Uniform networks are quite common in several (P2P) systems, where the software running on peers is configured to have a given number of links in the overlay. 
This is usually accomplished for load balancing purposes \cite{TDGsIMC07}.

As concerns the arrival of a novel node, a random set of neighbors was selected, 
whose size was equal to the initial degree parameter.
Of course, this causes an increment of nodes' degree that accept such a novel 
link. However, it does not alter the general idea of a network topology where 
all nodes have the same importance (uniform). 

\subsubsection{Clustered Networks}
The presented self-healing protocols are thought for those P2P overlays that have important links that connect different parts of the network; thus, it is interesting to observe how the protocol performs over nets composed of different connected clusters.
In these simulations, network clusters were set to be of the same size.

We set two different parameters to create the network. The first parameter is the probability $\gamma$ of creating a link among nodes of the same cluster. Each node is linked to another node of the same cluster with a probability $\gamma$; hence, inside a cluster, nodes are organized as a classic random graph.
As to inter-cluster links, the amount of links created between the two clusters 
was determined based on a certain probability $\omega$ times the number of nodes 
in the clusters (i.e.~each node has a probability $\omega$ of having a link with 
each external cluster). 

Upon a node arrival, the node was associated to a cluster and links with nodes in that cluster were randomly created based on the $\gamma$ probability, as in a classic random graph. 
Then, for each other cluster, the node creates, with probability $\omega$, a link with a random node of that cluster.

\subsubsection{Scale-Free Networks}
A scale-free network possesses the distinctive feature of having nodes with a degree distribution that can be well approximated by a power law function. Hence, the majority of nodes have a relatively low number of neighbors, while a non-negligible percentage of nodes (``hubs'') exists with higher degrees \cite{simutools}. 
The presence of hubs has an important impact on the connectivity of the net. 
In fact, the peculiarity of these networks is that they possess a very small diameter, thus allowing to propagate information in a low number of hops. 
To build scale-free networks, our simulator implements the construction method 
proposed in \cite{Aiello00arandom}; but we used also a classic 
preferential attachment generation approach, using a specific routine available 
in the Octave-network-toolbox \cite{nettoolbox,newmanHandbook}.

Upon a node arrival, a preferential attachment was utilized for scale-free networks.
That is, the higher the degree of a node the more likely it is to receive new
links. Thus, the more connected nodes have stronger ability to obtain novel 
links added to the network. This is the typical approach that leads to the 
formation of scale-free networks \cite{simutools,newmanHandbook}.
 
When not differently stated, we employ four different scale-free networks, with 
different characteristics. In fact, the first two networks are composed by a 
small amount of nodes (following a power-law degree distribution), that result 
in disconnected networks. Instead, the other two networks are composed of a 
main 
component, with the presence of important hubs that provide this connectivity.

\subsection{Simulation Scenarios}

We evaluated the presented approaches using different simulation modes, that basically differ in the way nodes were selected to be removed from the overlay, and if, during the simulation, novel nodes were allowed to enter the network or not.

\subsubsection{Evolution}
The first mode was based on a random selection of failed nodes, with an amount of failed nodes equal to the amount of joining nodes. This way, the network size remains stable during the simulation. 

\subsubsection{Targeted Attack to Nodes with Highest Degree}
In this case, at each step of the simulation the ``important'' nodes with some 
specific characteristics were selected to fail.  
In particular, as concerns uniform and scale-free networks, nodes with the 
higher degrees were selected to fail. Instead, in clustered networks the 
selected nodes were those with higher number of links connecting different 
clusters (i.e.~the highest inter-cluster degree); the rationale was to augment 
the probability of disconnecting the clusters.
In this scheme, as in the previous simulation mode, the amount of failed nodes 
per simulation time interval was kept equal to the amount of joining nodes.

\subsubsection{Targeted Attack to Nodes with Highest Betweenness}
This simulation type is similar to the targeted attack to nodes with highest degree. However, instead of selecting the node with highest degree (or highest inter-cluster degree in the case of clustered networks), the simulator detected the node to fail as that with highest node betweenness.

Betweenness is a centrality measure that, given a node in a network, calculates the number of shortest paths from all nodes to all others which pass through that node.
Thus, if a node $n$ has a high betweenness, it means that several paths in the 
overlay pass through $n$. Or, in other words, if you plan to go from a node to 
another in an overlay, it is quite probable that you will encounter $n$ during 
your path.
Nodes may have a low node degree but high betweenness.\footnote{E.g., imagine to have two separated clusters in a network and a single node $n$ that performs as intermediate, which is linked to a single node for each cluster. In this example, $n$ has a low degree (equal to $2$) but a high betweenness value, since all paths among two nodes in the different clusters have to pass through $n$.}

The formula for measuring the betweenness of a node $n$ is as follows. Assume 
that the amount of shortest paths between two nodes $m,p$ is denoted with 
$\sigma_{mp}$; with $\sigma_{mp}(n)$, we denote the amount of shortest paths 
between $m,p$ passing through $n$. 
Then, the betweenness of $n$ is measured as the fraction between the number of 
shortest paths passing through $n$, divided by the amount of shortest paths in 
the network, i.e.~$\text{bet}(n) = \sum_{m\neq n\neq p} 
\frac{\sigma_{mp}(n)}{\sigma_{mp}}$.

It should be clear that the removal of a node with high betweenness 
centra\-lity can lead to an increment of the path lengths and to network 
disconnections. Thus, this targeted attack is of main interest in our study. 

\subsubsection{Failure Churn}
In this case, during the simulation only failures occurred. Thus, each network 
started with all nodes active, which were (randomly) forced to fail until no 
active nodes remain in the network. This allows to understand if the 
self-healing protocols are able to react to situations with high failure rates. 
We refer to this simulation mode as ``failures only''.


\section{Results}\label{sec:res}

This section discusses on the results obtained in the simulation scenarios described above. 
A first comment worth of mention is that the considered approaches do not 
increase the connectivity of the network overlay being utilized. In fact, 
$P_{2n}$ and $P_{ECC}$ restore connections with lost 2nd neighbors, 
without looking for novel nodes.
Thus, the obtained connectivity is at most equal to the initial one (we will 
see that these two approaches are able to maintain it, while the ``none'' 
approach is not able to do it).

Another result is that the two approaches augment, in some cases, the 
average number of 2nd neighbors in the network. This happens especially 
upon removal of an important node (in terms of connectivity) $n$. In fact, in 
this case, the remaining nodes have to reorganize their connections. This might 
lead to the creation of multiple links (in spite of a single link) to connect 
to local clusters, previously reached through $n$. While the average amount 
of 1st neighbors is not particularly affected by the substitution of a single 
link to multiple ones, this multiplicative factor is more evident when counting 
the amount of novel 2nd neighbors (especially when the clustering coefficient is 
low).

While mentioned in the description of $P_{ECC}$, in these experiments the 
link reduction was not activated. The idea was to understand if that protocol is 
able to guarantee network connectivity. 
Thus, one should keep in mind that when the amount of added links becomes too 
high (and this is a metrics which depends on the specific application 
requirements), one can reduce it by removing unnecessary ones. 

\subsection{Evolution}

This is the simulation scenario where nodes enter and leave the network at the same rate. Leaving nodes are selected at random. Moreover, nodes that enter do respect the type of attachment related to the overlay topology. 
In fact, for uniform nets, neighbors are selected at random; for clustered nets, nodes are randomly assigned to a cluster and neighbors are randomly selected in that cluster (then, some links might be created among different clusters with a lower probability, as previously discussed); for scale-free nets, a preferential attachment is performed.
Thus, we do not expect that failures introduce relevant connectivity problems, 
and the use of $P_{2n}$, $P_{ECC}$ might be not necessary, in this case. In any 
case, we thought it would be interesting to understand how these self-healing 
protocols perform.

\subsubsection{Uniform Networks}
\begin{figure*}[t]
   \centering
   \subfigure[Main component size]{
     \includegraphics[width=.6\linewidth]{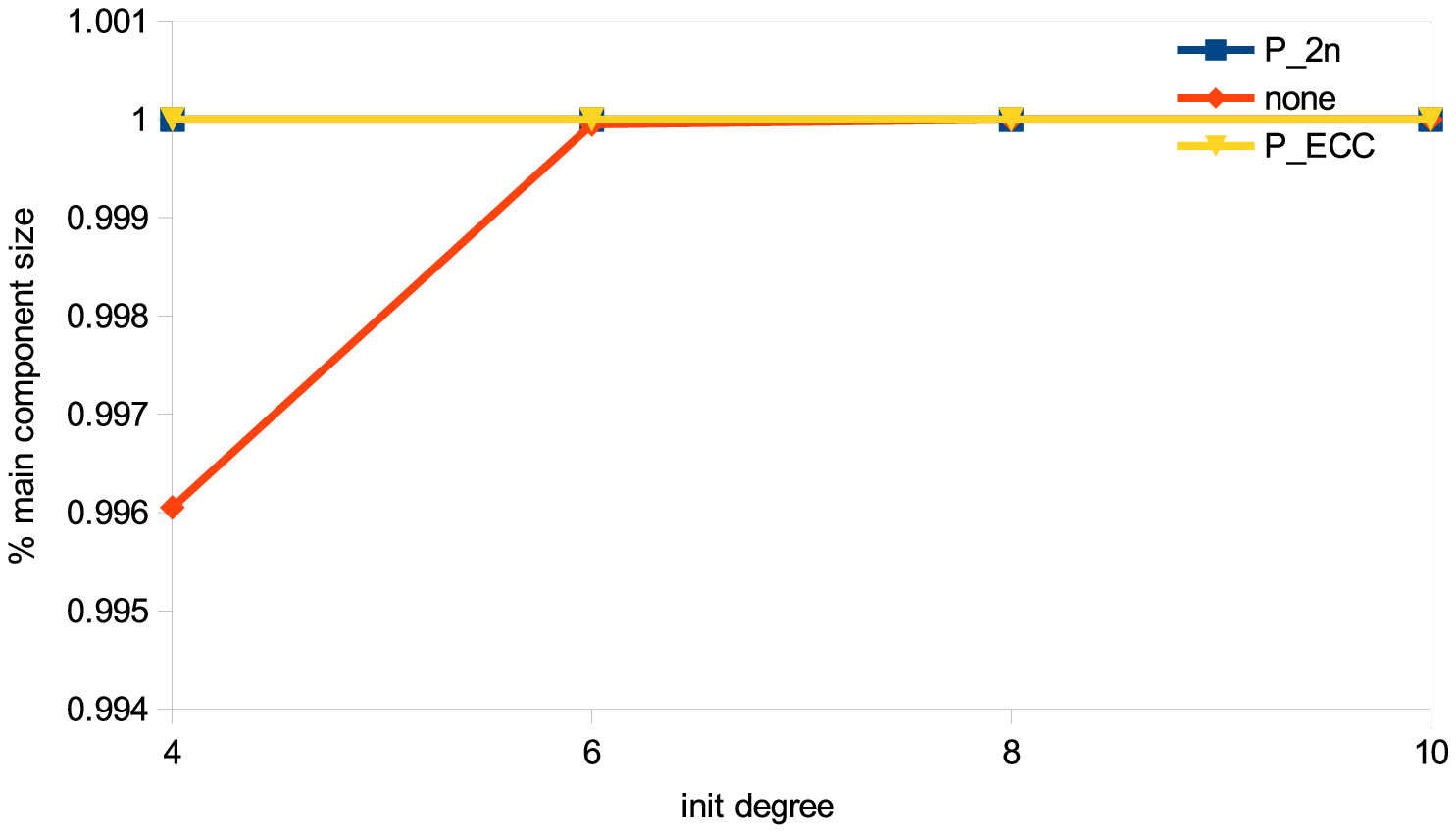}
   }
   \subfigure[Average amount of $1$st neighbors]{
     \includegraphics[width=.45\linewidth]{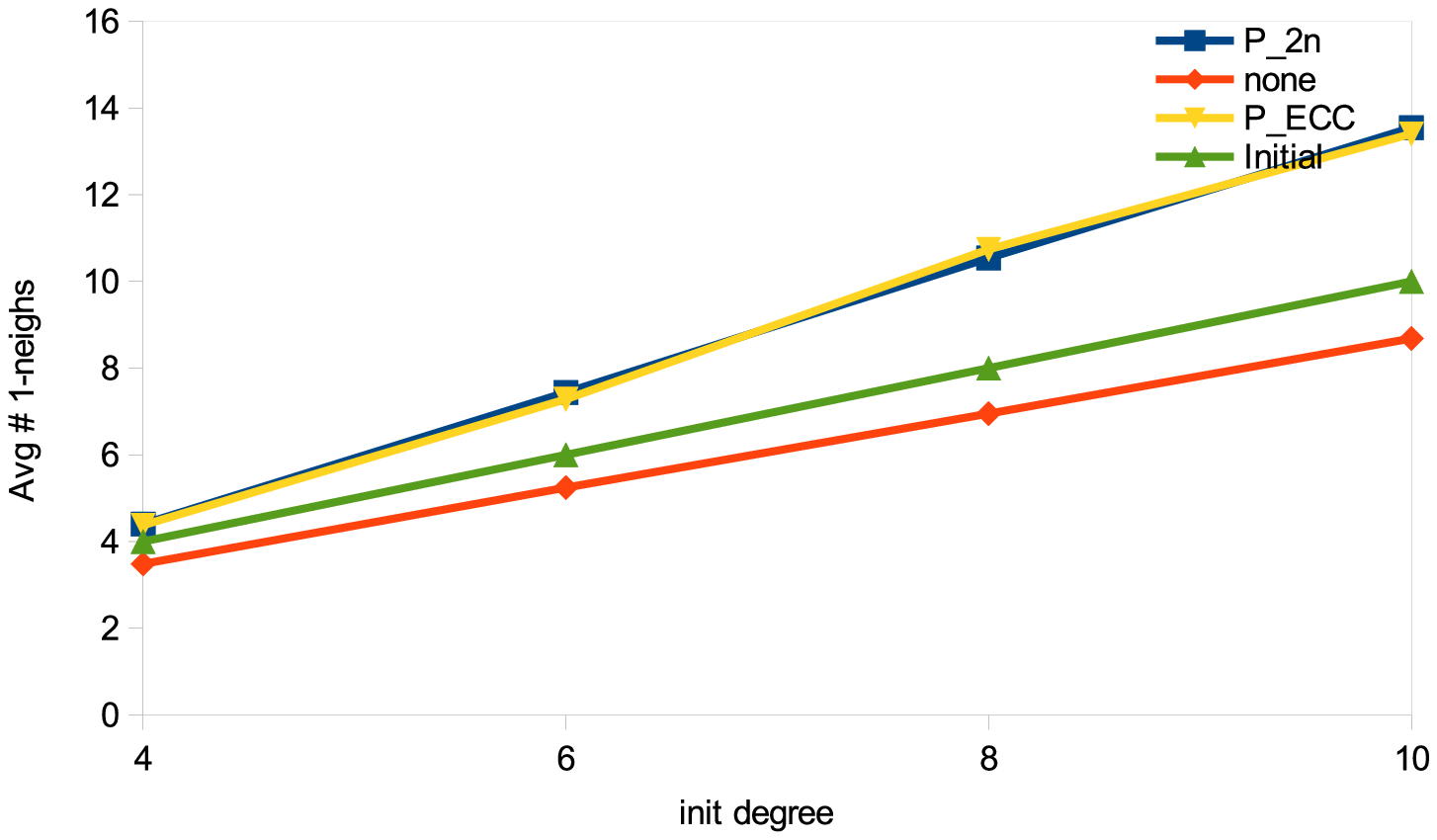}
   }
   \subfigure[Average amount of $2$nd neighbors]{
     \includegraphics[width=.45\linewidth]{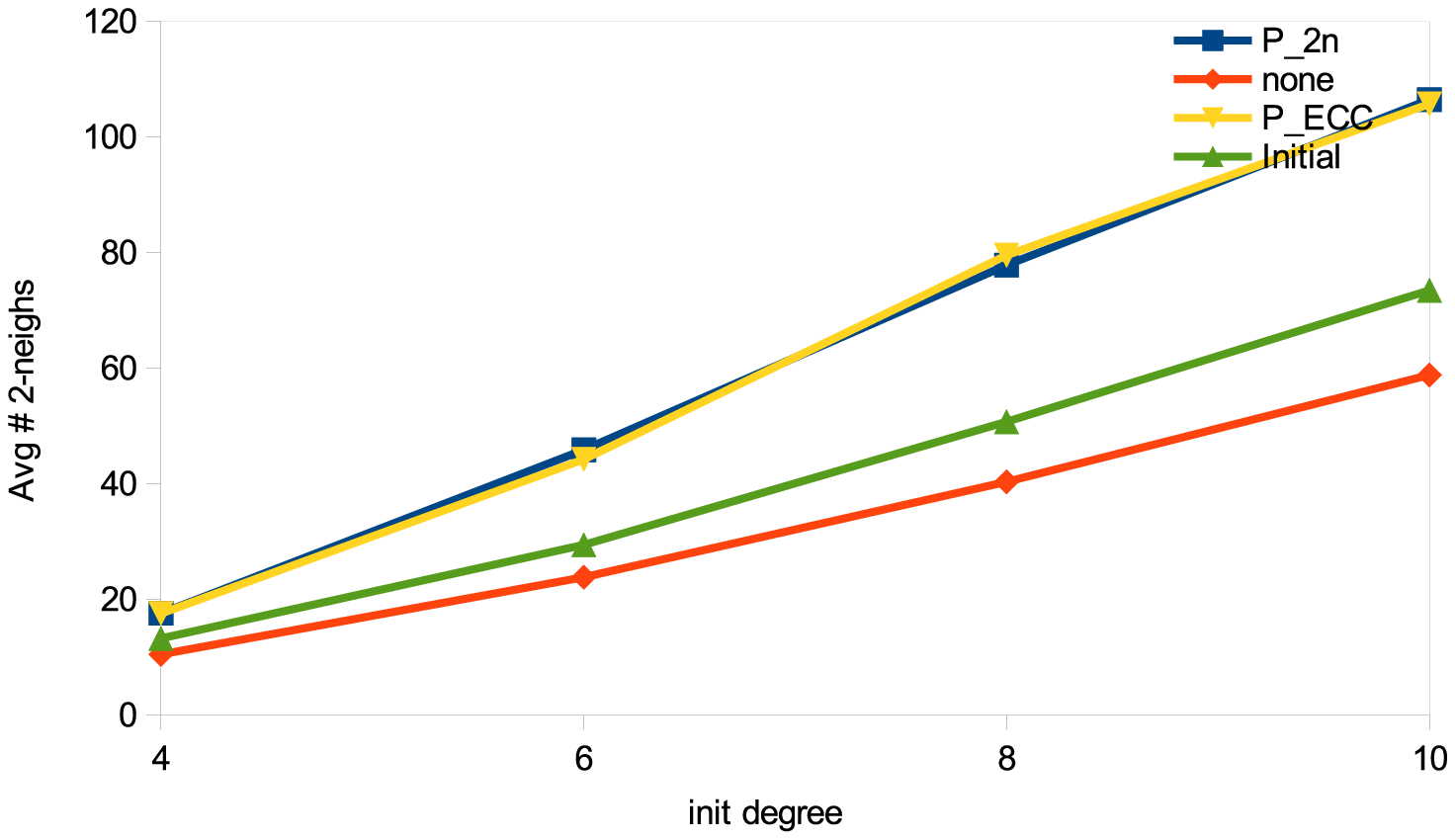}
   }
   \caption{Uniform networks -- evolution simulation mode.}
   \label{fig:unif_evol}
\end{figure*}

Figure \ref{fig:unif_evol} shows results for uniform networks. 
The top chart reports the average size of the main component for the three considered management protocols, while the other charts report the average amount of $1$st neighbors (bottom, left) and the amount of $2$nd neighbors (bottom, right). 

As expected the failure of nodes does not create particular problems, since others arrive in the meantime. Thus, the topology remains pretty much unvaried. It is interesting to observe that however, when the amount of links is low, a small portion of nodes of the network can remain outside the main component when no failure management mechanisms are employed (see ``none'' curve on the left chart). 

Another interesting aspect is that, while small variations on the average amount of $1$st neighbors is noticed for the three schemes (the ``none'' protocol has a slight lower average value than the other two approaches), the average amount $2$nd neighbors is significantly lower for the ``none'' protocol w.r.t.~$P_{2n}$, $P_{ECC}$. 
In particular, with respect to the initial value, this measure decreases, on average, for ``none'', while it increases with $P_{2n}$, $P_{ECC}$.
This increment was expected. We are running the protocols in the evolution mode, thus nodes leave and enter the overlay at the same pace. When entering the network, novel nodes randomly create their initial amount of links, by randomly selecting their neighbors. Hence, the general network topology remains unchanged during the evolution.

The two self-healing protocols are local. Hence, they are thought to avoid that a node loses connections with some nodes in its 2nd-neighborhood. When we add this kind of approaches to a network that evolves in a stable manner (on average), the amount of links in the network will increase.
Depending on the application requirements, whenever this property is undesired, one might couple the protocol with the mentioned link reduction process, or by employing a low threshold on the maximum degree.
Indeed, we will see in Section \ref{sec:eval_degree} that changing the threshold on the maximum degree can lower significantly the amount of 1st and 2nd neighbors, without evident differences on the size of the main component.

\subsubsection{Clustered Networks}
\begin{figure*}[t]
   \centering
   \subfigure[Main component size]{
     \includegraphics[width=.6\linewidth]{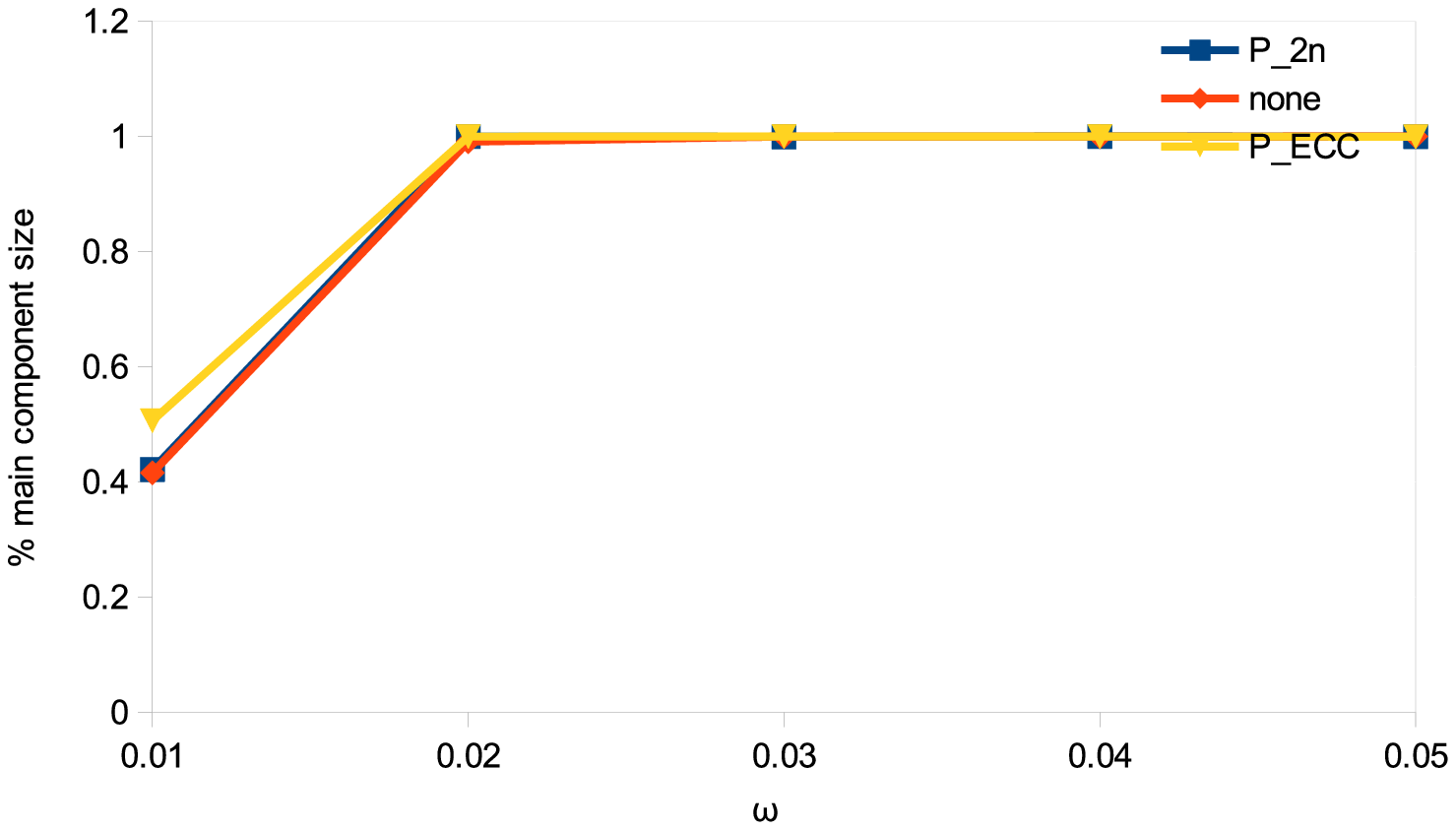}
   }
   \subfigure[Average amount of $1$st neighbors]{
     \includegraphics[width=.45\linewidth]{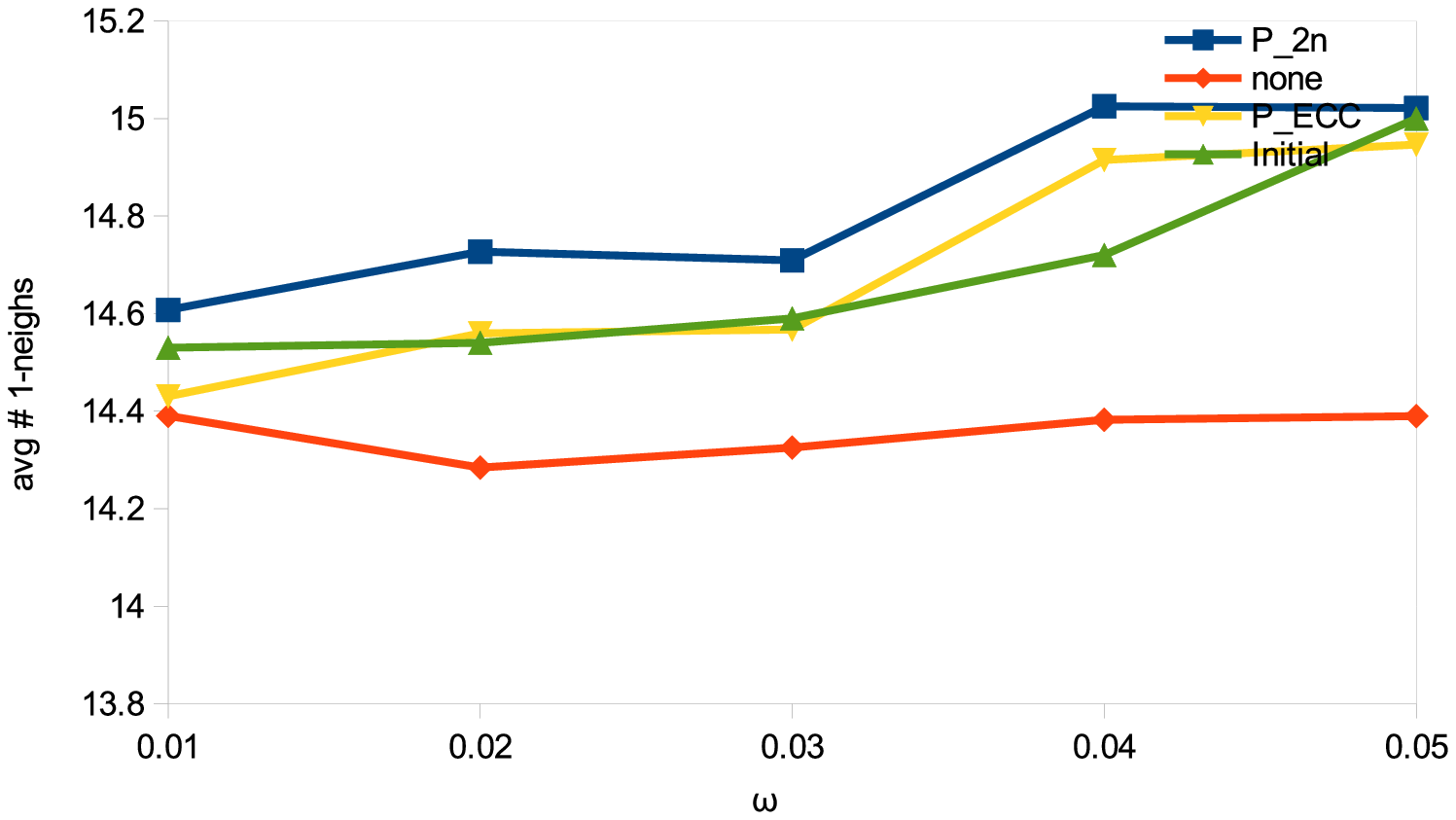}
   }
   \subfigure[Average amount of $2$nd neighbors]{
     \includegraphics[width=.45\linewidth]{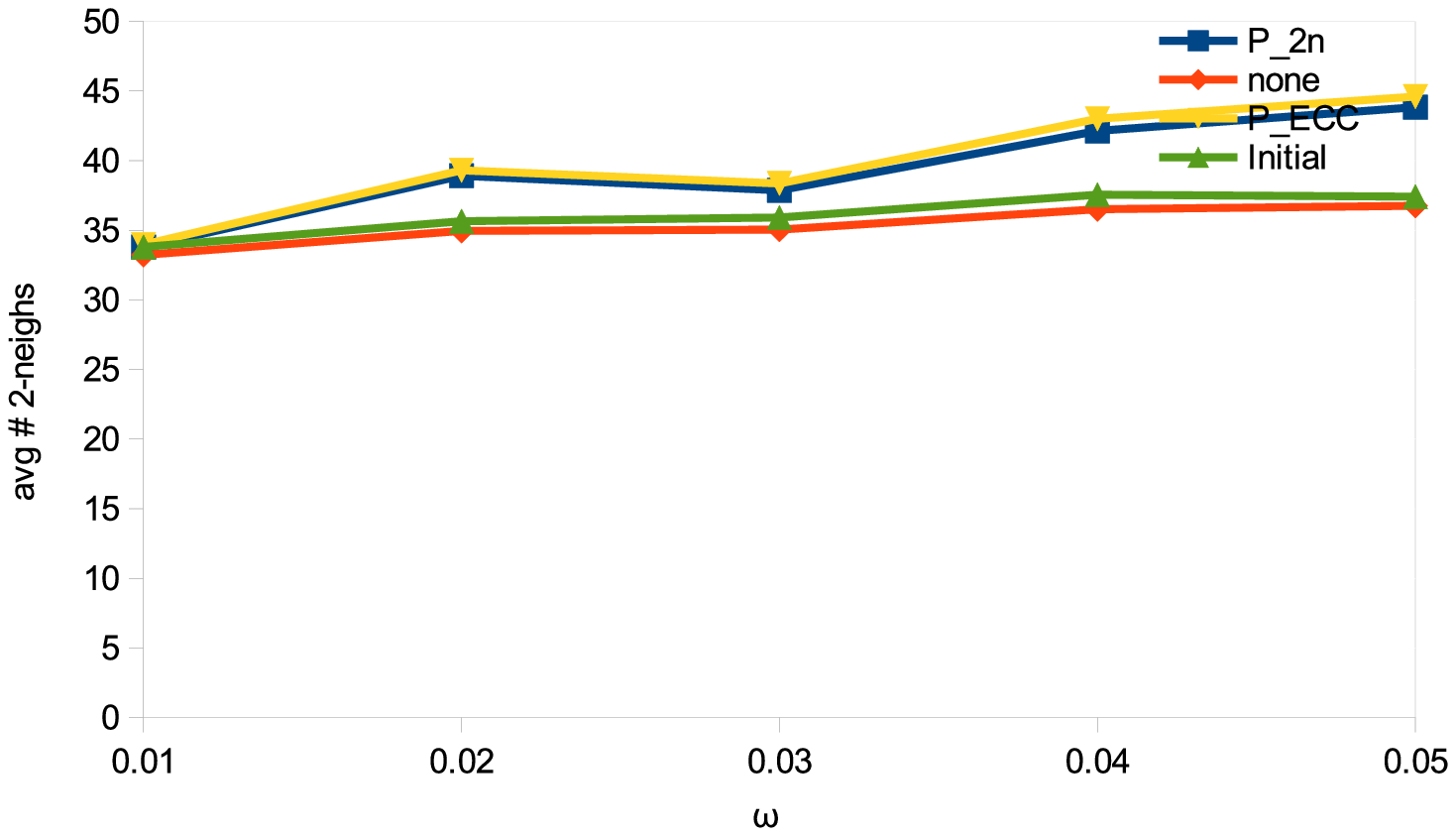}
   }
   \caption{Clustered networks -- evolution simulation mode.}
   \label{fig:clus_evol}
\end{figure*}

When dealing with clustered networks, also in this case a random removal of nodes (``evolution'' simulation mode) does not alter significantly the topology; hence, as concerns the main component size no particular benefits are evident from the use of $P_{2n}$ and $P_{ECC}$ w.r.t.~``none'' (see Figure \ref{fig:clus_evol}).

An interesting result is that with the ``none'' protocol a lower average node degree is measured, while higher values are obtained with $P_{2n}$ and $P_{ECC}$. In particular, $P_{ECC}$ provides values which are nearer the initial ones.
As for uniform networks, the amount of $2$nd-neighbors increases with $P_{2n}$ and $P_{ECC}$.

\subsubsection{Scale-Free Networks}
\begin{figure*}[t]
   \centering
   \subfigure[Main component size]{
     \includegraphics[width=.6\linewidth]{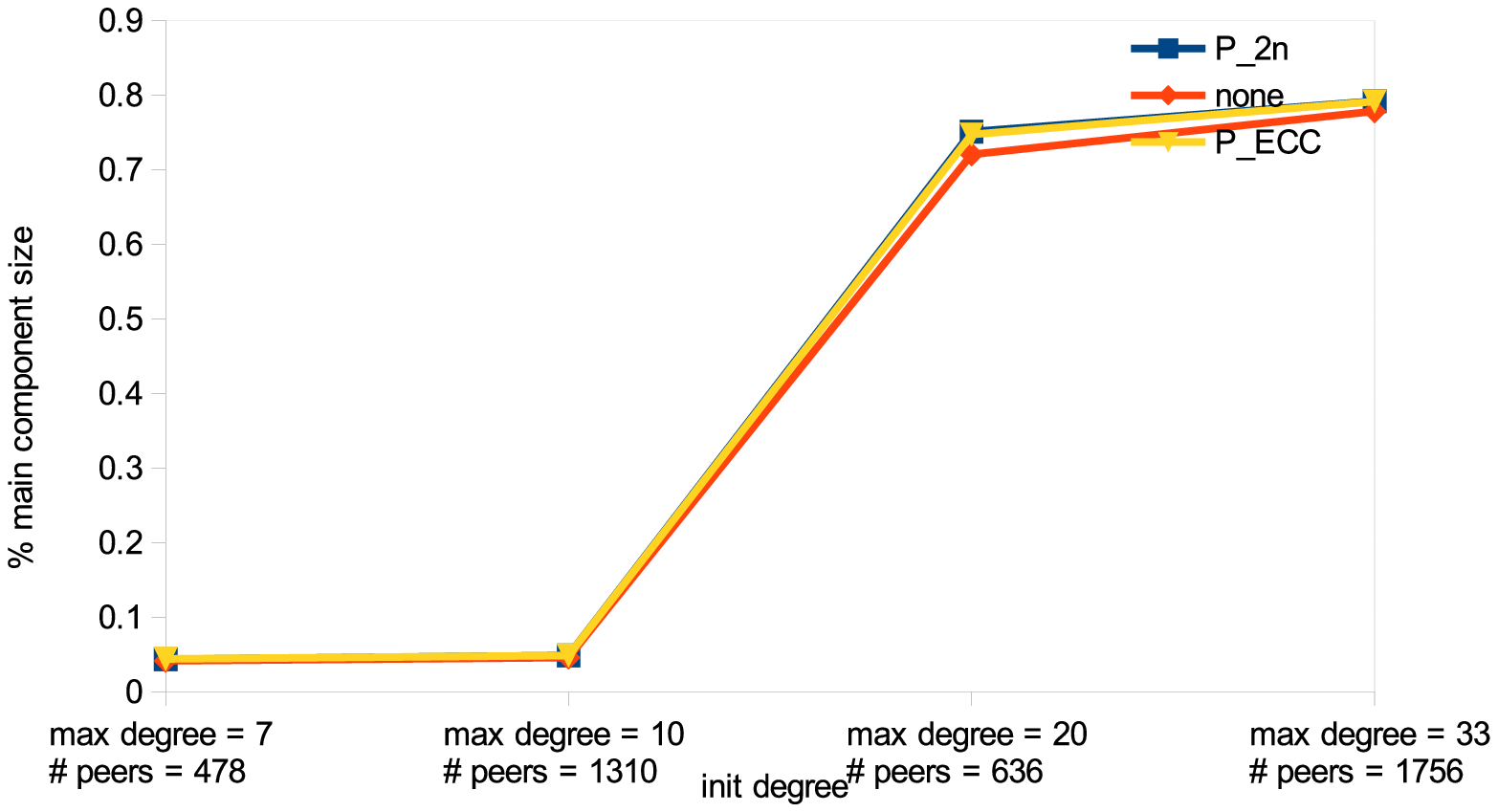}
   }
   \subfigure[Average amount of $1$st neighbors]{
     \includegraphics[width=.45\linewidth]{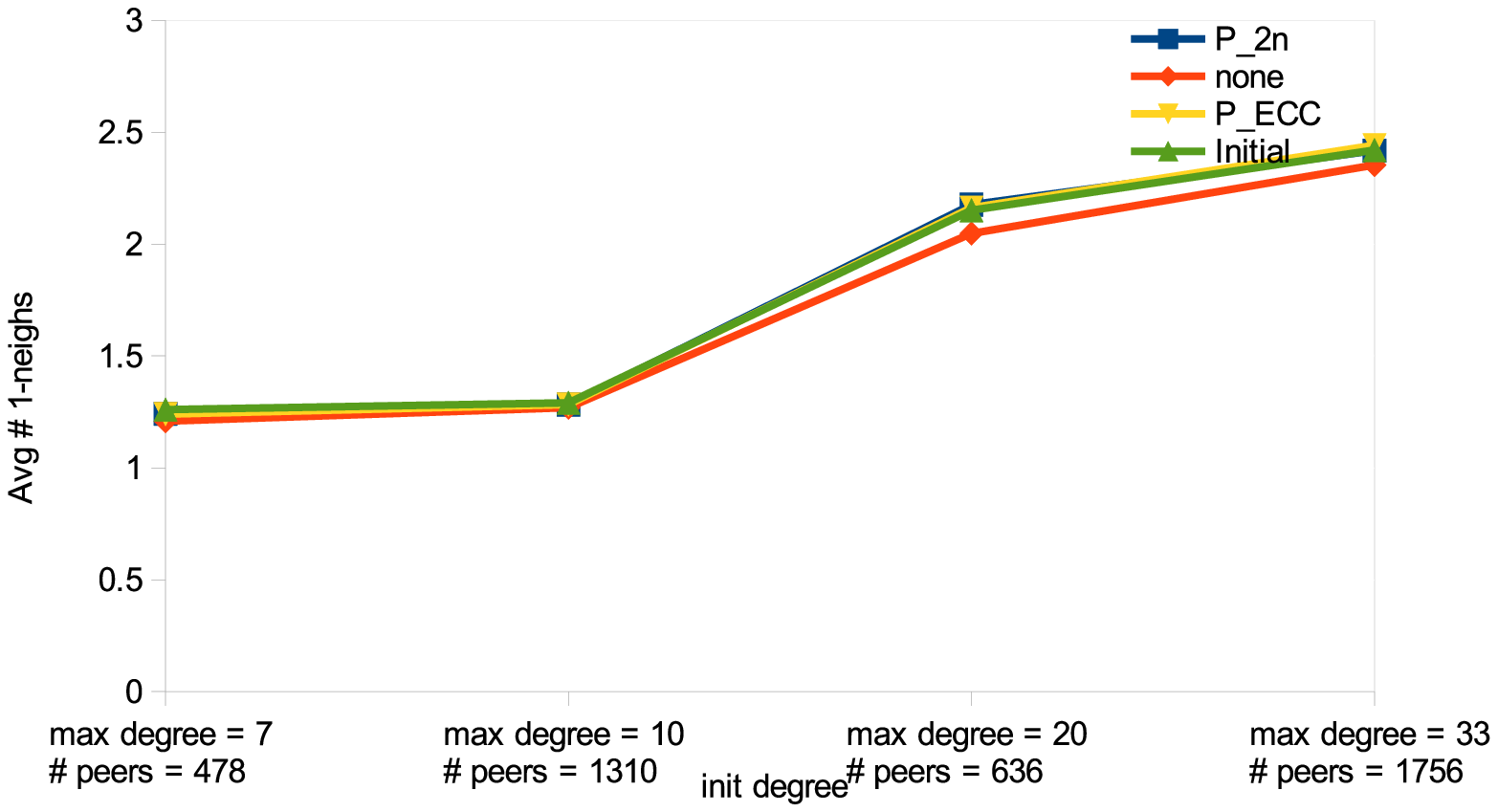}
   }
   \subfigure[Average amount of $2$nd neighbors]{
     \includegraphics[width=.45\linewidth]{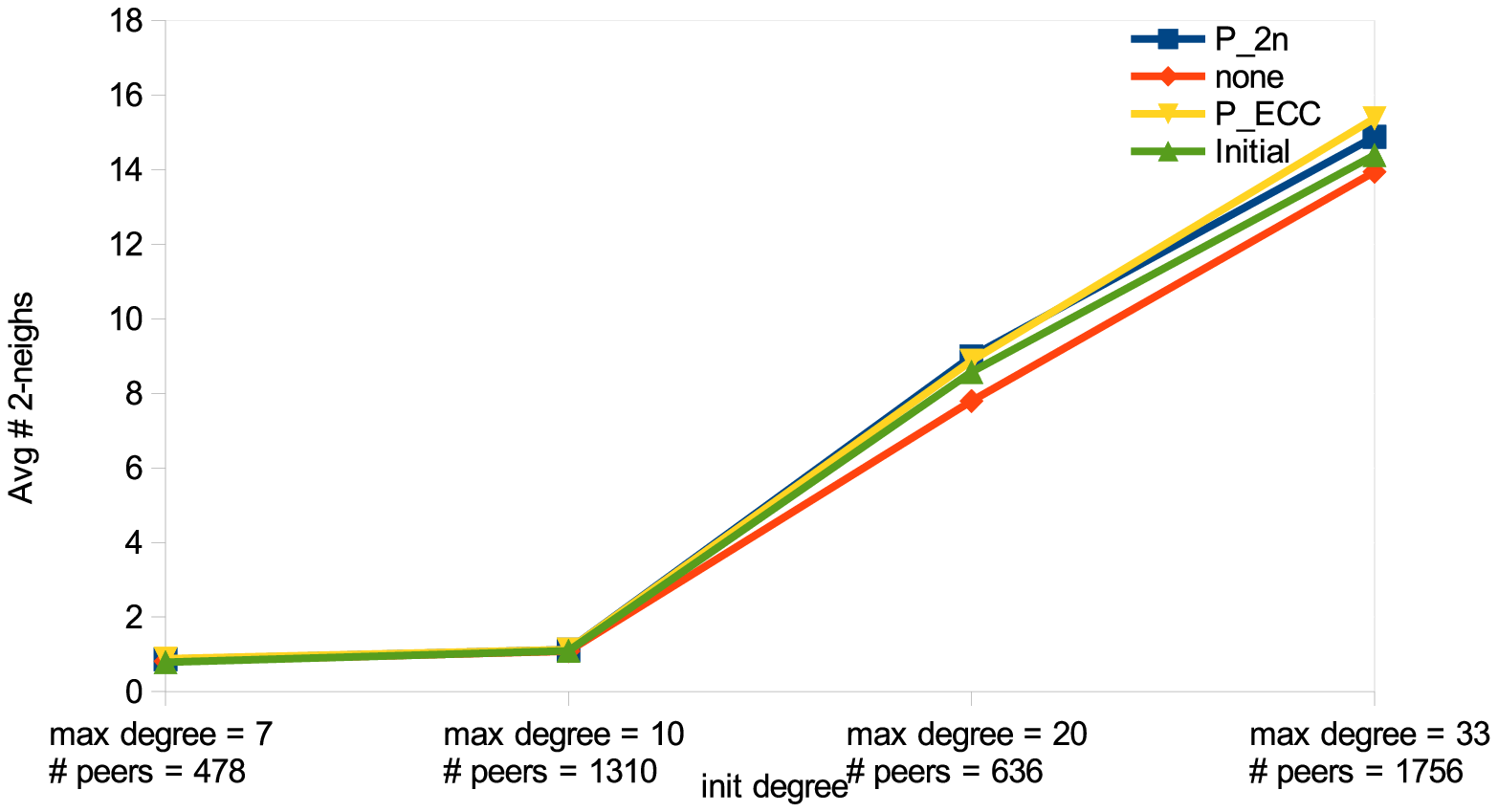}
   }
   \caption{Scale-free networks -- evolution simulation mode.}
   \label{fig:sf_evol}
\end{figure*}

Under the simulation evolution, no noticeable differences are evident for 
scale-free networks (see Figure \ref{fig:sf_evol}). One might notice that the 
first two considered scale-free networks are very disconnected ones. Hence, even 
if the degree distribution follows a power law, there are no real hubs that do 
connect all subnetworks.

\subsection{Targeted Attack to Node with Highest Degree}

\subsubsection{Uniform Networks}
\begin{figure*}[t]
   \centering
   \subfigure[Main component size]{
     \includegraphics[width=.6\linewidth]{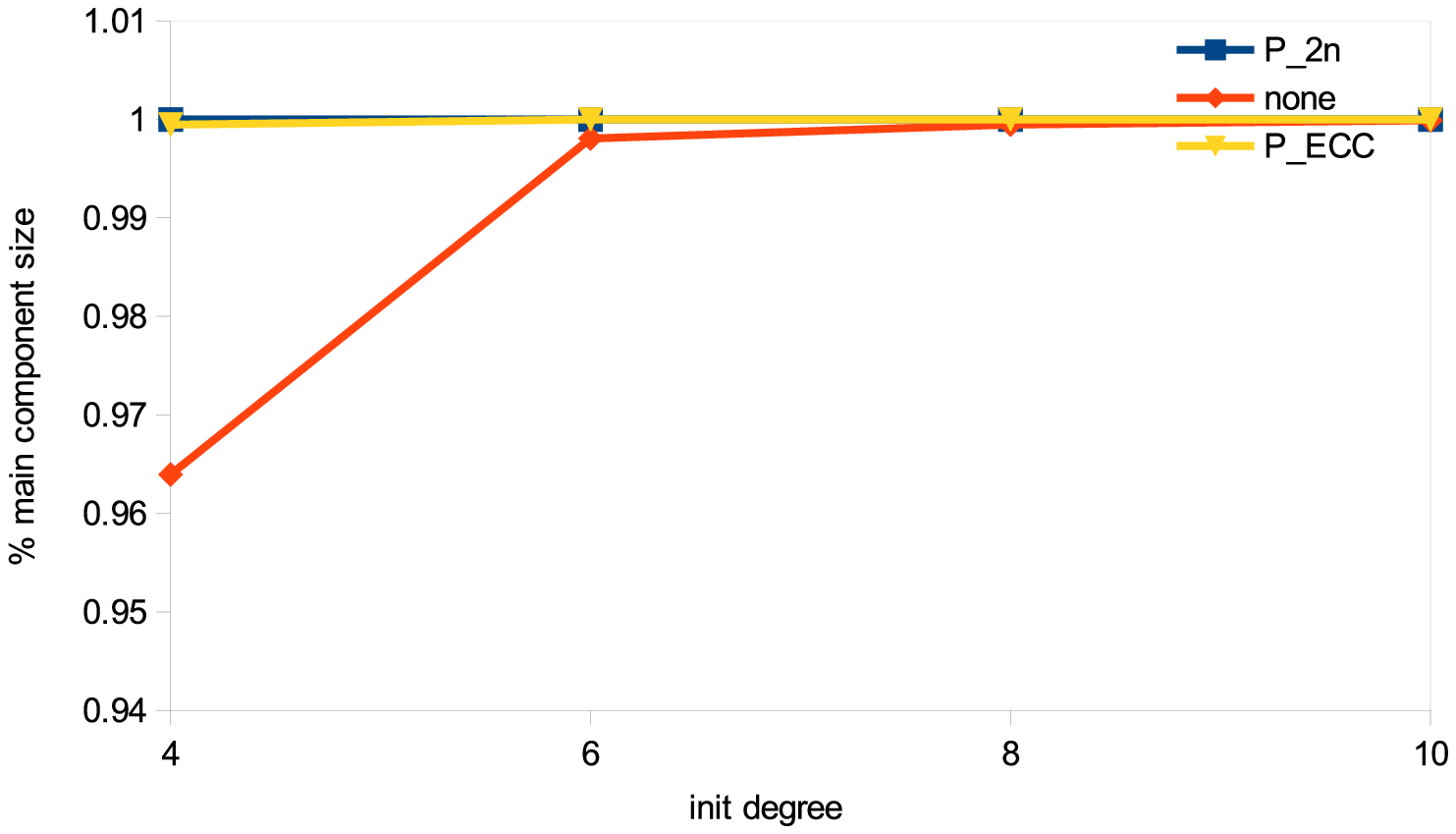}
   }
   \subfigure[Average amount of $1$st neighbors]{
     \includegraphics[width=.45\linewidth]{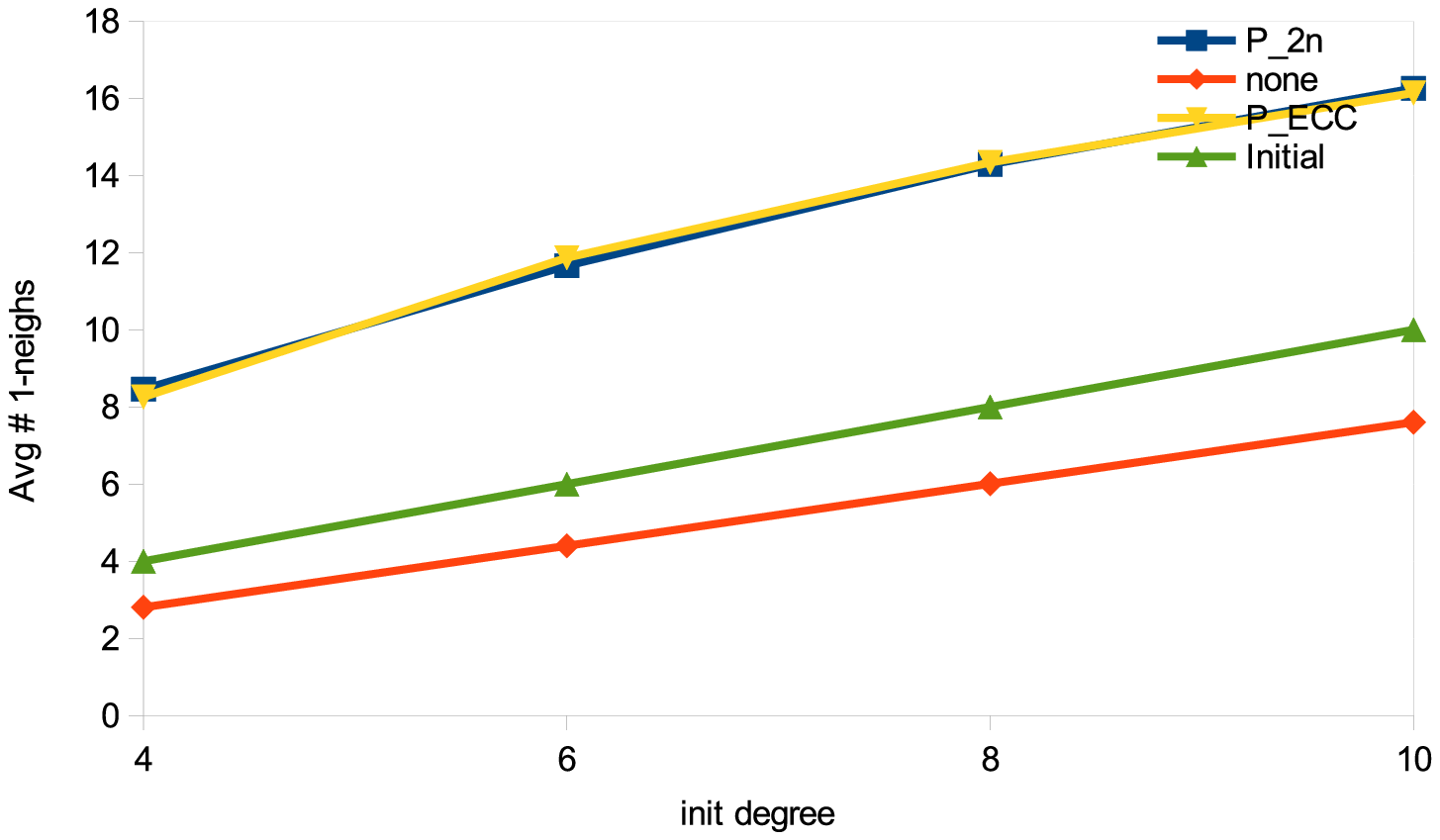}
   }
   \subfigure[Average amount of $2$nd neighbors]{
     \includegraphics[width=.45\linewidth]{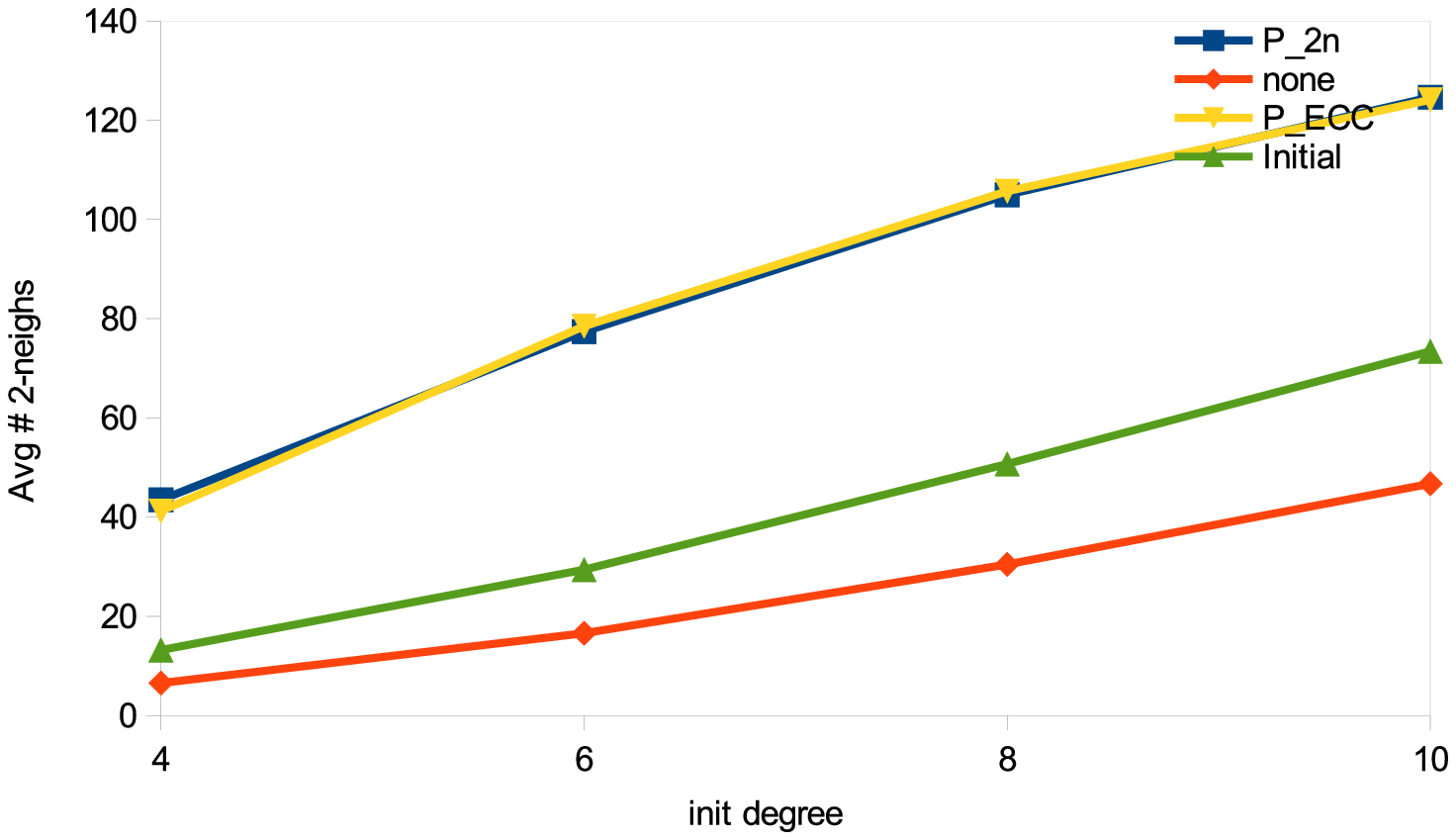}
   }
   \caption{Uniform networks -- Targeted attack simulation mode.}
   \label{fig:unif_targ}
\end{figure*}

When considering the targeted attack simulation mode with uniform networks results are not
that different to those performed during the evolution. Indeed, there are no important differences between nodes,
since all start with the same initial degree, during the network evolution links 
are established arbitrarily, and there are no important hubs in the network. 
Thus, the selection
of the node with highest degree has not a significant impact on
the topology (see Figure \ref{fig:unif_targ}). Nevertheless, it is possible to 
appreciate that the
numbers of 1st and 2nd neighbors decrease for the “none” protocol,
w.r.t. results obtained for the evolution simulation modes.
Similarly, in the “none” protocol the average size of the
main component results lower w.r.t. that obtained in the evolution simulations. 

Conversely, as concerns the average main component size, results remain 
unchanged for $P_{2n}$ and $P_{ECC}$.  
Instead, the numbers of 1st and 2nd neighbors increase. This can be explained as follows. 
Uniform networks are quite similar to random graphs, as links are established 
arbitrarily. Hence, there is a low clustering.  
Let consider a node $n$; upon failure of one of its neighbors, let say node $f$, 
due to the network topology it is unlikely that $n$ has as 1st neighbors the 
nodes that were connected to $f$. Thus, $P_{2n}$, and $P_{ECC}$ will force $n$ 
to create novel links with $f$'s neighbors. This is even more evident if we 
select the nodes with highest degree to fail. 

As previously stated, the approach to adopt, in order to cope with this possible 
issue, is application dependent. If the increment mentioned above is 
undesirable, one might employ a link reduction process, 
adding a limit on the maximum degree when creating links, or more drastically, 
turning off the self-healing protocols. 
As we will see in Section \ref{sec:eval_degree}, in some cases the introduction of a lower threshold on the maximum node degree does not alter the connectivity provided by $P_{2n}$ and $P_{ECC}$.

\subsubsection{Clustered Networks}
\begin{figure*}[t]
   \centering
   \subfigure[Main component size]{
     \includegraphics[width=.6\linewidth]{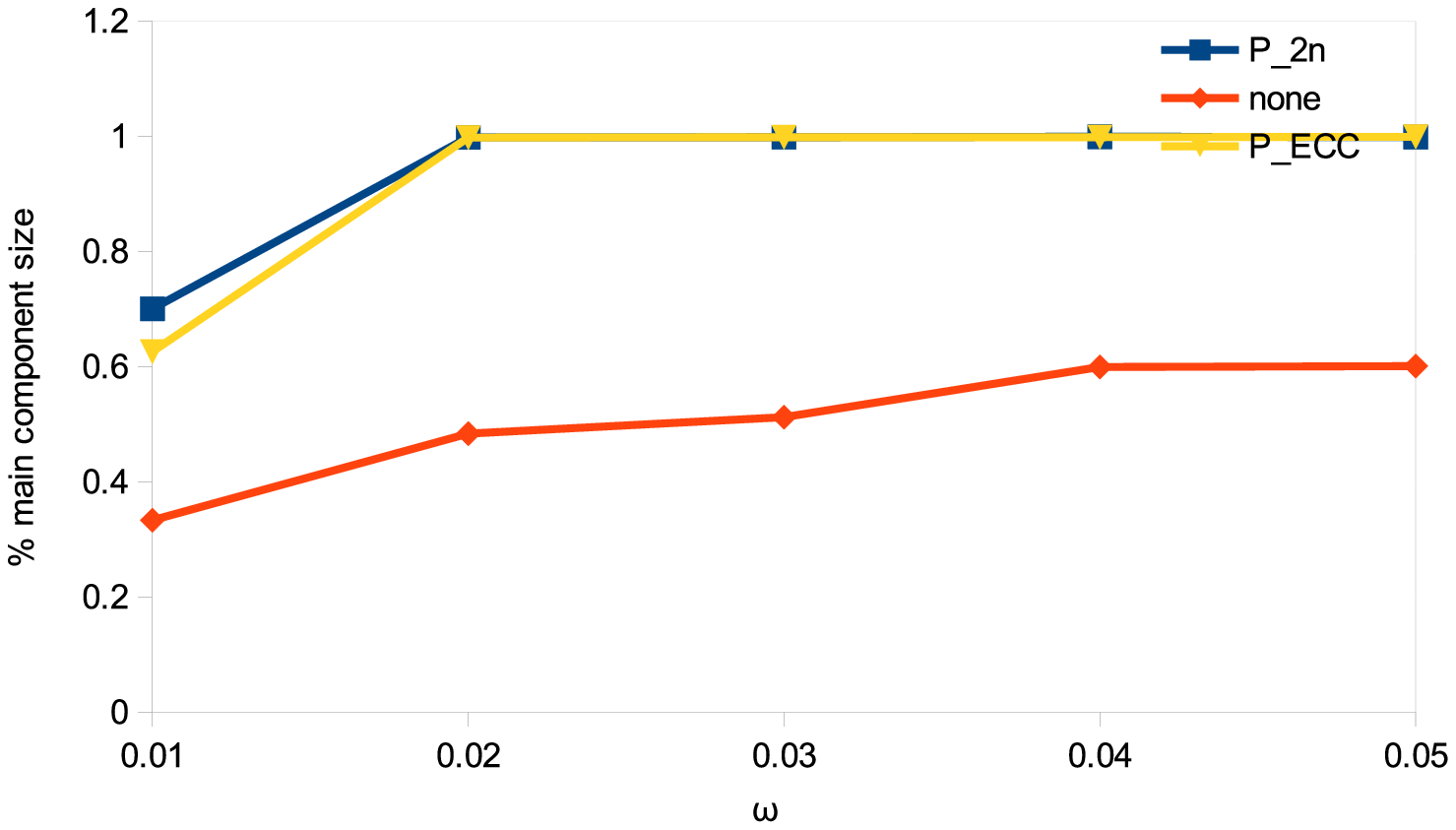}
   }
   \subfigure[Average amount of $1$st neighbors]{
     \includegraphics[width=.45\linewidth]{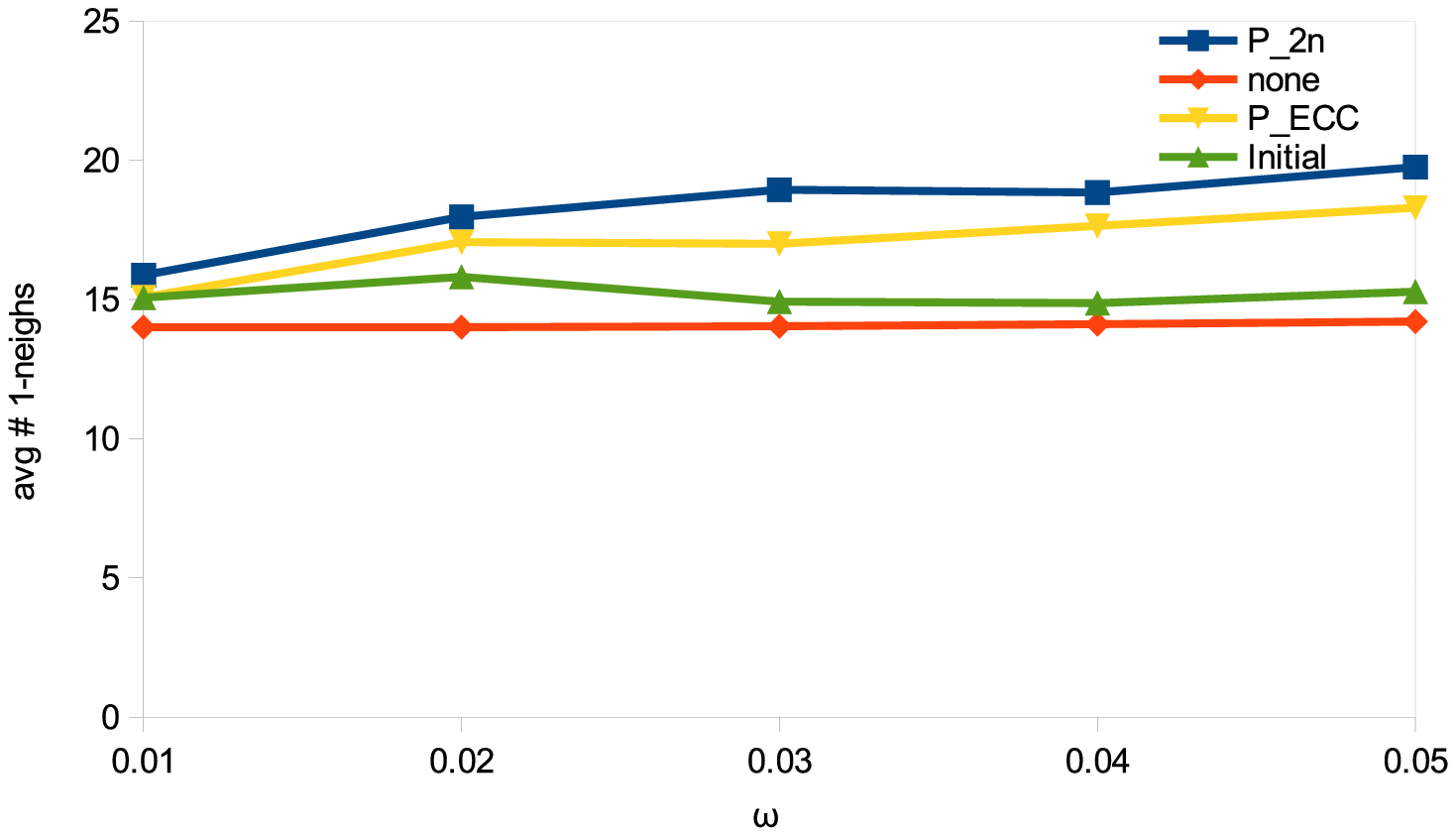}
   }
   \subfigure[Average amount of $2$nd neighbors]{
     \includegraphics[width=.45\linewidth]{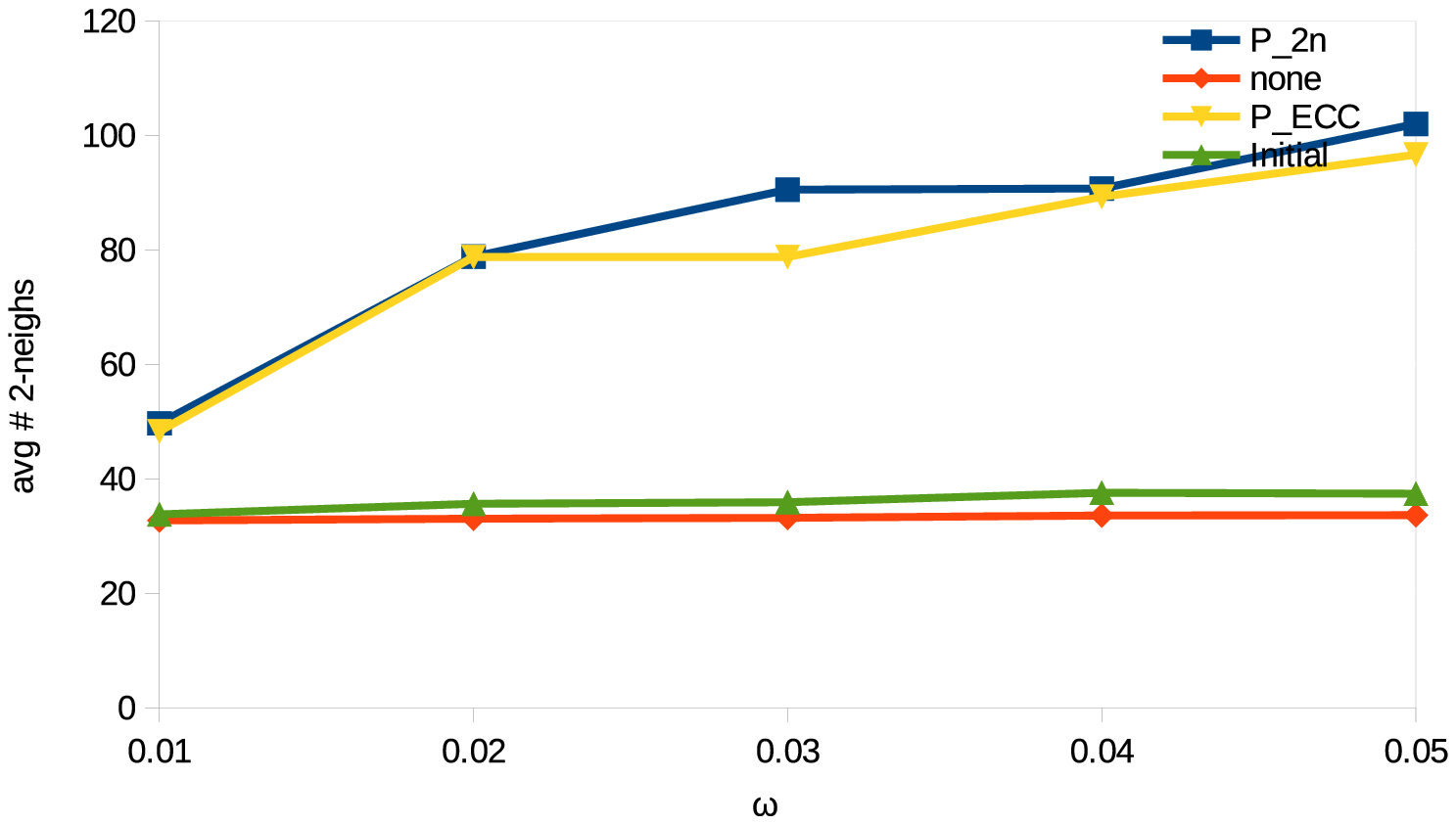}
   }
   \caption{Clustered networks -- targeted attack simulation mode.}
   \label{fig:clus_targ}
\end{figure*}

Clustered networks are particularly affected by the selection of targeted nodes 
with highest inter-cluster degrees. Indeed, with the ``none'' protocol the 
average size of the main component is highly reduced, while the two self-healing 
protocols $P_{2n}$, $P_{ECC}$ maintain a high (full) connectivity, as shown in 
Figure \ref{fig:clus_targ}.
This confirms the goodness and usefulness of the proposed protocols in these situations.
The increment on the average amount of 1st neighbors is limited, with $P_{ECC}$ that provides a slightly lower increment with respect to $P_{2n}$. 
Conversely, the use of the self-healing protocols causes an increment on the amount of 2nd 
neighbors (again, $P_{ECC}$ has a lower increment w.r.t.~$P_{2n}$). 
This is explained by the fact that, based on the clustered topology, only a 
limited amount of nodes have links with nodes in other clusters.
Without the self-healing protocols, these clusters become disconnected. Instead,
with the self-healing protocols the neighbors of the failed node share the task 
of replacing these inter-cluster connections. 
Thus, it is likely that multiple nodes create links towards other clusters (and 
also some links within the cluster). 

\subsubsection{Scale-Free Networks}
\begin{figure*}[t]
   \centering
   \subfigure[Main component size]{
     \includegraphics[width=.6\linewidth]{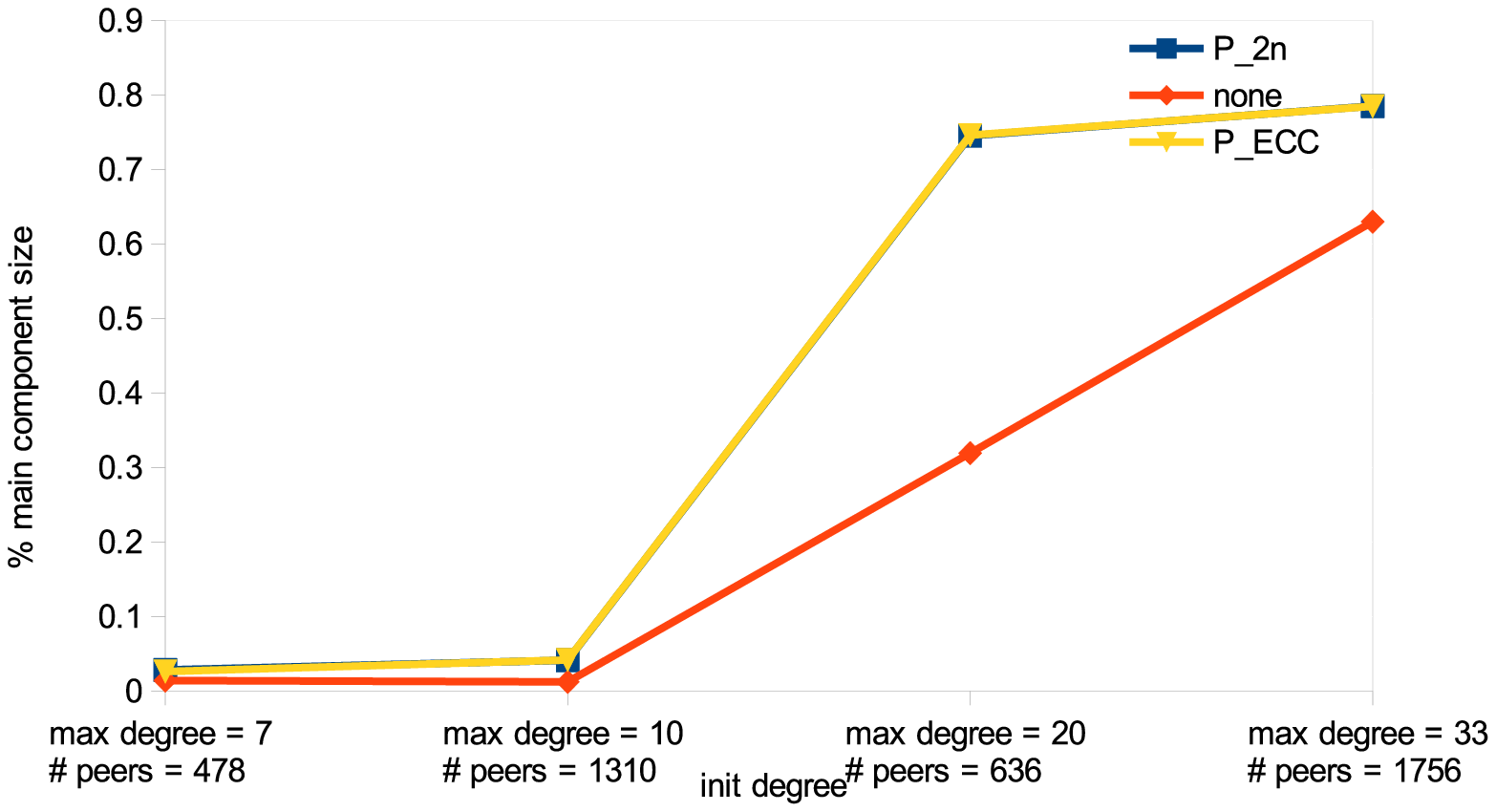}
   }
   \subfigure[Average amount of $1$st neighbors]{
     \includegraphics[width=.45\linewidth]{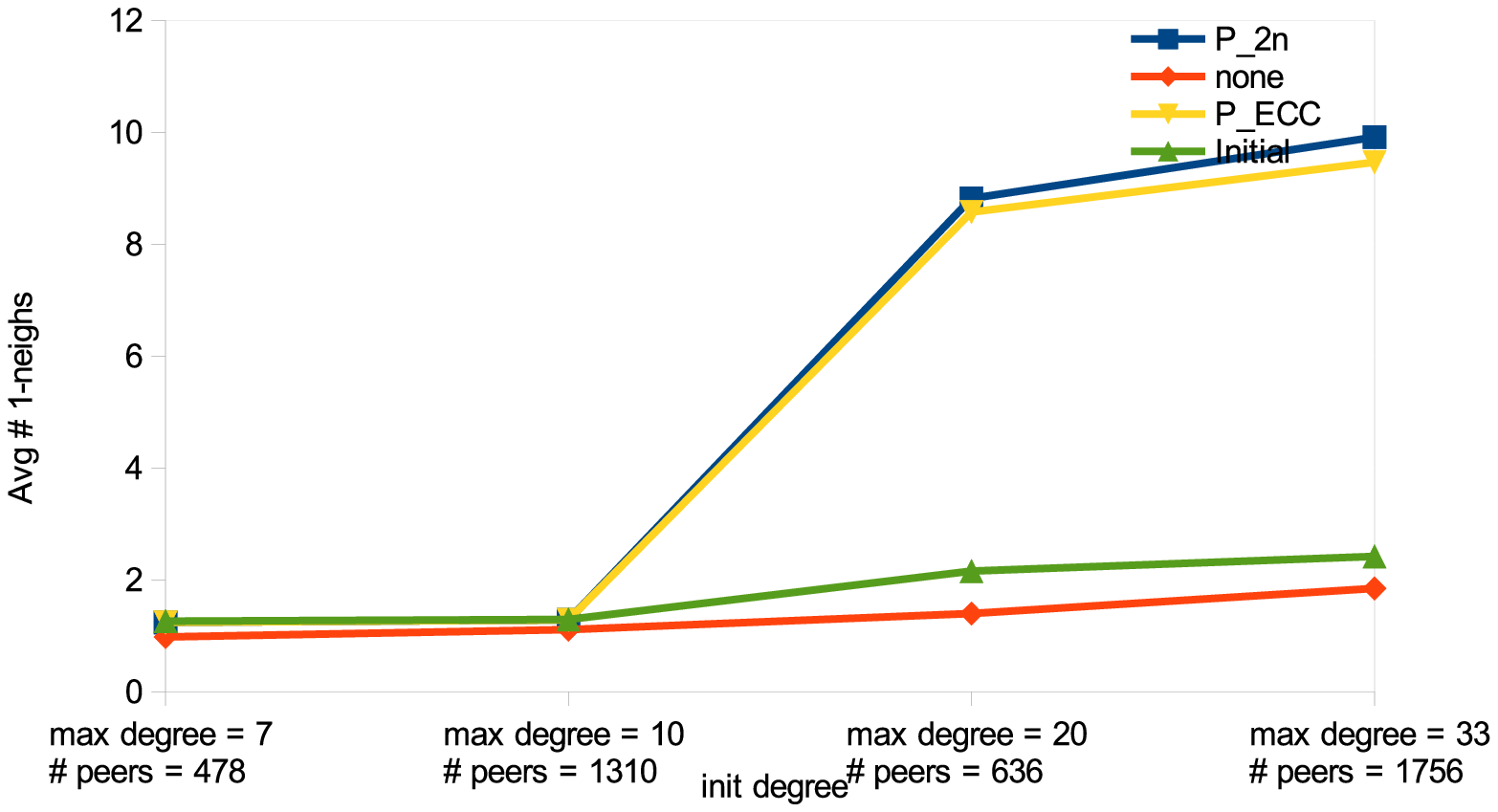}
   }
   \subfigure[Average amount of $2$nd neighbors]{
     \includegraphics[width=.45\linewidth]{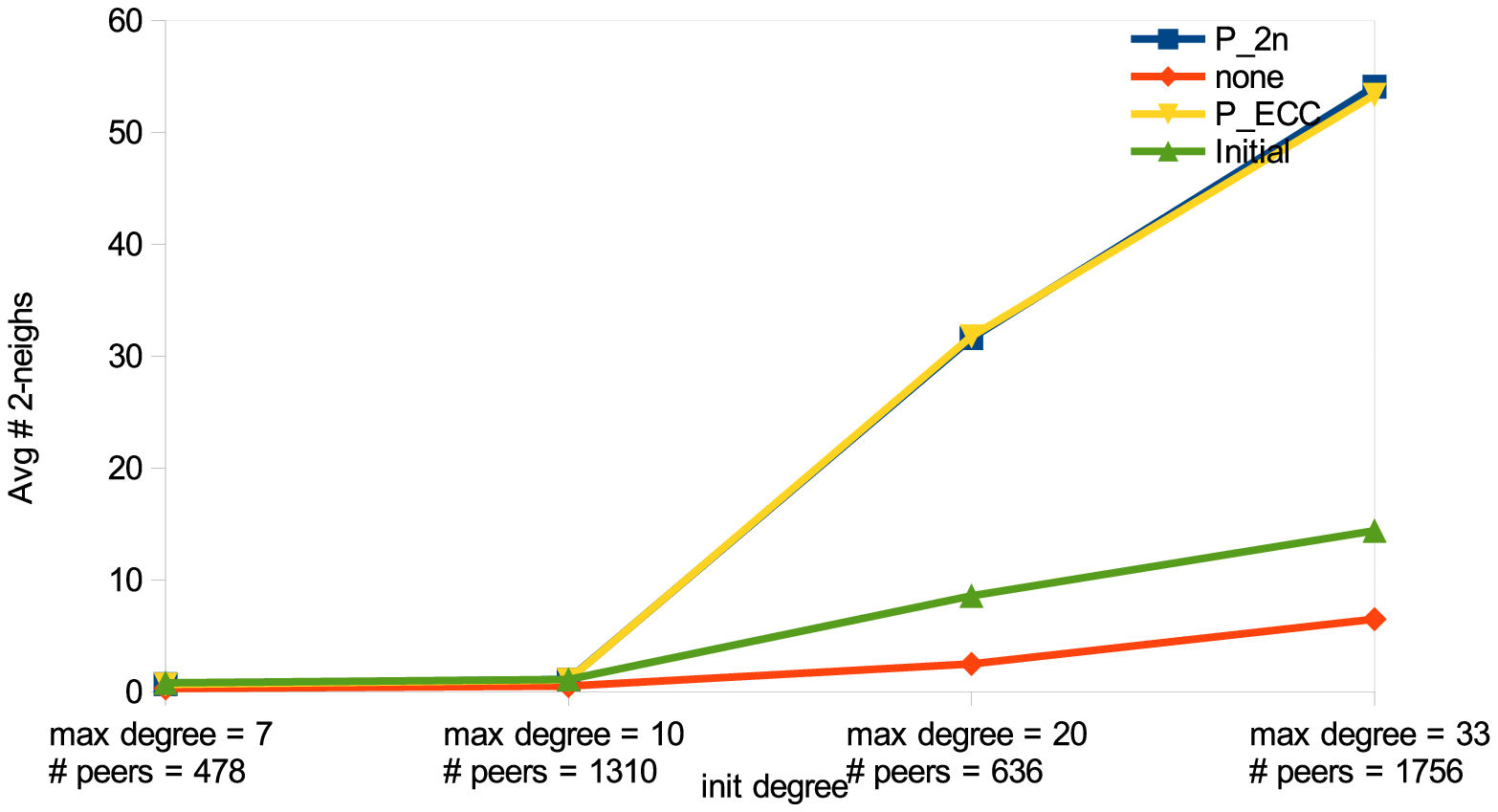}
   }
   \caption{Scale-free networks -- targeted attack simulation mode.}
   \label{fig:sf_targ}
\end{figure*}

The two protocols work well even for scale-free networks under the targeted 
attack. Figure \ref{fig:sf_targ} shows that $P_{2n}$ and $P_{ECC}$ guarantee 
high connectivity, at the cost of a little increment on the average degree. But 
again, this is expected, since while hubs fail, there are other nodes that enter 
the network at the same rate.
Conversely, the connectivity level decreases without the use of a self-healing 
protocol (i.e.,~``none'' protocol).
This is a well known result in the literature, as it has been recognized 
already that scale-free networks are not resilient to targeted attacks 
\cite{newmanHandbook}.

\subsection{Failure Churn}
As mentioned, this is the scenario where nodes progressively fail. It is an 
interesting experiment to assess whether the protocols are able to cope with 
extreme churn. 

\subsubsection{Uniform Networks}
\begin{figure*}[t]
   \centering
   \subfigure{\includegraphics[width=.45\linewidth]{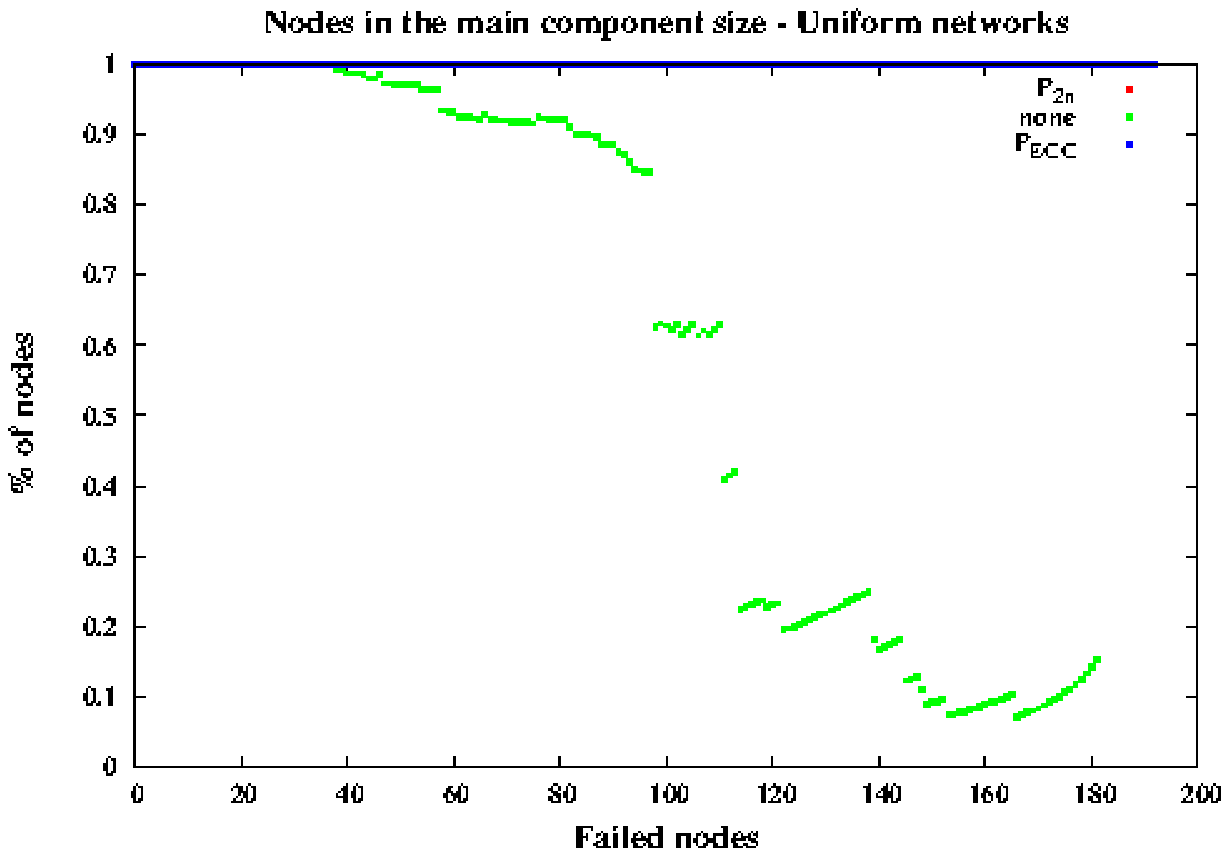}}
\subfigure{\includegraphics[width=.45\linewidth]{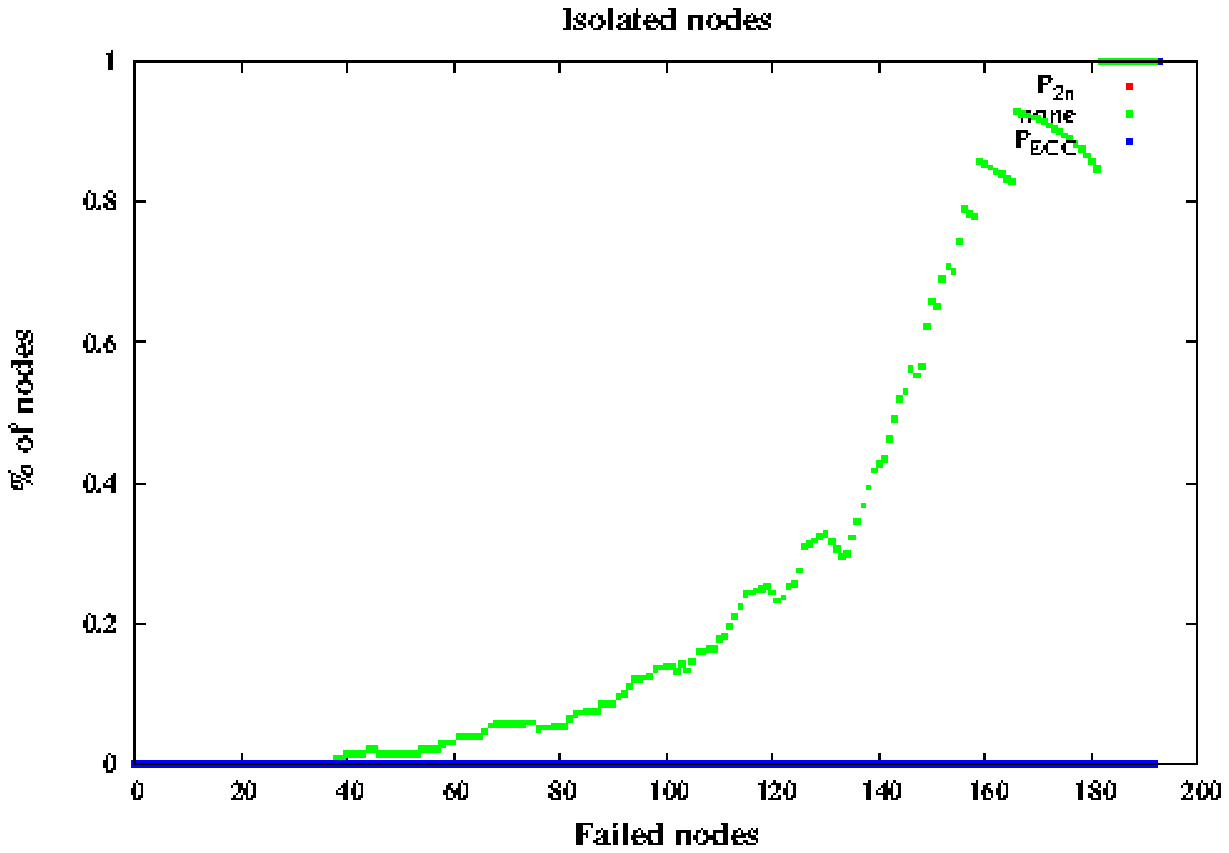}}
   \caption{Uniform networks -- progressive node failures: Amount of nodes in the main component, isolated nodes.}
   \label{fig:unif_fail}
\end{figure*}

Figure \ref{fig:unif_fail} shows results obtained under the ``failures only'' 
simulations with uniform networks. In particular, the chart on the left shows 
the amount of nodes that remain in the main component, while nodes continuously 
fail. (We repeated the same experiment multiple times, varying the network size, 
the initial nodes' degree, and the seed for random generations, obtaining 
comparable results.)
It is possible to see that, in the ``none'' protocol, at a certain point of the 
simulation the network gets disconnected and the percentage of active nodes in 
the main components decreases. Instead, in $P_{2n}$ and $P_{ECC}$, active nodes 
remain connected in the same, single component.  
This is confirmed by looking at the chart on the right in the same figure, which 
shows the amount of isolated nodes. While the percentage of isolated nodes 
increases in the ``none'' scheme, no nodes remain isolated for the other two 
protocols. 

\begin{figure*}[t]
   \centering
   \includegraphics[width=.45\linewidth]{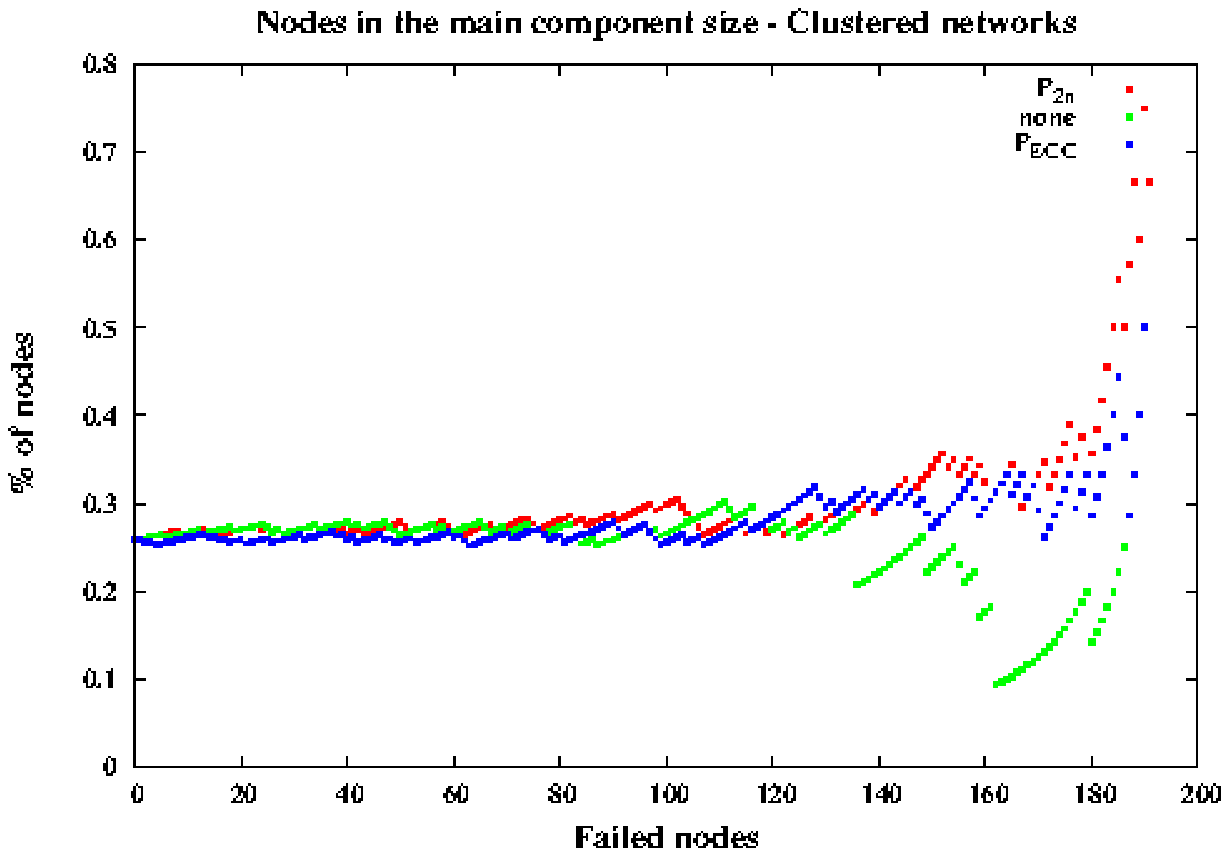}
   \includegraphics[width=.45\linewidth]{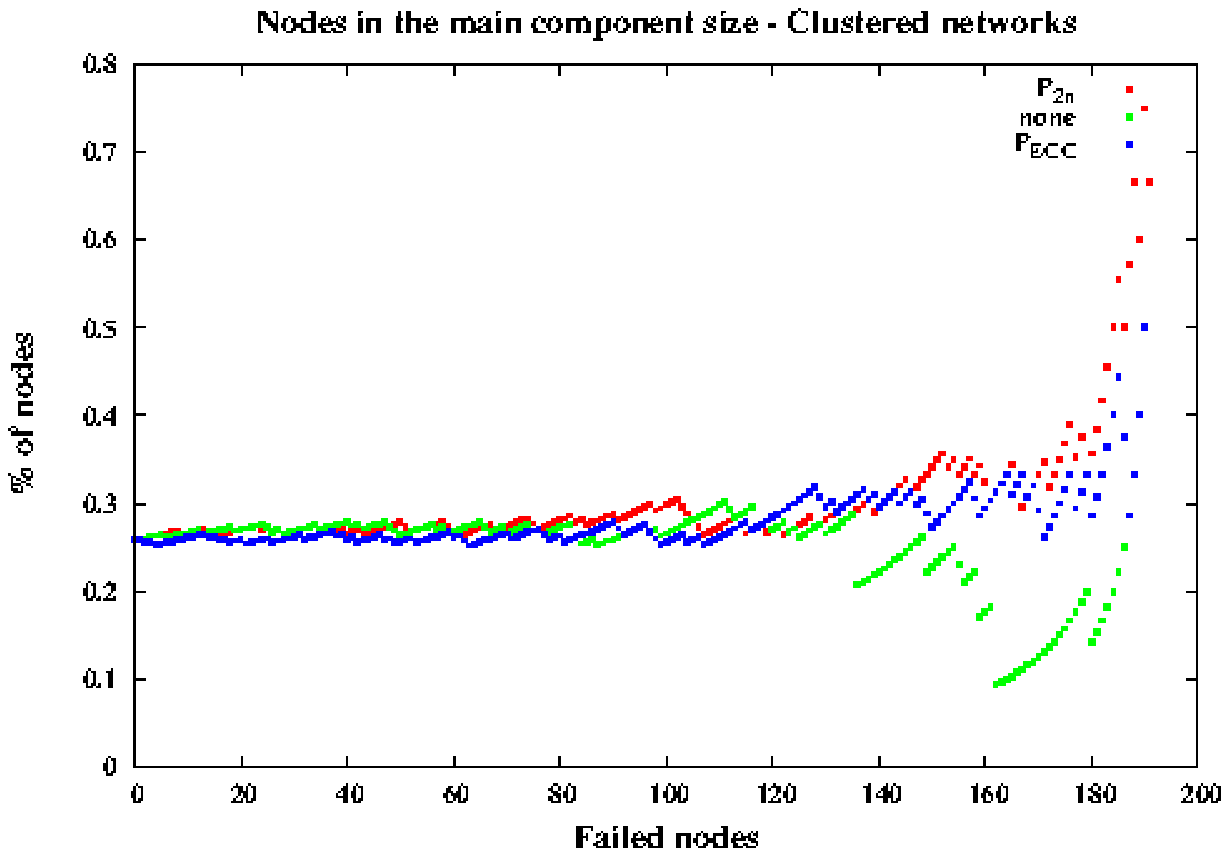}
   \caption{Clustered networks -- progressive node failures: Amount of nodes in the main component, isolated nodes.}
   \label{fig:clus_fail}
\end{figure*}

\subsubsection{Clustered Networks}
Figure \ref{fig:clus_fail} shows results obtained with clustered networks in the 
``failures only'' simulation mode. As for uniform networks, the chart on the 
left shows the amount of nodes that remain in the main component during the 
evolution, while nodes continuously fail. 
In this case, the network was disconnected, in the sense that the main component 
comprised only a percentage (slightly over $25\%$) of the whole set of nodes. 
We might see that in this case, the main component size remains almost stable, 
for all the three protocols, until a half of the nodes become disconnected. This 
is due to the fact that the random choice of the failing nodes would privilege 
those nodes that were not in the bigger component (that includes less than the 
$30\%$ of nodes). However, in the last part of the simulation run, the ``none'' 
protocol experiences a progressive decrement of nodes in the main component, 
since the main component is partitioned by the failures of its nodes. 
Conversely, the size of the main component increases for $P_{2n}$ and $P_{ECC}$. 
This is explained by the presence of the failure management protocols, that 
prevent the partition of the components. 
The chart on the right of the figure confirms this, by reporting the amount of 
isolated nodes. While this amount progressively increases with the ``none'' 
protocol, with $P_{2n}$, $P_{ECC}$ the percentage of isolated nodes remains 
negligible for the main part of the simulation. Only at the end of the 
simulation some non-negligible amount of isolated nodes appears. This is 
explained by the fact that after a while some (minor) component remained 
composed of a single node (all other nodes already failed). 

\begin{figure*}[t]
   \centering
   \includegraphics[width=.45\linewidth]{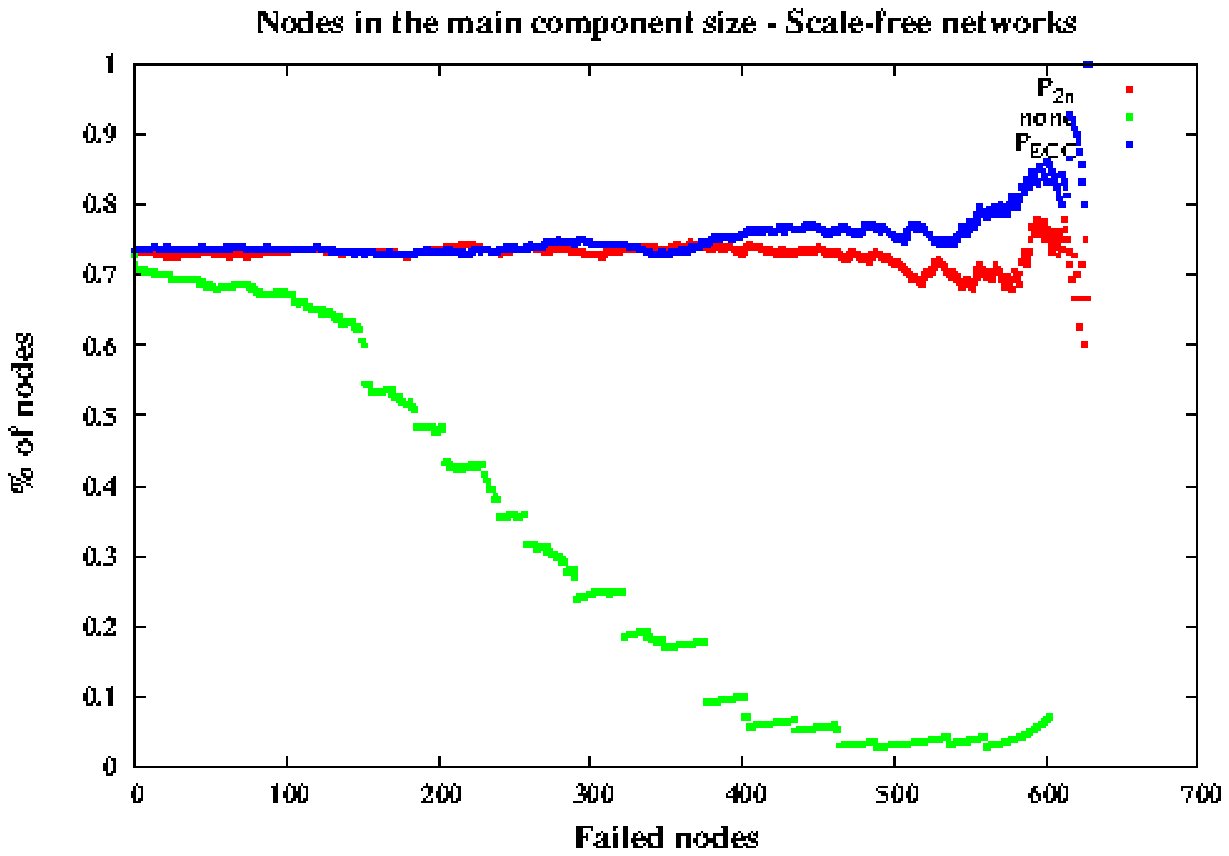}
   \includegraphics[width=.45\linewidth]{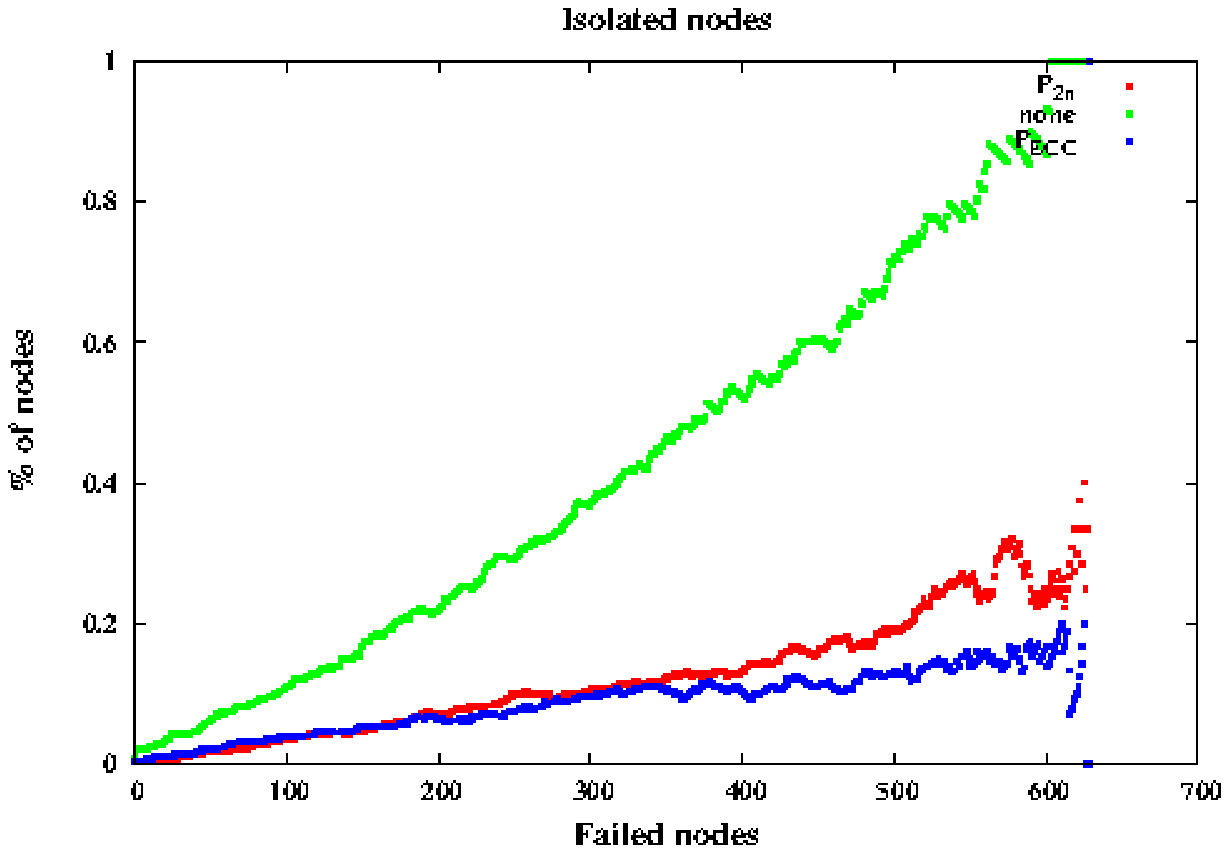}
   \caption{Scale-free networks -- progressive node failures: Amount of nodes in the main component, isolated nodes.}
   \label{fig:sf_fail}
\end{figure*}

\subsubsection{Scale-Free Networks}
Figure \ref{fig:sf_fail} reports results for a ``failures only'' simulation 
mode, run on a scale-free network composed of $636$ nodes, with a maximum degree 
of $20$ (for those interested in the specific construction method 
\cite{Aiello00arandom}, it employs two parameters that in this case were set to 
$a=6$, $b=2$).  
By looking at the chart on the right, it is possible to see that the simulation 
starts with a main component composed of more than the $70\%$ of the nodes. In 
the ``none'' mode, the component size progressively loses all its nodes, while 
in the $P_{2n}$ and $P_{ECC}$ protocols, the main component maintains its size 
(which actually increases in percentage, upon failure of nodes outside the main 
component). Actually, in this case $P_{ECC}$ outperforms $P_{2n}$. This is 
confirmed by the chart on the right in the figure, that reports the amount of 
isolated nodes.

\subsection{Targeted Attack to Nodes with Highest Betweenness}

It is generally accepted that in many networks the larger the degree the larger 
the betweenness \cite{holme2002attack}. The idea is that the higher the degree 
of a node the higher the probability that a path might pass through it. 
However, as previously stated this depends on the network topology. 

As concerns scale-free networks, for instance, it has been noticed that, unless 
the network has been built with a high level of disassortativity (i.e.~high 
repulsion between hubs), in general there is a high 
correlation between the degree of a node and its betweenness centrality 
\cite{PhysRevE.75.056115}. 
In this case, results obtained with a targeted attack to nodes with higher 
degrees are comparable to those obtained with targeted attacks to nodes 
with highest betweenness. 

When considering clustered networks, instead, nodes with higher betweenness 
might be those that are connected with different clusters. 
Thus it is important to study this kind of attack when dealing with clustered networks. 
For this reason and for the sake of conciseness, we focus here on clustered networks only.

\begin{figure*}[t]
   \centering
   \subfigure[Main component size]{
     \includegraphics[width=.6\linewidth]{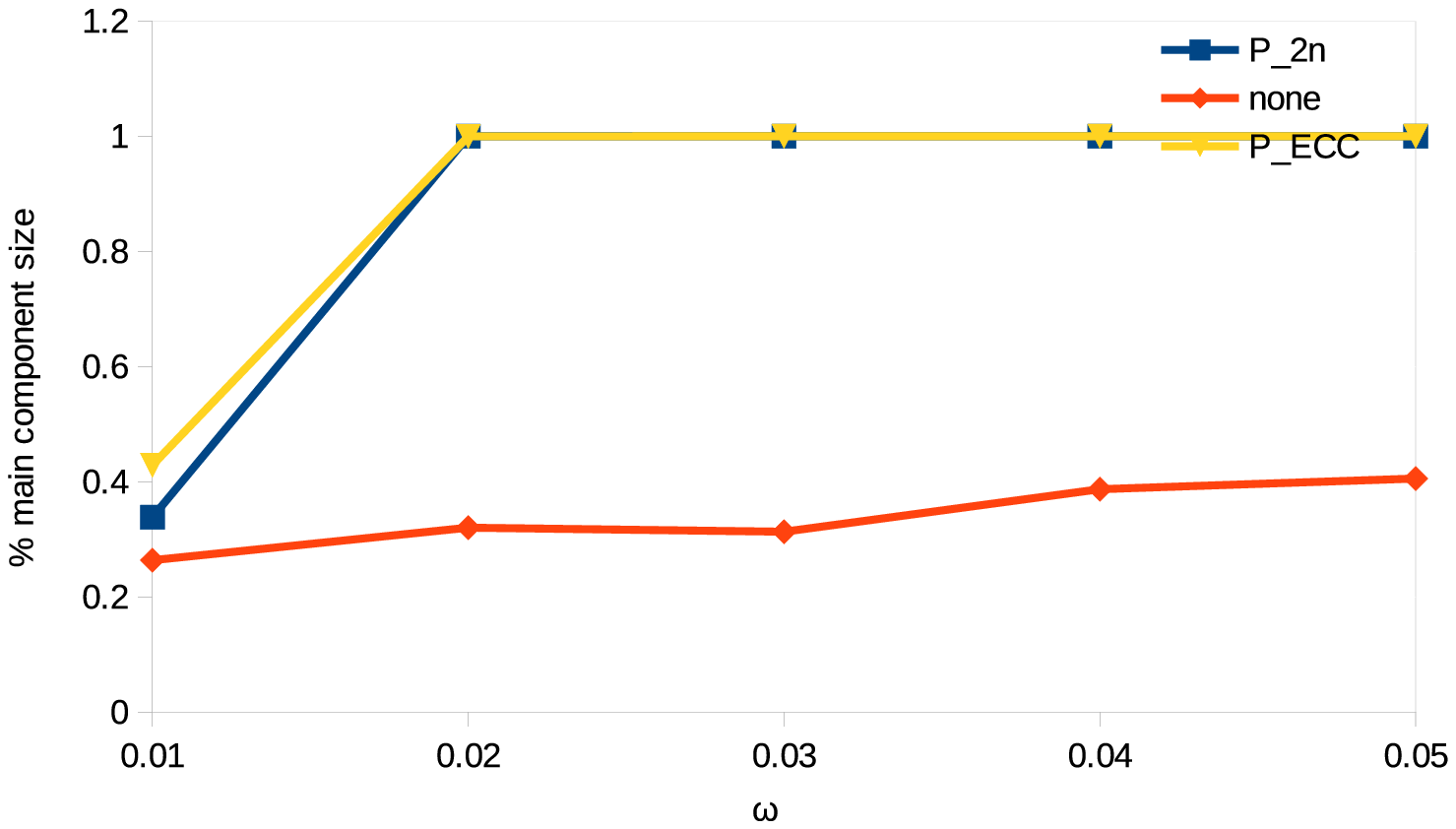}
   }
   \subfigure[Average amount of $1$st neighbors]{
     \includegraphics[width=.45\linewidth]{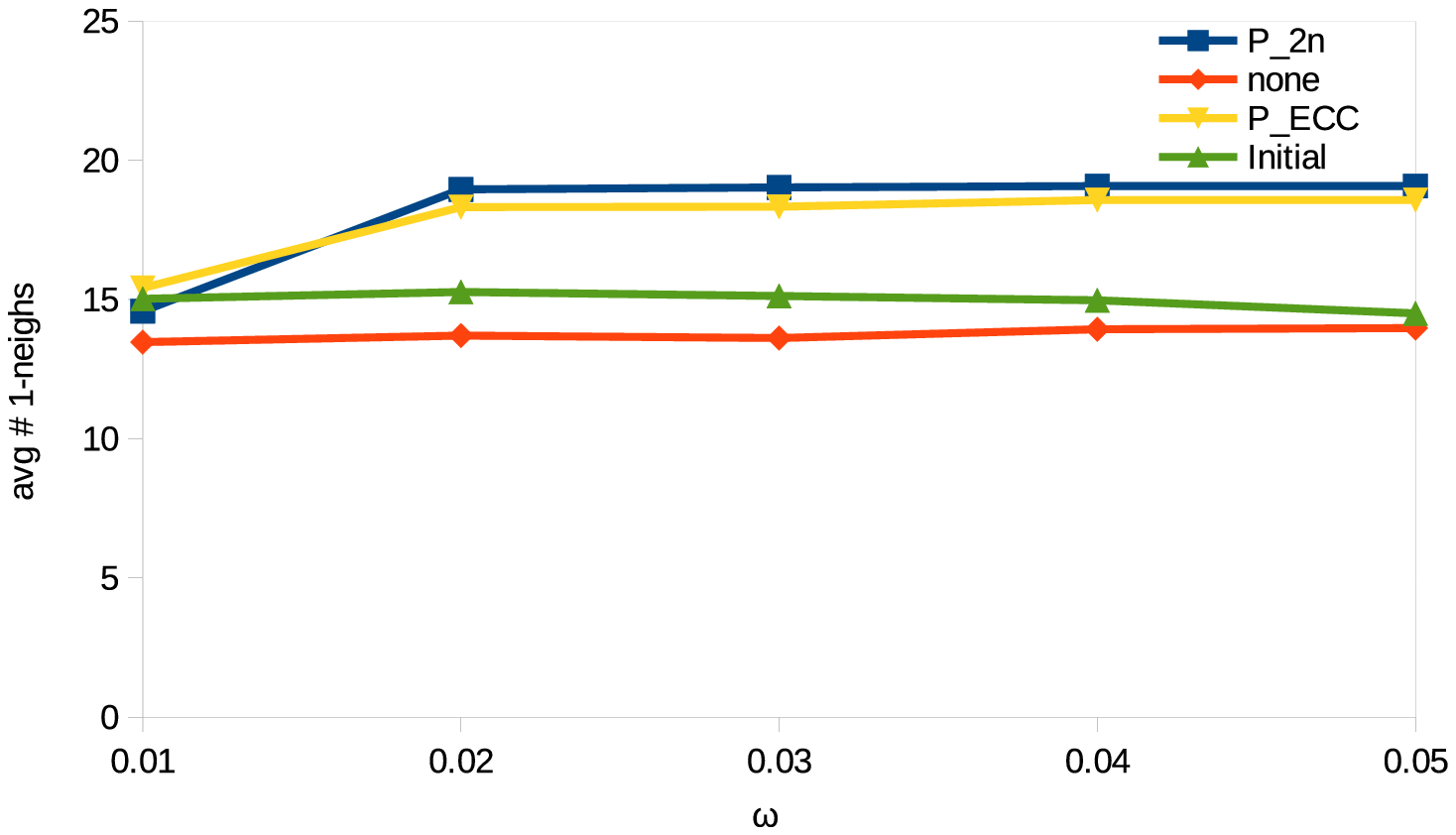}
   }
   \subfigure[Average amount of $2$nd neighbors]{
     \includegraphics[width=.45\linewidth]{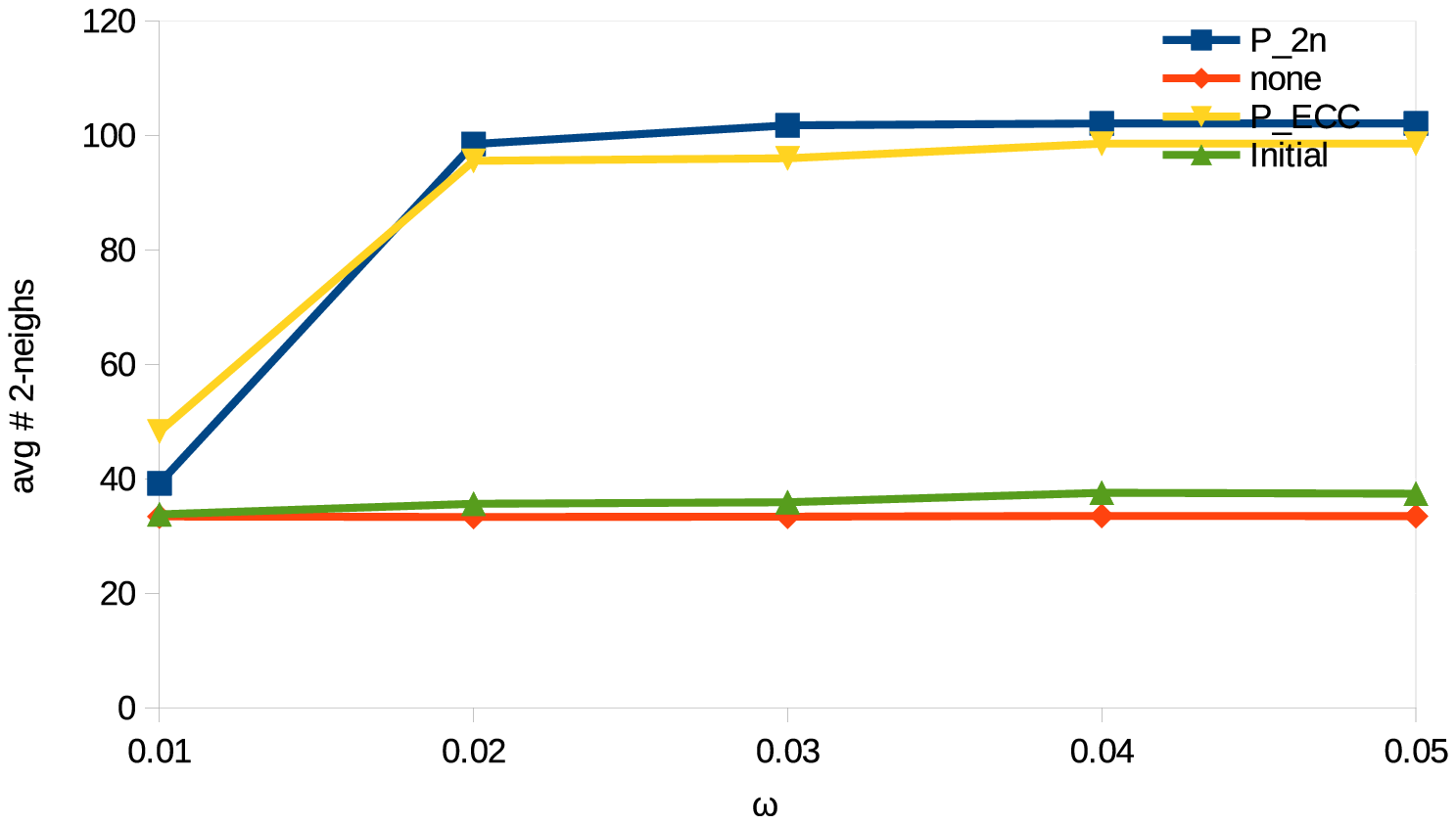}
   }
   \caption{Clustered networks -- targeted attack to nodes with 
highest betweenness simulation mode.}
   \label{fig:clus_bet}
\end{figure*}

\subsubsection{Clustered Networks}
In this case, the discrepancy between the two self-healing protocols $P_{2n}$, 
$P_{ECC}$ and ``none'' is even more evident than in other cases. In 
particular, the connectivity provided by ``none'' is significantly lower than 
the other two approaches (see Figure \ref{fig:clus_bet}).
The average amounts of 1st and 2nd neighbors decrease with ``none'' with 
respect to the original topologies, while these values increase with $P_{2n}$ 
and $P_{ECC}$. However, the increment with $P_{ECC}$ is lower than with 
$P_{2n}$.

\begin{figure*}[t]
   \centering
   \subfigure[Uniform networks]{
     \includegraphics[width=.45\linewidth]{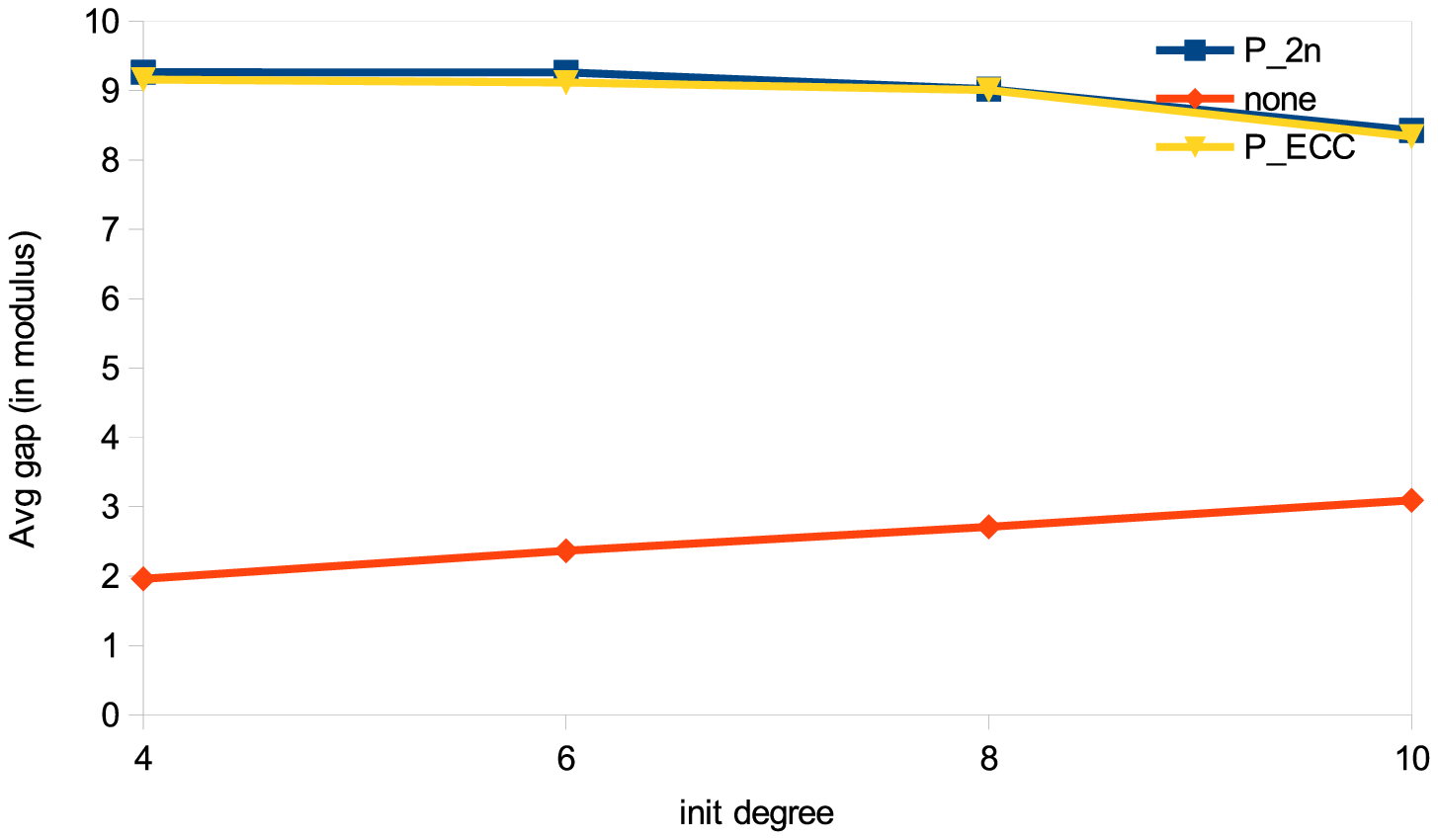}
   }
   \subfigure[Clustered networks]{
     \includegraphics[width=.45\linewidth]{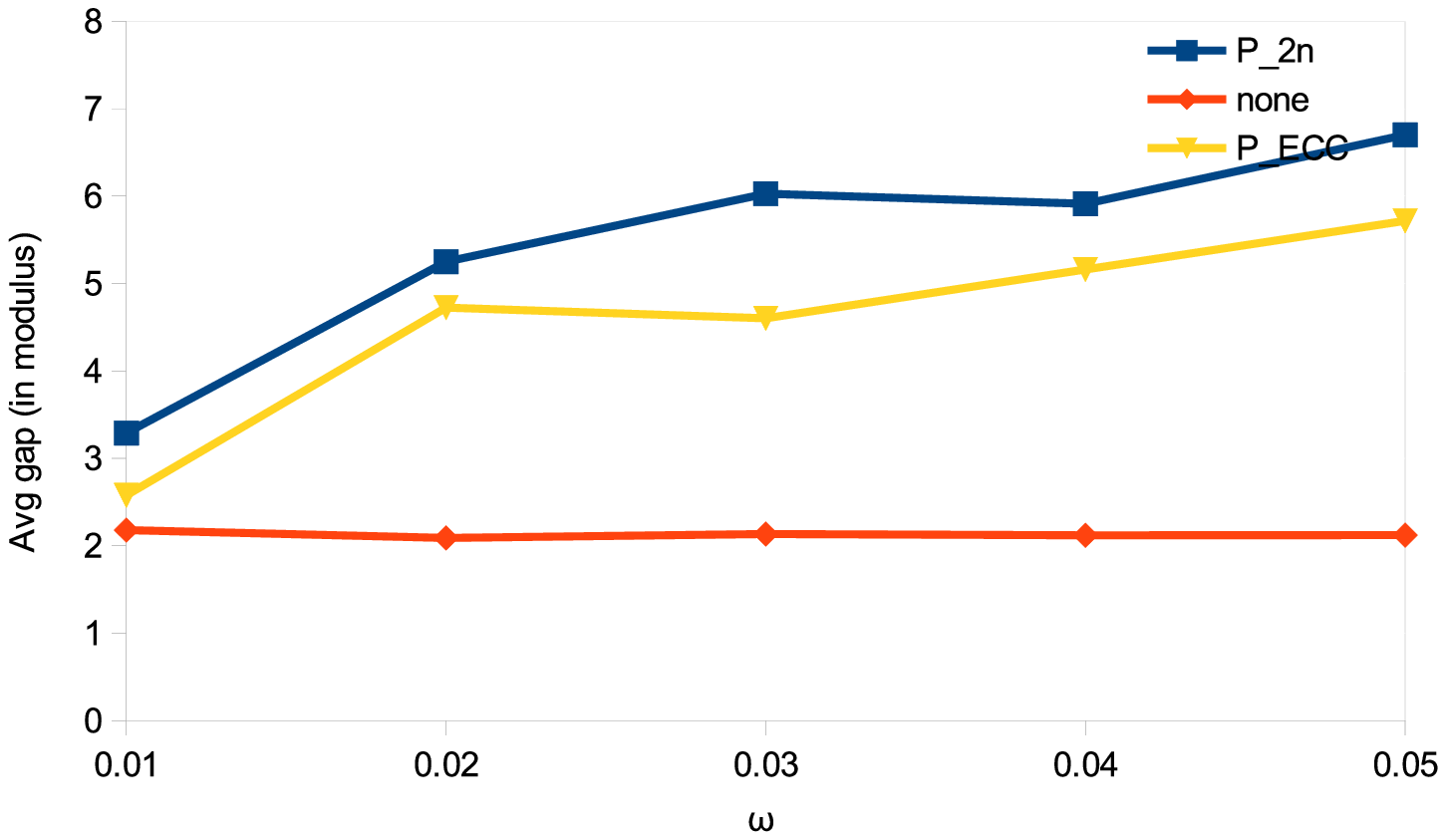}
   }
   \subfigure[Scale-free networks]{
     \includegraphics[width=.45\linewidth]{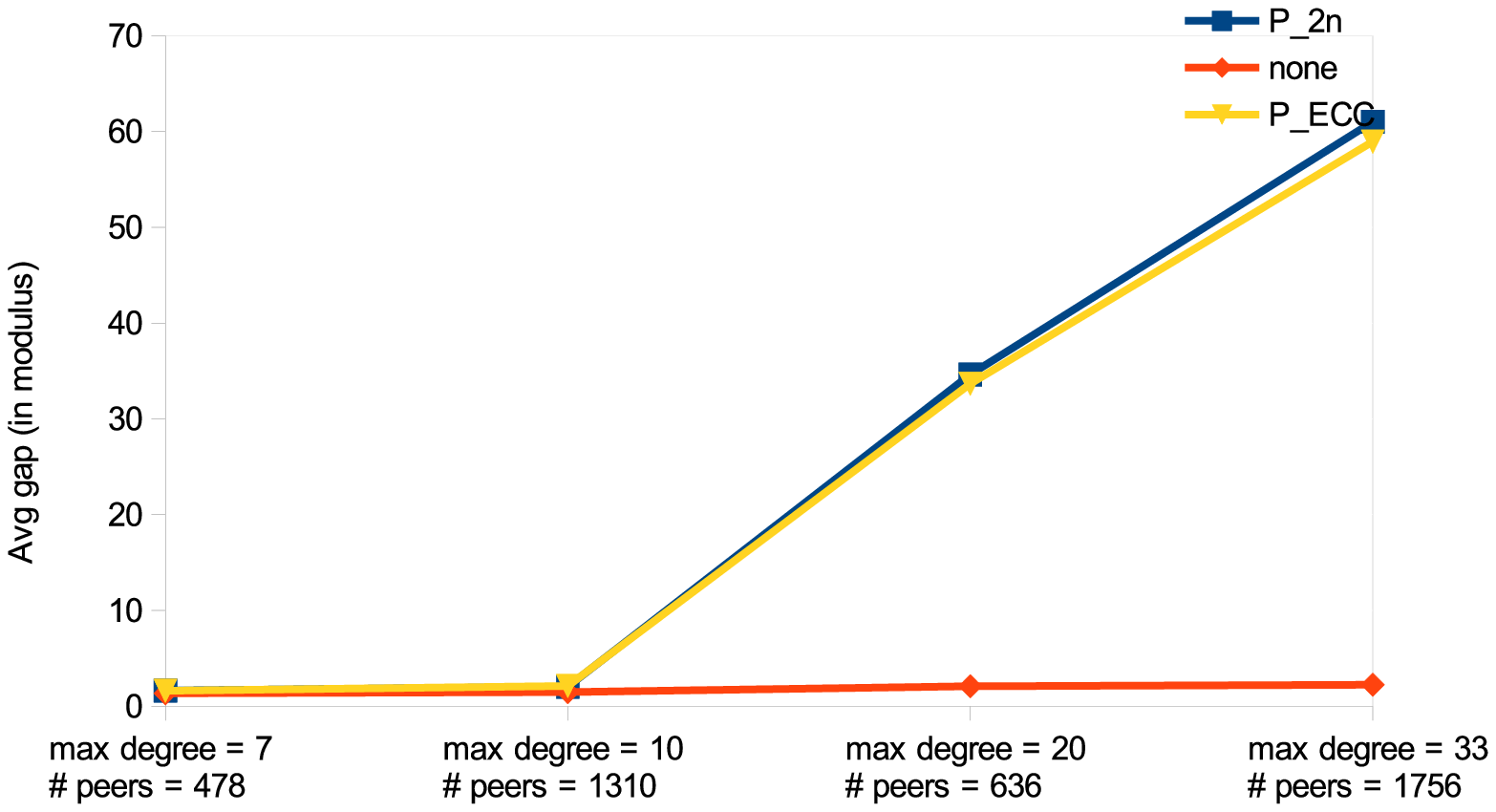}
   }
   \caption{Average gap on degrees, for those nodes that experienced some 
alterations in their neighborhood.}
   \label{fig:gap_deg}
\end{figure*}

\subsection{Variation of the Node Degree}
In order to assess how the node degree is altered by the use/non-use of the 
self-healing protocols, we report in this subsection how the nodes' degree  
changes, on average. 
 
We consider only those nodes that experience a degree variation during the 
simulation. Hence, this is not an average of all nodes (the average 
variation of the node degree on the whole peer set results quite lower).  
However, this measure gives an idea on local alterations in the networks.
For the sake of conciseness, we consider the targeted attack simulation mode only.

Figure \ref{fig:gap_deg} shows the variations of node degrees, in modulus, with 
different configurations of the three considered types of network topologies.  
It is possible to notice that, as expected, since the network evolves, the node 
degree varies, and this is more evident with the use of the self-healing 
protocols.  
It seems also that $P_{ECC}$ has slightly lower variations, with respect to 
$P_{2n}$.

\subsection{Impact of the Threshold on the Maximum Node Degree}\label{sec:eval_degree}
We mentioned that the two self-healing protocols $P_{2n}$, $P_{ECC}$ employ a 
threshold on the maximum degree a node might have. In the previous subsections, 
this parameter was set equal to $100$, which might be a high (considering the 
sizes of 
the employed networks) but quite reasonable value for P2P systems. 
In this section, we study the impact of this threshold. In fact, this parameter 
can be tuned to obtain a good trade-off between the ability of the protocol to 
guarantee connectivity, and imposing a limit on the variation of the node 
degrees. 

\begin{figure*}[t]
   \centering
   \subfigure[Main component size]{
     \includegraphics[width=.6\linewidth]{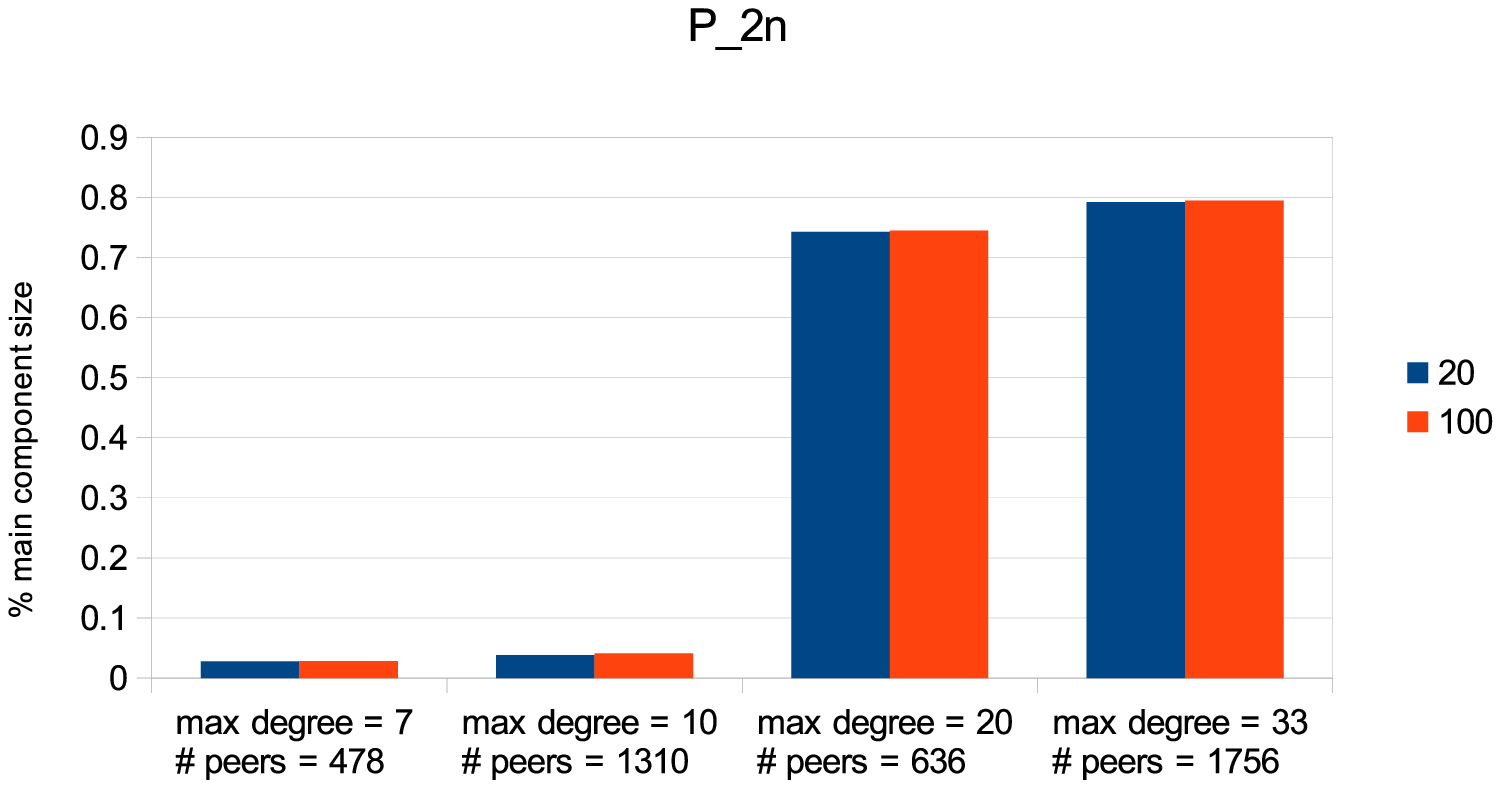}
   }
   \subfigure[Average amount of $1$st neighbors]{
     \includegraphics[width=.45\linewidth]{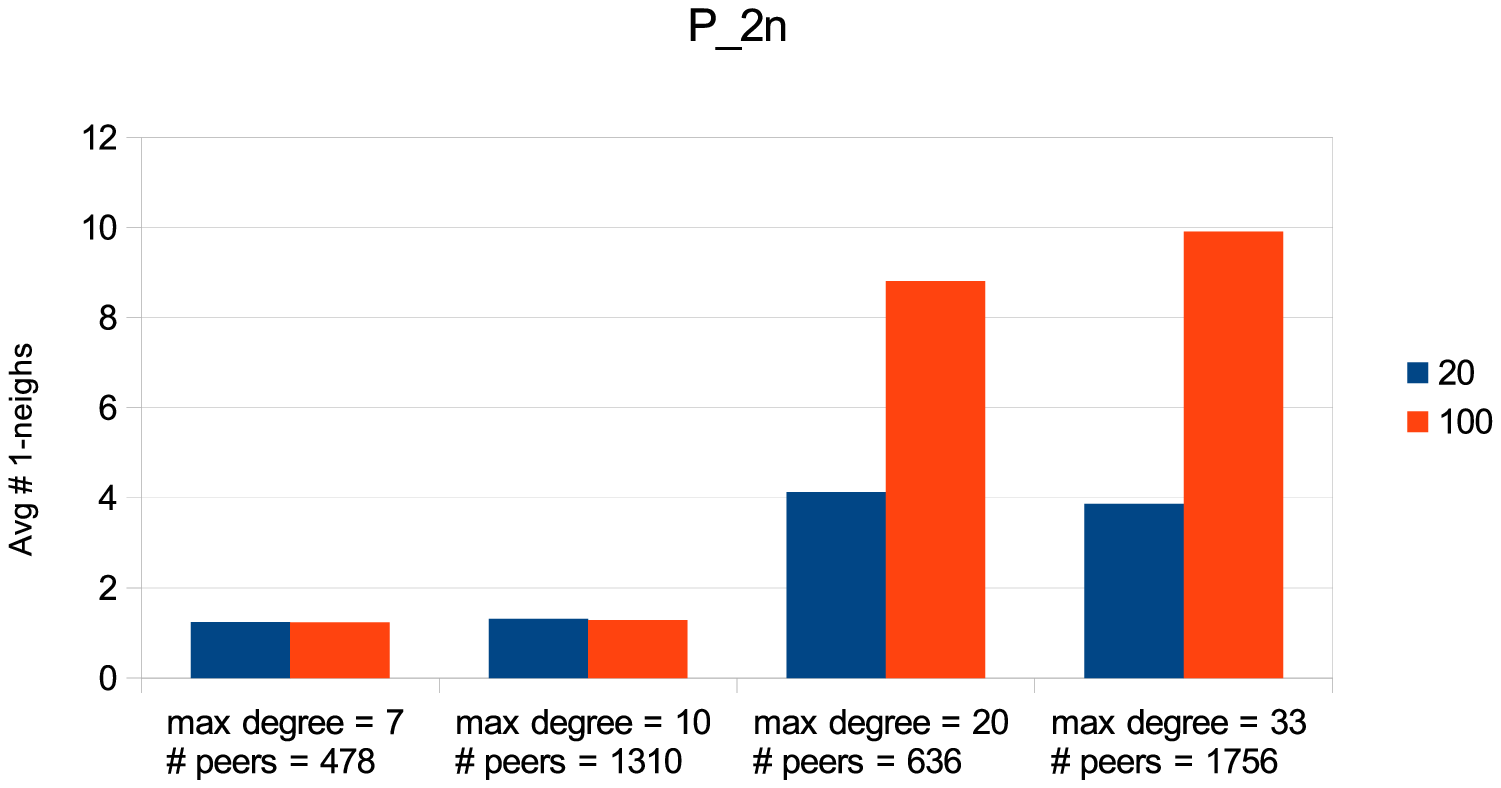}
   }
   \subfigure[Average amount of $2$nd neighbors]{
     \includegraphics[width=.45\linewidth]{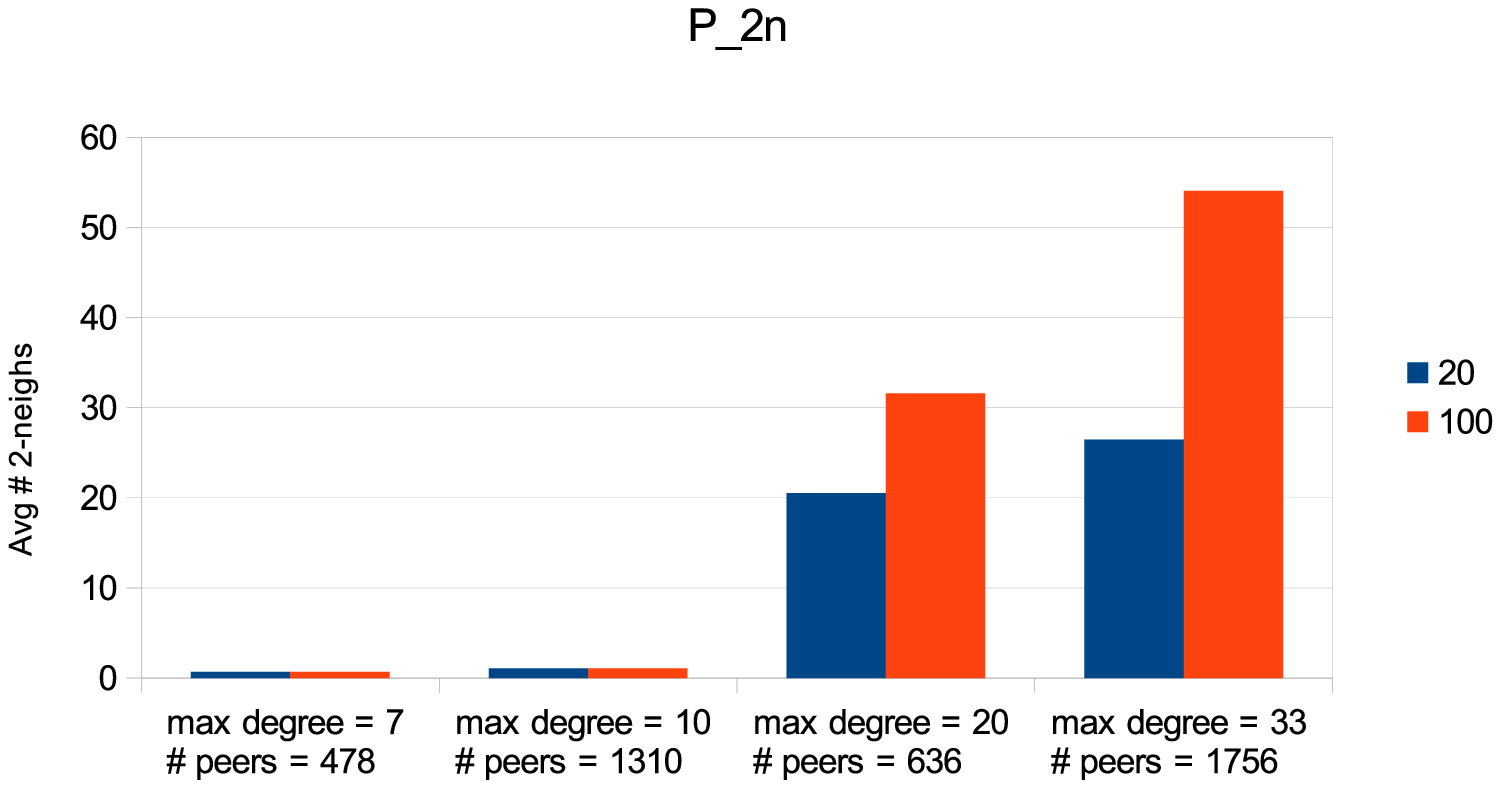}
   }
   \caption{Scale-free networks, targeted attack -- impact of the threshold on 
maximum node degree with $P_{2n}$.}
   \label{fig:deg_p2n}
\end{figure*}

\begin{figure*}[t]
   \centering
   \subfigure[Main component size]{
     \includegraphics[width=.6\linewidth]{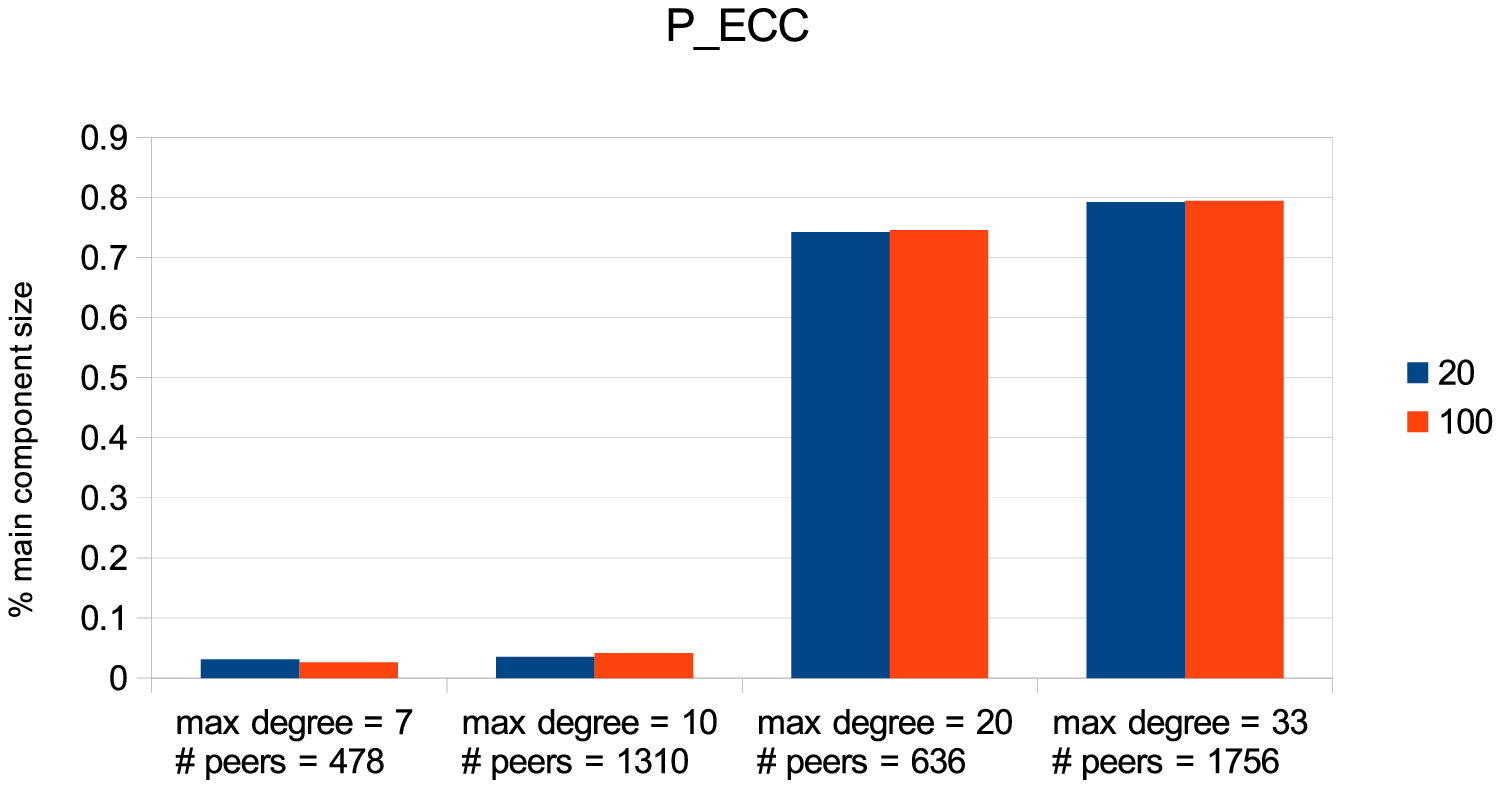}
   }
   \subfigure[Average amount of $1$st neighbors]{
     \includegraphics[width=.45\linewidth]{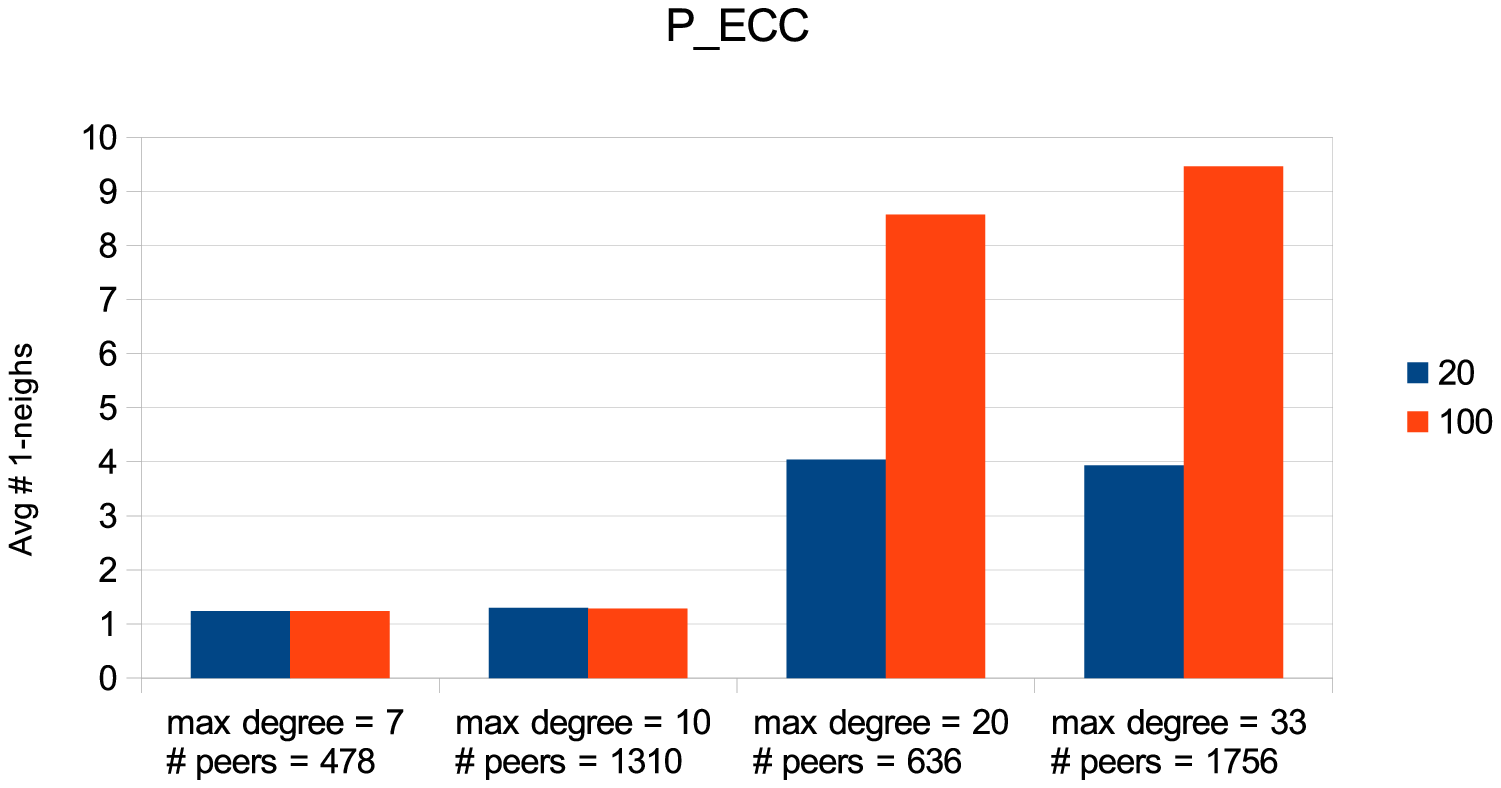}
   }
   \subfigure[Average amount of $2$nd neighbors]{
     \includegraphics[width=.45\linewidth]{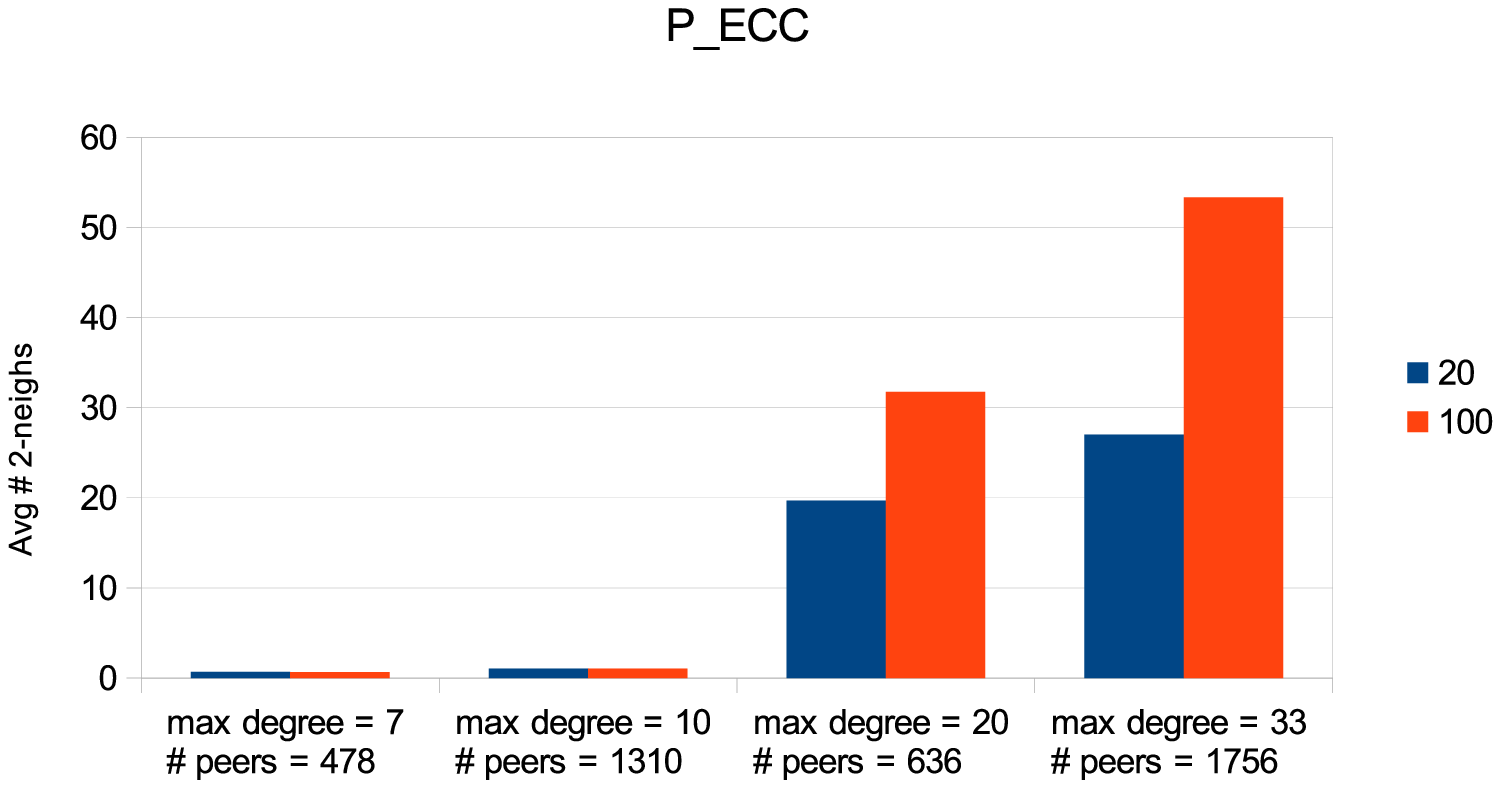}
   }
   \caption{Scale-free networks, targeted attack -- impact of the threshold on 
maximum node degree with $P_{ECC}$.}
   \label{fig:deg_pecc}
\end{figure*}

In this case, for the sake of conciseness we will focus on scale-free networks 
under targeted attack only. The choice of this topology is due to the presence 
of hubs that have a node degree much higher than the majority of other nodes. 
When this network is under a targeted attack, hubs are removed, causing network partitions. Thus, the self-healing approaches become very important in this case.

Figures \ref{fig:deg_p2n} and \ref{fig:deg_pecc} show, for $P_{2n}$ and 
$P_{ECC}$, respectively, the differences on the use of a threshold on maximum 
node degree set equal to $20$ and $100$.
Note that a low threshold value, such in the case of $20$, means that upon 
failure of a hub, no node will be available to replace its role, since the 
amount of novel connections it can create is limited. Thus, novel links, created
to maintain network connectivity, must be shared among nodes.
It is possible to notice that while the average amounts of 1st and 2nd neighbors 
decrease with a lower threshold, the connectivity of the network remains almost 
unchanged. 
This is a very important result, confirming that the tuning of the parameters in 
$P_{2n}$, $P_{ECC}$, depending on the topology in use, can guarantee the 
effectiveness of the self-healing protocols, without altering that much the 
nodes workload.

\begin{figure*}[t]
   \centering
   \subfigure[$P_{2n}$]{
     \includegraphics[width=.45\linewidth]{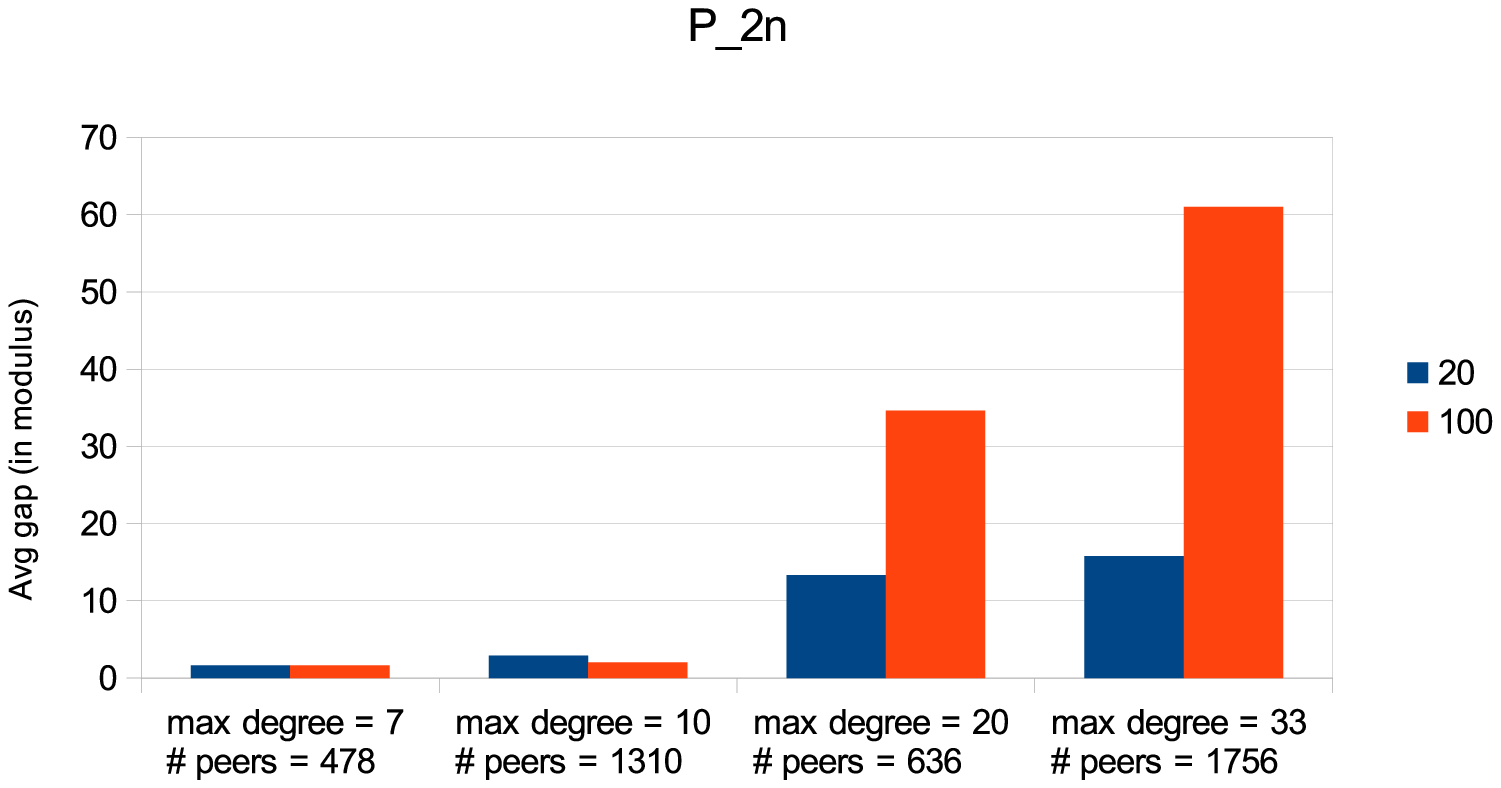}
   }
   \subfigure[$P_{ECC}$]{
     \includegraphics[width=.45\linewidth]{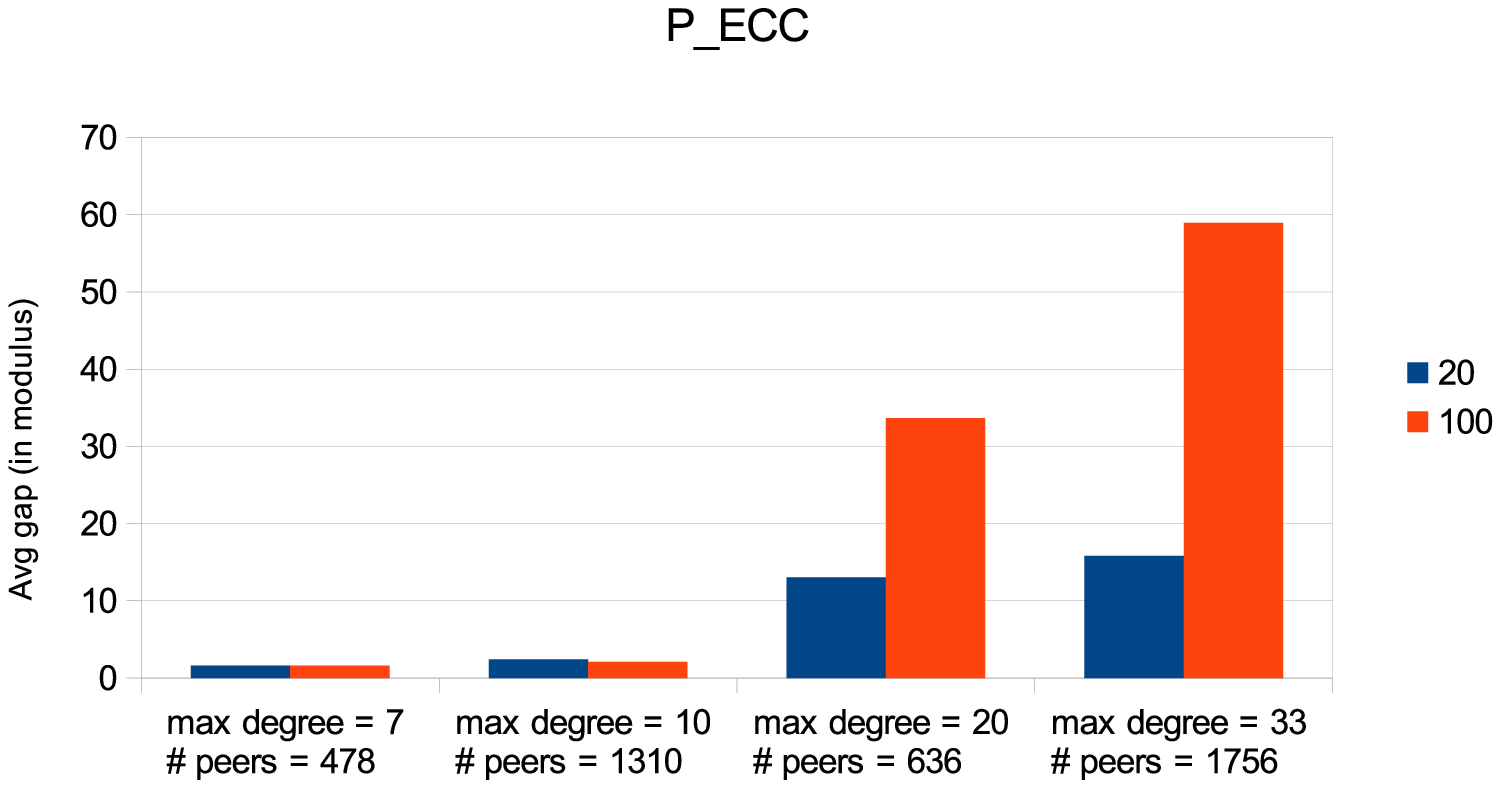}
   }
   \caption{Scale-free networks, targeted attack -- Average gap on degrees, for 
those nodes that experienced some alterations in 
their neighborhood, with different thresholds 
on the maximum degree.}
   \label{fig:deg_gap_sf}
\end{figure*}

Figure \ref{fig:deg_gap_sf} shows how the variation on the degree of a network 
changes when the threshold is modified. 
It is possible to observe an important reduction of this gap with a lower 
threshold. 

To conclude this discussion, it is worth mentioning that the tuning of this threshold parameter is not the sole option to control the growth of the node degrees in presence of a churn. 
The self-healing protocols can be coupled with a link reduction approach, that 
might remove redundant links (e.g., those with high ECC). It is important to 
notice that this would alter the clustering of the overlay. 
Another option can be to avoid the use of a fixed threshold on the node 
degrees, but rather to set the threshold based on the variation of the actual 
degree of a node, w.r.t.~its initial/target degree. The idea is that the 
fluctuations of the nodes degree should not surpass some limit. 
However, this might be a problem in certain topologies. For instance, if we 
consider a scale-free network, the failure of a hub means that several links are 
removed from the network. If remaining nodes want to maintain network 
connectivity, they need to replace in some way these lost links, and this would 
likely result (in some cases) in an increment of node degrees.  
The use of this hypothetical approach could be in contrast with this issue.

\subsection{On the Clustering Coefficient and Network 
Diameter}\label{sec:eval_cc_diam}

In this subsection, we will look at the influence of $P_{2n}$ and $P_{ECC}$ on 
the network clustering coefficient and on the network diameter.
The idea was to analyze the resulting networks when the self-healing protocols 
are executed on an evolving P2P system. 

As to the clustering coefficient, previous works assert that it is 
undesirable for an unstructured P2P overlay to have high clustering 
\cite{voulgaris.jnsm.2005}. In fact, clustering reduces the connectivity of a 
cluster to the rest of the net, increases the probability of partitioning, and 
it  may cause redundant message delivery. 

As to the network diameter, it is evident that the lower the 
diameter the faster the message dissemination in the overlay.

\subsubsection{Uniform Networks}
\begin{figure*}[t]
   \centering
   \subfigure[Clustering coefficient]{
     \includegraphics[width=.45\linewidth]{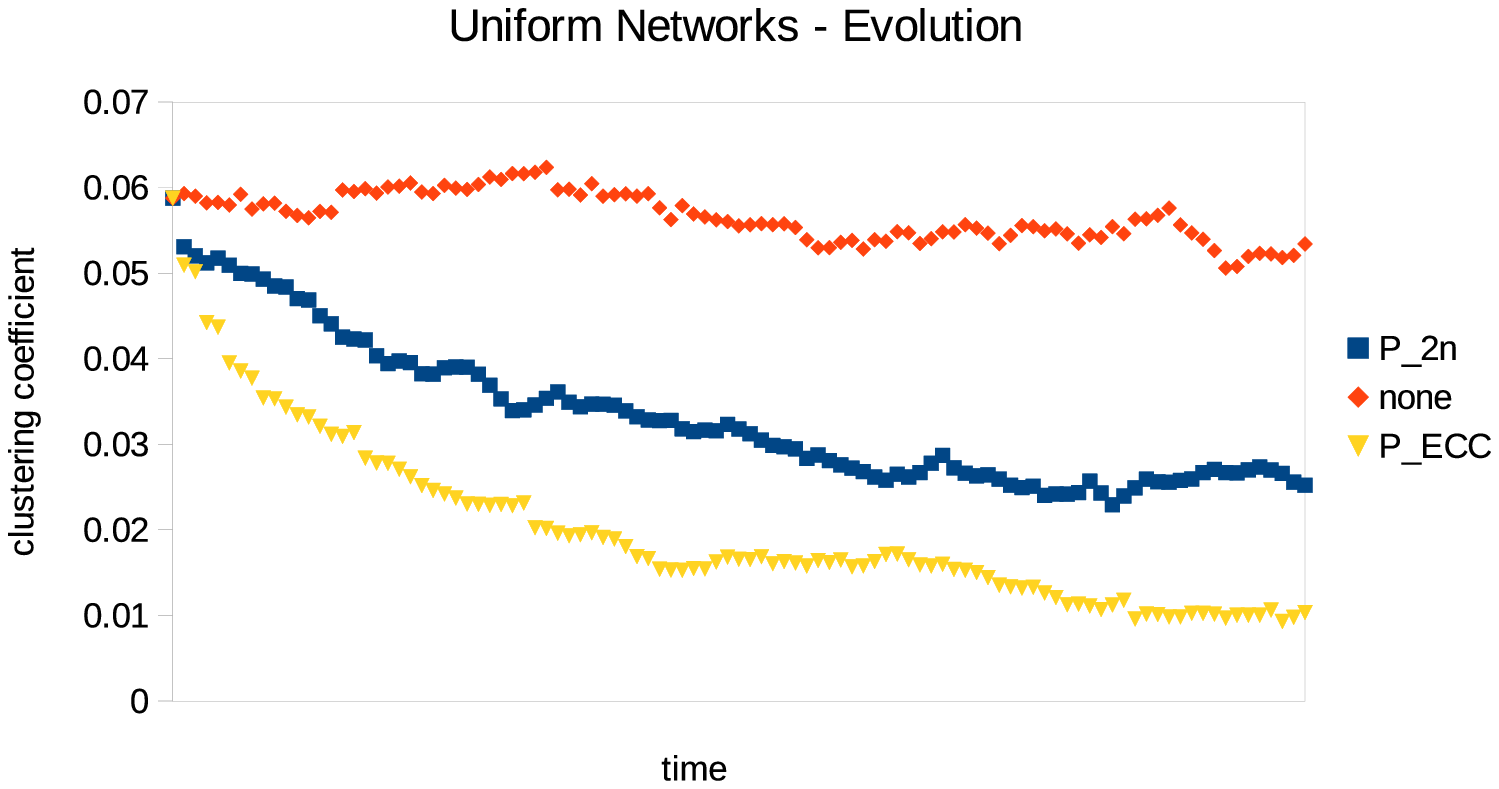}
   }
   \subfigure[Diameter]{
     \includegraphics[width=.45\linewidth]{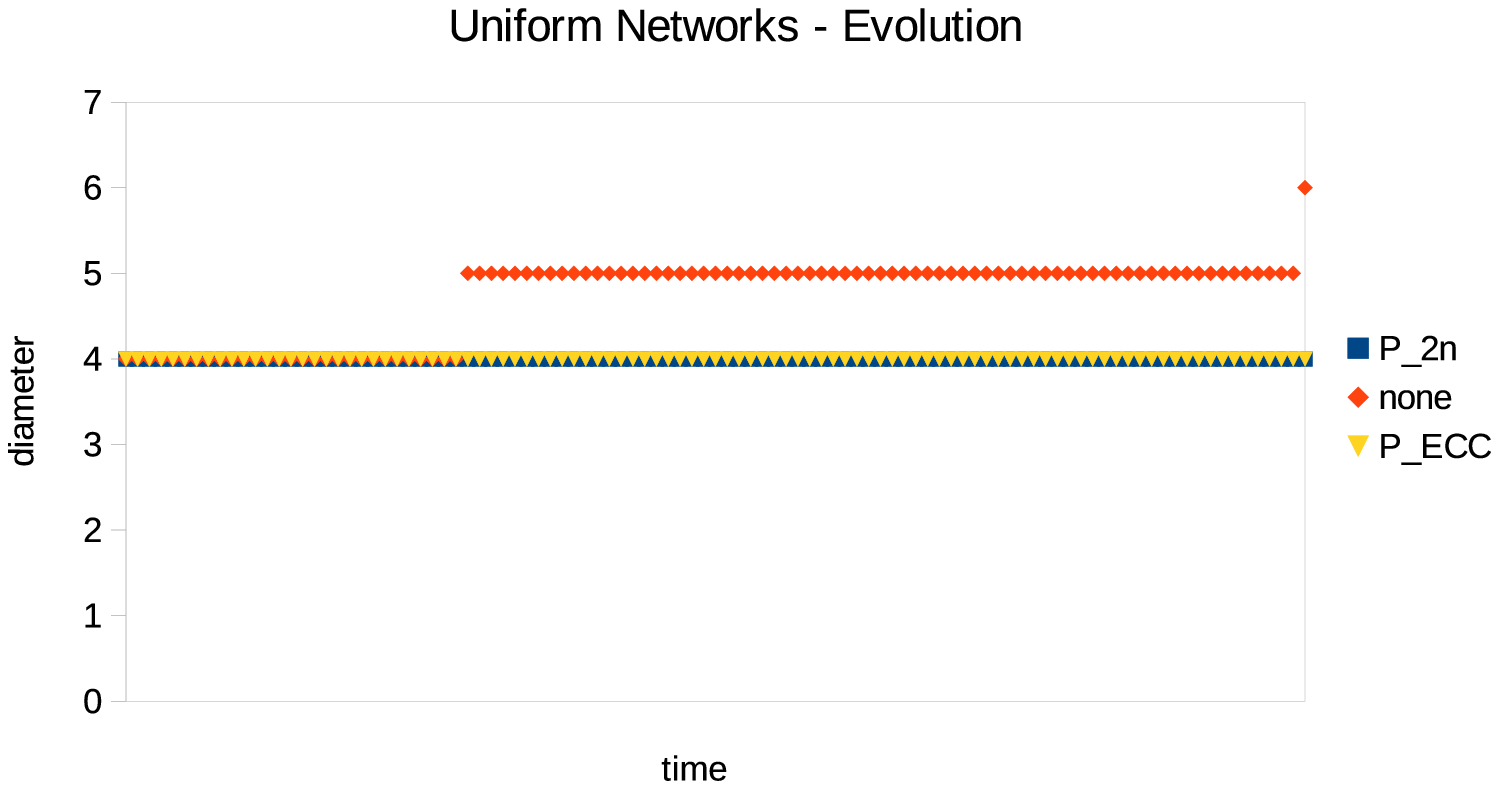}
   }
   \caption{Uniform networks -- clustering coefficient and diameter of a 
typical network during the evolution simulation mode.}
   \label{fig:cc_d_unif}
\end{figure*}

Figure \ref{fig:cc_d_unif} shows how the clustering coefficient and the 
diameter change in a typical uniform network when the evolution simulation 
mode is employed (the test was repeated multiple times with different networks, 
obtaining the same qualitative results).
It is possible to observe that the use of $P_{2n}$ and $P_{ECC}$ lowers the 
clustering coefficient, as the uniform network evolves.
Moreover, $P_{ECC}$ has a higher decrement. Conversely, as expected 
``none'' protocol maintains a stable clustering coefficient, since the network 
evolves as a typical unstructured uniform network, i.e., nodes enter and 
randomly select a fixed amount of novel neighbors.

As shown in the figure, with the ``none'' approach, the network experiences 
an increment of the network diameter, while $P_{2n}$ and $P_{ECC}$ allow 
maintaining a constant diameter. 

This confirms the viability of the two proposals for the support of P2P 
overlays. Moreover, $P_{ECC}$ allows differentiating the links created 
by neighbor nodes.

\subsubsection{Clustered Networks}
\begin{figure*}[t]
   \centering
   \subfigure[Clustering coefficient]{
     \includegraphics[width=.45\linewidth]{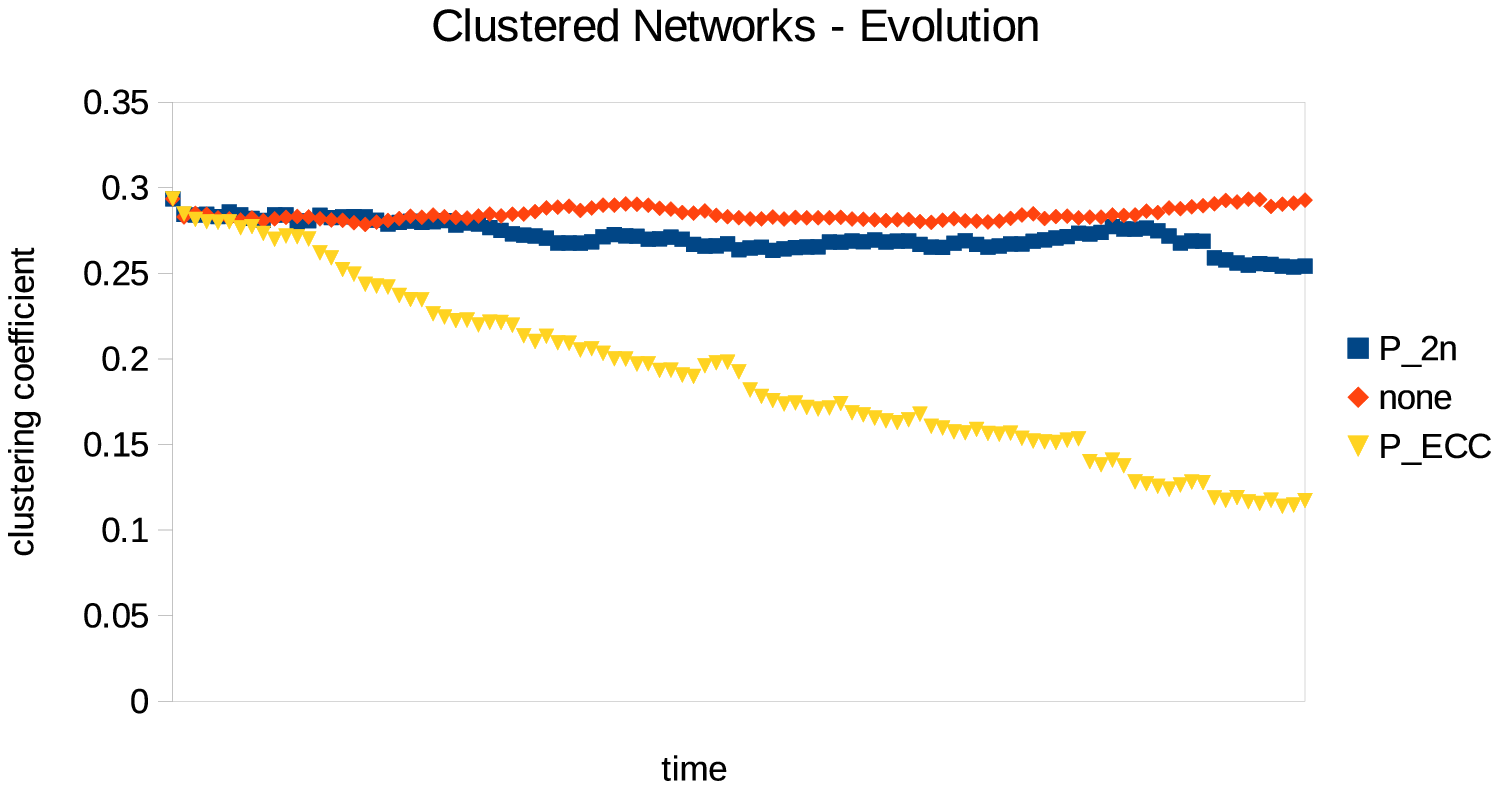}
   }
   \subfigure[Diameter]{
     \includegraphics[width=.45\linewidth]{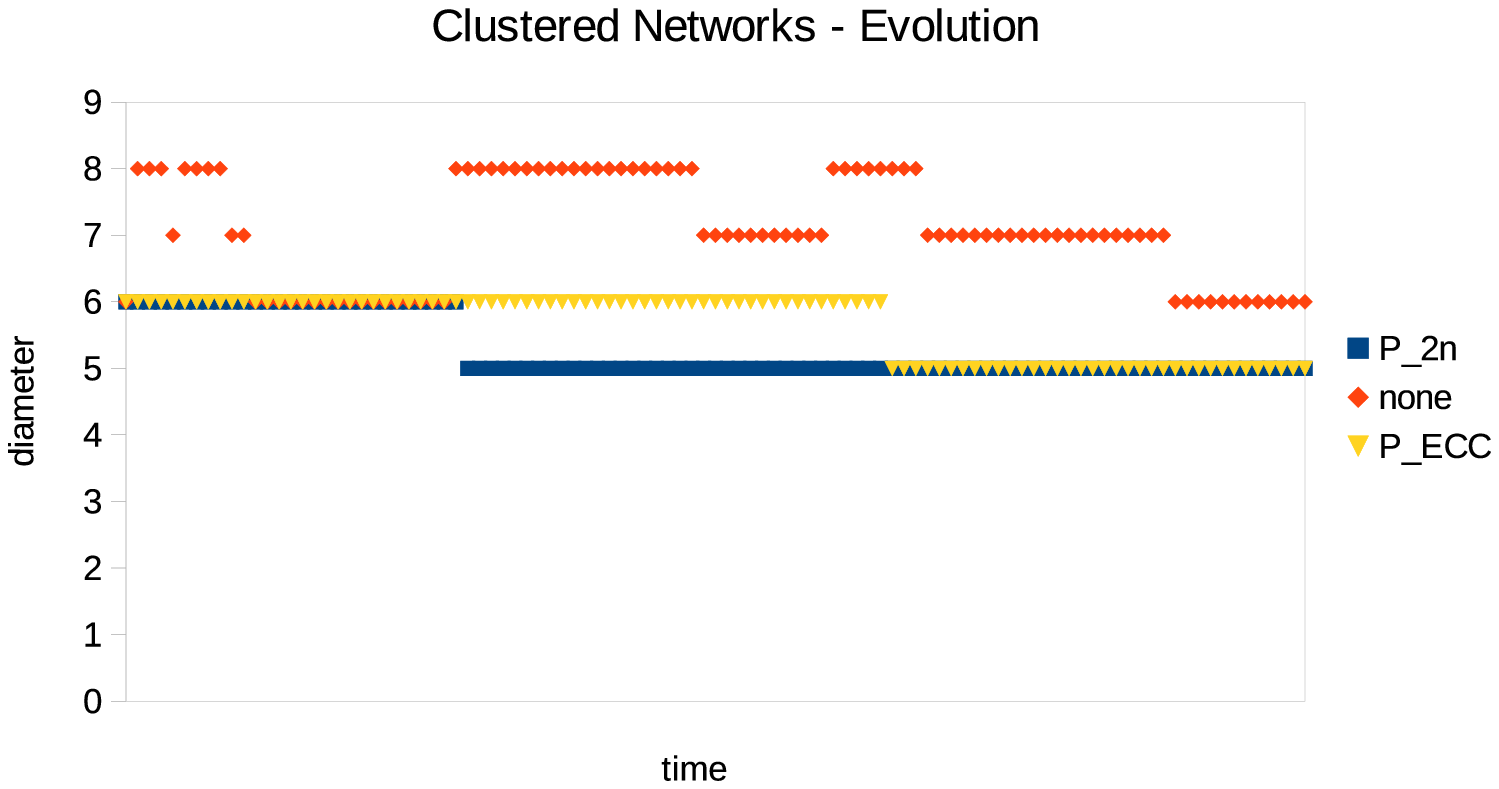}
   }
   \caption{Clustered networks -- clustering coefficient and diameter of a 
typical network during the evolution simulation mode.}
   \label{fig:cc_d_clus}
\end{figure*}

Figure \ref{fig:cc_d_clus} shows the variation of the clustering coefficient 
and 
diameter during an exemplar evolution of the simulation with a clustered 
network. In this case, the decrement of the clustering coefficient is sensible 
for $P_{ECC}$, while $P_2n$ has a minor impact on this metric.
The diameter of the network decreases with both protocols.
This allows concluding that one might decide if turning to $P_{ECC}$ or 
$P_{2n}$ if such reduction of the clustering coefficient is a desired effect (as 
commonly stated \cite{voulgaris.jnsm.2005}) or not.

\subsubsection{Scale-Free Networks}
\begin{figure*}[t]
   \centering
   \subfigure[Clustering coefficient]{
     \includegraphics[width=.45\linewidth]{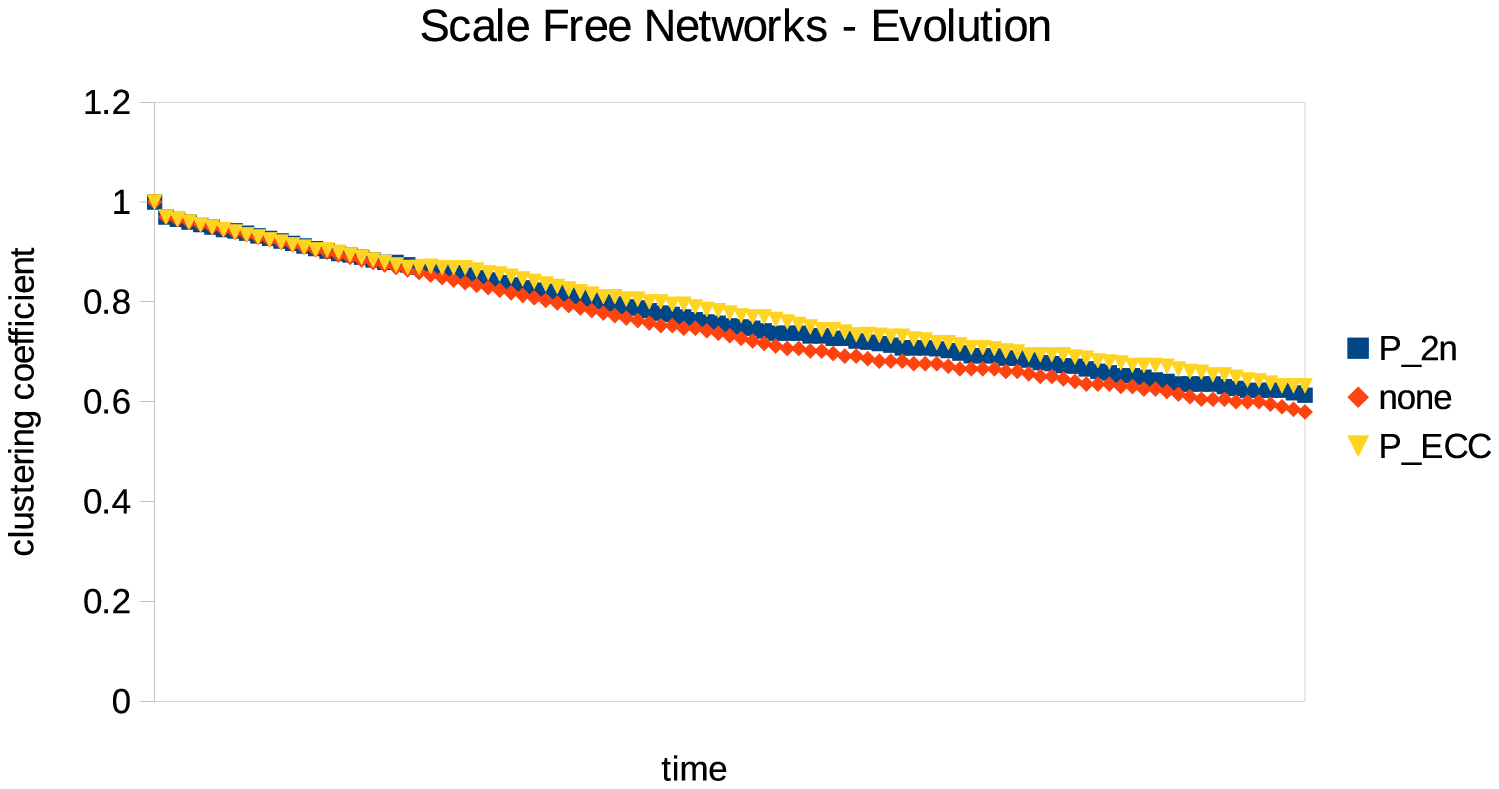}
   }
   \subfigure[Diameter]{
     \includegraphics[width=.45\linewidth]{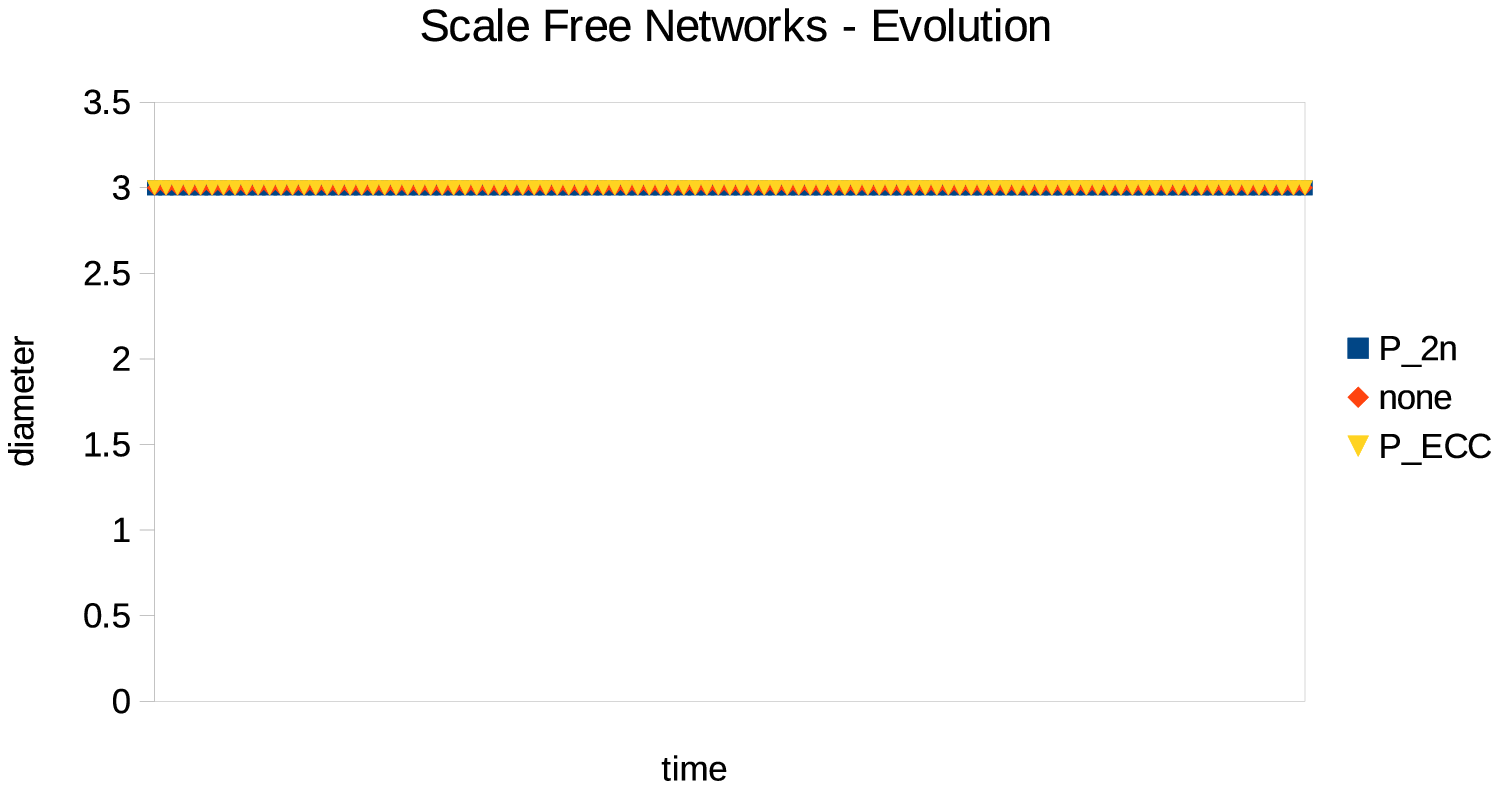}
   }
   \caption{Scale-Free networks -- clustering coefficient and diameter of a 
typical network during the evolution simulation mode.}
   \label{fig:cc_d_sf}
\end{figure*}

The impact noticed for other networks is not evident in scale-free networks, 
under the evolution simulation mode. In fact, in this case the hubs do maintain 
their main role in the network. 
The scale-free networks were generated using a classic 
preferential attachment approach, using a specific routine available in the 
Octave-network-toolbox \cite{nettoolbox,newmanHandbook}.
Figure \ref{fig:cc_d_sf} shows the clustering 
coefficient and diameter variations that, in this case, are negligible.
Different scale-free networks with varying network sizes were considered; 
results showed the same trend in all cases.

\section{Conclusions}\label{sec:conc}
This paper focused on two distributed mechanisms that can be executed locally by peers in an unstructured P2P overlay, in order to cope with node failures and augment the resilience of the network.
The two self-healing protocols require knowledge of $1$st and $2$nd neighbors. 
Outcomes confirm that it is possible to augment resilience and avoid disconnections in unstructured P2P overlay networks. 

In particular, while both schemes help to avoid network disconnections, our 
results suggest that the use of the Edge Clustering Coefficient (ECC) 
provides some additional advantages during the self-healing phase.
In fact, ECC provides an idea of how much inter-communitarian a link is. It can 
be thus exploited to: i) replace lost important links with novel ones after some 
failures, ii) (if needed) remove those (novel) links that might augment 
excessively the degree of some node and the amount of triangles it belongs, 
iii) reduce the clustering coefficient of the overlay (depending on its 
topology).  

The two self-healing protocols are local. They avoid that a node loses 
connections in its 2nd-neighborhood. When we employ them in a network that 
evolves in a stable manner (on average), the amount of links in the network will 
increase, as noticed in the evaluation assessment. However, it is possible to 
limit this increment, while maintaining network connectivity. 
Thus, depending on the application requirements, whenever such an increment is 
undesired, it is possible to couple the self-healing protocols with a link 
reduction process, or by setting a low threshold on the maximum 
degree. 

The employed system model assumes that only nodes can fail; hence there are no 
single links removals. 
This simplification does not introduce important limitations, since the protocol 
can be easily upgraded (without any substantial modifications) to handle single 
link failures.

Moreover, the model assumes that network changes (in a given neighborhood) are 
slower than the execution of a step of the self-healing protocols. 
This is a common assumption, that enables nodes self-repairing network 
partitions through local interactions only.
However, scenarios are not considered when a network is partitioned by the 
simultaneous failure of a node set, so that nodes in the remaining components 
have no information about other components (i.e., given two nodes in different 
components after the churn, the distance between these two nodes before the 
churn was higher than $2$).
This prevents the creation of novel links to repair the partition.
This is an uncommon situation, that can be faced in different ways. Increasing 
the local knowledge at peers would be of help. For example, peers could store 
in their caches a subset of $k$th neighbors, so that the amount 
of node entries at distance $k$ is inversely proportional to $k$. (This to 
avoid that the global amount of stored data increases 
exponentially.)
This approach, coupled with a gossip protocol, might help to find novel 
connections that would repair such kinds of partitions.


\begin{thebibliography}{10}
\providecommand{\url}[1]{{#1}}
\providecommand{\urlprefix}{URL }
\expandafter\ifx\csname urlstyle\endcsname\relax
  \providecommand{\doi}[1]{DOI~\discretionary{}{}{}#1}\else
  \providecommand{\doi}{DOI~\discretionary{}{}{}\begingroup
  \urlstyle{rm}\Url}\fi

\bibitem{nettoolbox}
octave-networks-toolbox.
\newblock \urlprefix\url{http://aeolianine.github.io/octave-networks-toolbox/}

\bibitem{planetsim}
Ahull\'{o}, J.P., L\'{o}pez, P.G.: Planetsim: An extensible framework for
  overlay network and services simulations.
\newblock In: Proceedings of the 1st International Conference on Simulation
  Tools and Techniques for Communications, Networks and Systems \& Workshops,
  Simutools '08, pp. 45:1--45:1. ICST (Institute for Computer Sciences,
  Social-Informatics and Telecommunications Engineering), ICST, Brussels,
  Belgium, Belgium (2008).
\newblock \urlprefix\url{http://dl.acm.org/citation.cfm?id=1416222.1416274}

\bibitem{Aiello00arandom}
Aiello, W., Chung, F., Lu, L.: A random graph model for power law graphs.
\newblock Experimental Math \textbf{10}, 53--66 (2000)

\bibitem{Bader:2007}
Bader, D.A., Kintali, S., Madduri, K., Mihail, M.: Approximating betweenness
  centrality.
\newblock In: Proc. of the 5th international conference on Algorithms and
  models for the web-graph, WAW'07, pp. 124--137. Springer (2007)

\bibitem{Baraglia:2013}
Baraglia, R., Dazzi, P., Mordacchini, M., Ricci, L.: A peer-to-peer recommender
  system for self-emerging user communities based on gossip overlays.
\newblock J. Comput. Syst. Sci. \textbf{79}(2), 291--308 (2013).
\newblock \doi{10.1016/j.jcss.2012.05.011}.
\newblock \urlprefix\url{http://dx.doi.org/10.1016/j.jcss.2012.05.011}

\bibitem{baset}
Baset, S., Schulzrinne, H.: An analysis of the skype peer-to-peer internet
  telephony protocol.
\newblock In: INFOCOM 2006. 25th IEEE International Conference on Computer
  Communications. Proceedings, pp. 1--11 (2006).
\newblock \doi{10.1109/INFOCOM.2006.312}

\bibitem{Basu05nodewiz}
Basu, S., Banerjee, S., Sharma, P., ju~Lee, S.: Nodewiz: Peer-to-peer resource
  discovery for grids.
\newblock In: Proc. of IEEE/ACM GP2PC’05, pp. 213--220 (2005)

\bibitem{Cai03maan}
Cai, M., Frank, M., Chen, J., Szekely, P.: Maan: A multi-attribute addressable
  network for grid information services.
\newblock In: Journal of Grid Computing, p. 184. IEEE Computer Society (2003)

\bibitem{simgrid}
Casanova, H., Giersch, A., Legrand, A., Quinson, M., Suter, F.: Versatile,
  scalable, and accurate simulation of distributed applications and platforms.
\newblock Journal of Parallel and Distributed Computing \textbf{74}(10),
  2899--2917 (2014).
\newblock \urlprefix\url{http://hal.inria.fr/hal-01017319}

\bibitem{chaudhry}
Chaudhry, J., Park, S.: Ahsen – autonomic healing-based self management
  engine for network management in hybrid networks.
\newblock In: C.~Cérin, K.C. Li (eds.) Advances in Grid and Pervasive
  Computing, \emph{Lecture Notes in Computer Science}, vol. 4459, pp. 193--203.
  Springer Berlin Heidelberg (2007).
\newblock \doi{10.1007/978-3-540-72360-8_17}.
\newblock \urlprefix\url{http://dx.doi.org/10.1007/978-3-540-72360-8_17}

\bibitem{Costa:2003}
Costa, P., Migliavacca, M., Picco, G.P., Cugola, G.: Introducing reliability in
  content-based publish-subscribe through epidemic algorithms.
\newblock In: Proceedings of the 2nd international workshop on Distributed
  event-based systems, DEBS '03, pp. 1--8. ACM, New York, NY, USA (2003).
\newblock \doi{10.1145/966618.966629}.
\newblock \urlprefix\url{http://doi.acm.org/10.1145/966618.966629}

\bibitem{simutools}
D'Angelo, G., Ferretti, S.: Simulation of scale-free networks.
\newblock In: Simutools '09: Proc.~of the 2nd International Conference on
  Simulation Tools and Techniques, pp. 1--10. ICST, ICST, Brussels, Belgium
  (2009).
\newblock \doi{http://dx.doi.org/10.4108/ICST.SIMUTOOLS2009.5672}

\bibitem{lunes}
D'Angelo, G., Ferretti, S.: {LUNES}: Agent-based simulation of {P2P} systems.
\newblock In: Proceedings of the International Workshop on Modeling and
  Simulation of Peer-to-Peer Architectures and Systems (MOSPAS 2011). IEEE
  (2011)

\bibitem{doerr}
Doerr, C., Hernandez, J.: A computational approach to multi-level analysis of
  network resilience.
\newblock In: Dependability (DEPEND), 2010 Third International Conference on,
  pp. 125--132 (2010).
\newblock \doi{10.1109/DEPEND.2010.27}

\bibitem{gridpeer}
Ferretti, S.: Modeling faulty, unstructured p2p overlays.
\newblock In: Proc.~of the 19th International Conference on Computer
  Communications and Networks (ICCCN 2010). IEEE (2010)

\bibitem{ferretti_trans.cs.2012.10-12.e2}
Ferretti, S.: On the degree distribution of faulty peer-to-peer overlay
  networks.
\newblock EAI Endorsed Transactions on Complex Systems \textbf{12}(1) (2012).
\newblock \doi{10.4108/trans.cs.2012.10-12.e2}

\bibitem{simplex}
Ferretti, S.: Publish-subscribe systems via gossip: a study based on complex
  networks.
\newblock In: Proc.~of the 4th Annual Workshop on Simplifying Complex Networks
  for Practitioners, SIMPLEX '12, pp. 7--12. ACM, New York, NY, USA (2012).
\newblock \doi{10.1145/2184356.2184359}.
\newblock \urlprefix\url{http://doi.acm.org/10.1145/2184356.2184359}

\bibitem{Ferretti-fgcs}
Ferretti, S.: Gossiping for resource discovering: An analysis based on complex
  network theory.
\newblock Future Generation Computer Systems \textbf{29}(6), 1631 -- 1644
  (2013).
\newblock \doi{http://dx.doi.org/10.1016/j.future.2012.06.002}.
\newblock Including Special sections: High Performance Computing in the Cloud
  and Resource Discovery Mechanisms for P2P Systems

\bibitem{simplex13}
Ferretti, S.: Resilience of dynamic overlays through local interactions.
\newblock In: 22nd International World Wide Web Conference, WWW '13, Rio de
  Janeiro, Brazil, May 13-17, 2013, Companion Volume, pp. 813--820.
  International World Wide Web Conferences Steering Committee / ACM (2013)

\bibitem{networking14}
Ferretti, S.: On the topology maintenance of dynamic p2p overlays through
  self-healing local interactions.
\newblock In: Networking Conference, 2014 IFIP, pp. 1--9 (2014).
\newblock \doi{10.1109/IFIPNetworking.2014.6857126}

\bibitem{complenet}
Ferretti, S.: Searching in unstructured overlays using local knowledge and
  gossip.
\newblock In: P.~Contucci, R.~Menezes, A.~Omicini, J.~Poncela-Casasnovas (eds.)
  Complex Networks V, \emph{Studies in Computational Intelligence}, vol. 549,
  pp. 63--74. Springer International Publishing (2014).
\newblock \doi{10.1007/978-3-319-05401-8_7}.
\newblock \urlprefix\url{http://dx.doi.org/10.1007/978-3-319-05401-8_7}

\bibitem{forestiero}
Forestiero, A., Mastroianni, C., Papuzzo, G., Spezzano, G.: Towards a
  self-structured grid: An ant-inspired p2p algorithm.
\newblock In: C.~Priami, F.~Dressler, O.~Akan, A.~Ngom (eds.) Transactions on
  Computational Systems Biology X, \emph{Lecture Notes in Computer Science},
  vol. 5410, pp. 1--19. Springer Berlin Heidelberg (2008).
\newblock \doi{10.1007/978-3-540-92273-5_1}.
\newblock \urlprefix\url{http://dx.doi.org/10.1007/978-3-540-92273-5_1}

\bibitem{Ganesh:2003}
Ganesh, A.J., Kermarrec, A.M., Massouli\'{e}, L.: Peer-to-peer membership
  management for gossip-based protocols.
\newblock IEEE Trans. Comput. \textbf{52}, 139--149 (2003).
\newblock \doi{10.1109/TC.2003.1176982}.
\newblock \urlprefix\url{http://dl.acm.org/citation.cfm?id=642778.642782}

\bibitem{p2psim}
Gil, T.M., Kaashoek, F., Li, J., Morris, R., Stribling, J.: p2psim: a simulator
  for peer-to-peer ({P2P}) protocols.
\newblock http://pdos.csail.mit.edu/p2psim/ (2009).
\newblock \urlprefix\url{http://pdos.csail.mit.edu/p2psim/}

\bibitem{Giordanelli:2012}
Giordanelli, R., Mastroianni, C., Meo, M.: Bio-inspired p2p systems: The case
  of multidimensional overlay.
\newblock ACM Trans. Auton. Adapt. Syst. \textbf{7}(4), 35:1--35:28 (2012).
\newblock \doi{10.1145/2382570.2382571}.
\newblock \urlprefix\url{http://doi.acm.org/10.1145/2382570.2382571}

\bibitem{girvan}
Girvan, M., Newman, M.E.: Community structure in social and biological
  networks.
\newblock Proc Natl Acad Sci U S A \textbf{99}(12), 7821--7826 (2002)

\bibitem{Goncalves:2012}
Gon\c{c}alves, G.D., Guimar\~{a}es, A., Vieira, A.B., Cunha, I., Almeida, J.M.:
  Using centrality metrics to predict peer cooperation in live streaming
  applications.
\newblock In: Proc.~of the 11th Int.~IFIP TC6 Conference on Networking,
  IFIP'12, pp. 84--96. Springer-Verlag (2012)

\bibitem{Hidalgo:2011}
Hidalgo, N., Rosas, E., Arantes, L., Marin, O., Sens, P., Bonnaire, X.: Dring:
  A layered scheme for range queries over dhts.
\newblock In: Proc.~of the 2011 IEEE 11th International Conference on Computer
  and Information Technology, CIT '11, pp. 29--34. IEEE (2011)

\bibitem{holme2002attack}
Holme, P., Kim, B.J., Yoon, C.N., Han, S.K.: Attack vulnerability of complex
  networks.
\newblock Physical Review E \textbf{65}(5), 056,109 (2002)

\bibitem{huan}
Huan, W., Hidenori, N.: Failure detection in p2p-grid environments.
\newblock In: Distributed Computing Systems Workshops (ICDCSW), 2012 32nd
  International Conference on, pp. 369--374 (2012).
\newblock \doi{10.1109/ICDCSW.2012.18}

\bibitem{TDGsIMC07}
Iliofotou, M., Pappu, P., Faloutsos, M., Mitzenmacher, M., Singh, S., Varghese,
  G.: Network monitoring using traffic dispersion graphs (tdgs).
\newblock In: Proceedings of the 7th ACM SIGCOMM Internet Measurement
  Conference, pp. 315--320. ACM, New York, NY, USA (2007).
\newblock \doi{http://doi.acm.org/10.1145/1298306.1298349}

\bibitem{jelasity2003newscast}
Jelasity, M., Kowalczyk, W., Van~Steen, M.: Newscast computing.
\newblock Tech. rep., Technical Report IR-CS-006, Vrije Universiteit Amsterdam,
  Department of Computer Science, Amsterdam, The Netherlands (2003)

\bibitem{PhysRevE.75.056115}
Kitsak, M., Havlin, S., Paul, G., Riccaboni, M., Pammolli, F., Stanley, H.:
  Betweenness centrality of fractal and nonfractal scale-free model networks
  and tests on real networks.
\newblock Phys. Rev. E \textbf{75}, 056,115 (2007).
\newblock \doi{10.1103/PhysRevE.75.056115}.
\newblock \urlprefix\url{http://link.aps.org/doi/10.1103/PhysRevE.75.056115}

\bibitem{massoulie}
Massoulie, L., Kermarrec, A.M., Ganesh, A.: Network awareness and failure
  resilience in self-organizing overlay networks.
\newblock In: Reliable Distributed Systems, 2003. Proceedings. 22nd
  International Symposium on, pp. 47--55 (Oct.).
\newblock \doi{10.1109/RELDIS.2003.1238054}

\bibitem{Melliar-Smith:2012}
Melliar-Smith, P.M., Moser, L.E., Michel~Lombera, I., Chuang, Y.T.: itrust:
  trustworthy information publication, search and retrieval.
\newblock In: Proc.~of the 13th Int.~Conf.~on Distributed Computing and
  Networking, ICDCN'12, pp. 351--366. Springer (2012)

\bibitem{peersim}
Montresor, A., Jelasity, M.: {PeerSim}: A scalable {P2P} simulator.
\newblock In: Proc. of the 9th Int. Conference on Peer-to-Peer (P2P'09), pp.
  99--100. Seattle, WA (2009)

\bibitem{newmanHandbook}
Newman, M.E.J.: Random graphs as models of networks, pp. 35--68.
\newblock Wiley-VCH Verlag GmbH and Co. KGaA (2005).
\newblock \doi{10.1002/3527602755.ch2}.
\newblock \urlprefix\url{http://dx.doi.org/10.1002/3527602755.ch2}

\bibitem{Newman200539}
Newman, M.J.: A measure of betweenness centrality based on random walks.
\newblock Social Networks \textbf{27}(1), 39 -- 54 (2005).
\newblock \doi{10.1016/j.socnet.2004.11.009}.
\newblock
  
\urlprefix\url{
http://www.sciencedirect.com/science/article/pii/S0378873304000681}

\bibitem{pandurangan}
Pandurangan, G., Raghavan, P., Upfal, E.: Building low-diameter peer-to-peer
  networks.
\newblock Selected Areas in Communications, IEEE Journal on \textbf{21}(6),
  995--1002 (2003).
\newblock \doi{10.1109/JSAC.2003.814666}

\bibitem{Pournarasb}
Pournaras, E., Warnier, M., Brazier, F.M.T.: Adaptive agent-based
  self-organization for robust hierarchical topologies.
\newblock In: Proc.~of the Int.~Conf.~on Adaptive and Intelligent Systems
  (ICAIS'09), pp. 69--76. IEEE (2009)

\bibitem{qiu}
Qiu, T., Chan, E., Chen, G.: Overlay partition: Iterative detection and
  proactive recovery.
\newblock In: Communications, 2007. ICC '07. IEEE International Conference on,
  pp. 1854--1859 (2007).
\newblock \doi{10.1109/ICC.2007.309}

\bibitem{radicchi2004}
Radicchi, F., Castellano, C., Cecconi, F., Loreto, V., Parisi, D.: {Defining
  and identifying communities in networks}.
\newblock Proceedings of the National Academy of Sciences \textbf{101}(9), 2658
  (2004)

\bibitem{vanRenesse:2009}
van Renesse, R., Minsky, Y., Hayden, M.: A gossip-style failure detection
  service.
\newblock In: Proceedings of the IFIP International Conference on Distributed
  Systems Platforms and Open Distributed Processing, Middleware '98, pp.
  55--70. Springer-Verlag, London, UK, UK (1998).
\newblock \urlprefix\url{http://dl.acm.org/citation.cfm?id=1659232.1659238}

\bibitem{SimontonCS06}
Simonton, E., Choi, B.K., Seidel, S.: Using gossip for dynamic resource
  discovery.
\newblock In: Proceedings of the 2006 International Conference on Parallel
  Processing (ICPP 2006), pp. 319--328 (2006)

\bibitem{stavrou}
Stavrou, A., Rubenstein, D., Sahu, S.: A lightweight, robust p2p system to
  handle flash crowds.
\newblock In: Network Protocols, 2002. Proceedings. 10th IEEE International
  Conference on, pp. 226--235 (2002).
\newblock \doi{10.1109/ICNP.2002.1181410}

\bibitem{gems}
Subramaniyan, R., Raman, P., George, A., Radlinski, M.: Gems: Gossip-enabled
  monitoring service for scalable heterogeneous distributed systems.
\newblock Cluster Computing \textbf{9}(1), 101--120 (2006).
\newblock \doi{10.1007/s10586-006-4900-5}.
\newblock \urlprefix\url{http://dx.doi.org/10.1007/s10586-006-4900-5}

\bibitem{tarkoma}
Tarkoma, S.: Overlay Networks - Toward Information Networking.
\newblock {CRC} Press (2010)

\bibitem{Terpstra:2007}
Terpstra, W.W., Kangasharju, J., Leng, C., Buchmann, A.P.: Bubblestorm:
  resilient, probabilistic, and exhaustive peer-to-peer search.
\newblock SIGCOMM Comput. Commun. Rev. \textbf{37}, 49--60 (2007).
\newblock \doi{http://doi.acm.org/10.1145/1282427.1282387}.
\newblock \urlprefix\url{http://doi.acm.org/10.1145/1282427.1282387}

\bibitem{voulgaris.jnsm.2005}
Voulgaris, S., Gavidia, D., {van Steen}, M.: Cyclon: Inexpensive membership
  management for unstructured p2p overlays.
\newblock Journal of Network and Systems Management \textbf{13}(2), 197--217
  (2005).
\newblock \doi{10.1007/s10922-005-4441-x}

\bibitem{Vu:2010}
Vu, L., Gupta, I., Nahrstedt, K., Liang, J.: Understanding overlay
  characteristics of a large-scale peer-to-peer iptv system.
\newblock ACM Trans. Multimedia Comput. Commun. Appl. \textbf{6}(4),
  31:1--31:24 (2010).
\newblock \doi{10.1145/1865106.1865115}.
\newblock \urlprefix\url{http://doi.acm.org/10.1145/1865106.1865115}

\bibitem{Wong:2008}
Wong, B., Guha, S.: Quasar: a probabilistic publish-subscribe system for social
  networks.
\newblock In: Proceedings of the 7th international conference on Peer-to-peer
  systems, IPTPS'08, pp. 2--2. USENIX Association, Berkeley, CA, USA (2008).
\newblock \urlprefix\url{http://dl.acm.org/citation.cfm?id=1855641.1855643}

\bibitem{phenix}
Wouhaybi, R., Campbell, A.: Phenix: supporting resilient low-diameter
  peer-to-peer topologies.
\newblock In: INFOCOM 2004. Twenty-third AnnualJoint Conference of the IEEE
  Computer and Communications Societies, vol.~1, pp. --119 (2004).
\newblock \doi{10.1109/INFCOM.2004.1354486}

\bibitem{zhuang}
Zhuang, S., Geels, D., Stoica, I., Katz, R.: On failure detection algorithms in
  overlay networks.
\newblock In: INFOCOM 2005. 24th Annual Joint Conference of the IEEE Computer
  and Communications Societies. Proceedings IEEE, vol.~3, pp. 2112--2123 vol. 3
  (2005).
\newblock \doi{10.1109/INFCOM.2005.1498487}

\end{thebibliography}

\end{document}